\documentclass[fleqn,usenatbib,useAMS]{mnras}

\usepackage{graphicx}    
\usepackage{amssymb,amsmath}    
\usepackage{multicol}        
\usepackage{bm}        
\usepackage{pdflscape}    
\usepackage{subfig}
\usepackage{caption}
\usepackage[inline]{enumitem}
\usepackage{geometry}
\usepackage{float}
\usepackage{hyperref}
\usepackage{breqn}
\usepackage{multirow}
\usepackage{cite}
\usepackage{mwe}

\newcommand{\kms}{\,km\,s$^{-1}$} 

\newcommand{\velociraptor}{\textsc{VELOCIraptor}}
\newcommand{\hi}{\textsc{Hi}}
\newcommand{\hmol}{H$_{\rm 2}$}

\newcommand{\shark}{\textsc{Shark}}
\newcommand{\lcdm}{$\Lambda \rm CDM$}
\newcommand{\surfs}{\textsc{surfs}}
\newcommand{\M}{${\rm M}_{\odot}$}
\newcommand{\solarValue}[1]{$10^{\rm #1}$\M}
\newcommand{\subsuperscript}[3]{$#1^{\rm #3}_{\rm #2}$}
\newcommand{\agefifty}{$z_{50}$}
\newcommand{\fracmvir}{$M_{\rm h}^{\rm sat}/M_{\rm vir}$ }

\usepackage[T1]{fontenc}
\usepackage{ae,aecompl}

\usepackage{newtxtext}

\title[The H\,{\textsc i}--halo Mass relation in \shark]{The physical drivers of the atomic hydrogen--halo mass relation}

\author[G. Chauhan et al.]{Garima Chauhan,$^{1,2}$\thanks{Contact e-mail: \href{mailto:garima.chauhan@icrar.org}{garima.chauhan@icrar.org}} Claudia del P. Lagos,$^{1,2}$ Adam R. H. Stevens,$^{1,2}$ 
\newauthor{Danail Obreschkow,$^{1,2}$ Chris Power,$^{1,2}$ Martin Meyer$^{1,2}$}
\\
$^{1}$International Centre for Radio Astronomy Research, The University of Western Australia, 35 Stirling Highway, Crawley, WA 6009, Australia \\
$^{2}$ARC Centre of Excellence for All Sky Astrophysics in 3 Dimensions (ASTRO 3D)}
\date{Last updated 2015 May 22; in original form 2013 September 5}

\pubyear{2020}

\begin{document}
\label{firstpage}
\pagerange{\pageref{firstpage}--\pageref{lastpage}}
\maketitle

\begin{abstract}

We use the state-of-the-art semi-analytic galaxy formation model, \shark, to investigate the physical processes involved in dictating the shape, scatter and evolution of the \hi--halo mass (HIHM) relation at $0 \leq z \leq 2$. We compare \shark\ with \hi~clustering and spectral stacking of the HIHM relation derived from observations finding excellent agreement with the former and a deficiency of \hi\ in \shark\ at $M_{\rm vir}\approx$ \solarValue{12-13} in the latter. In \shark, we find that the \hi\ mass increases with the halo mass up to a critical mass of $\approx$ \solarValue{11.8}; between $\approx$ \solarValue{11.8} and \solarValue{13}, the scatter in the relation increases by $0.7$~dex and the \hi\ mass decreases with the halo mass on average (till $M_{\rm vir}\sim 10^{12.5}\,\rm M_{\odot}$, after which it starts increasing); at $M_{\rm vir}\gtrsim 10^{13}\,\rm M_{\odot}$, the \hi\ content continues to increase with increasing halo mass, as a result of the increasing \hi\ contribution from satellite galaxies. We find that the critical halo mass of $\approx$ \solarValue{12} is set by feedback from Active Galactic Nuclei (AGN) which affects both the shape and scatter of the HIHM relation, with other physical processes playing a less significant role. We also determine the main secondary parameters responsible for the scatter of the HIHM relation, namely the halo spin parameter at \subsuperscript{M}{vir}{} $<$ \solarValue{11.8}, and the fractional contribution from substructure to the total halo mass (\fracmvir) for \subsuperscript{M}{vir}{} $>$ \solarValue{13}. The scatter at \solarValue{11.8} $<$ \subsuperscript{M}{vir}{} $<$ \solarValue{13} is best described by the black-hole-to-stellar mass ratio of the central galaxy, reflecting the relevance of AGN feedback. We present a numerical model to populate dark matter-only simulations with \hi\ at $0\le z\le 2$ based solely on halo parameters that are measurable in such simulations.  

\end{abstract}





\section{Introduction}
\label{sec:Introduction}

Understanding the distribution and evolution of neutral atomic hydrogen (\hi) in the Universe provides key insights into cosmology, galaxy formation and the epoch of cosmic reionisation \citep{Blanton2009-Environment_physical_properties,Pritchard2012,Somerville2015,Rhee2017}. A long-standing challenge in galaxy formation and evolution is addressing the relationship between stars, gas and metals in galaxies, haloes and the large-scale structure. \hi\ is a primary ingredient for star formation and a key input to understand how various processes govern galaxy formation and evolution. The \hi\ content of dark matter (DM) haloes forms an intermediate state in the baryon cycle that connects the largely ionised gas in the intergalactic medium (IGM), the shock heated gas at the virial radius and the star-forming, cold gas in the interstellar medium (ISM) of galaxies \citep{Putman2012GaseousHalos, Krumholz2012METALLICITY-DEPENDENTGALAXIES}. Constraints on \hi\ at all relevant scales (IGM, halo and galaxy scales) are therefore key to reveal the role of gas dynamics, cooling and regulatory processes such as stellar feedback, gas inflows and outflows \citep{Prochaska2009-HIgas_evolution, VandeVoort2011}, and the effect of environment in galaxy formation \citep{Fabello2012-environment_HIstacking, Zhang2013-cluster_gas_depletion}. 

When studying galaxy formation and evolution, the exploration of scaling relations is particularly useful as a way of reducing the inherent complexity of the process and providing a quantitative means of examining physical properties of galaxies. The dependence of the abundance of baryons on the host halo mass is considered one of the most fundamental scaling relations \citep{Wechsler2018-galaxyDMhaloes}. In particular, the stellar--halo mass relation has been studied in detail, and has been shown to have little scatter ($\approx 0.2$~dex, see \citealp{Behroozi2010-Stellar-halo-mass, Moster2010-StellarHalo-mass}) and a shape that reflects the mismatch between the halo and stellar mass functions - the latter has a much shallower low-mass end slope and a more abrupt break at the high-mass end than the former (see review \citealt{Wechsler2018-galaxyDMhaloes}). The scatter around these scaling relations is particularly useful because it helps to pinpoint how a halo's assembly history affects its baryon content \citep{Kulier2019-BaryonFraction-EAGLE,Mitchell2016-stellarHalo,Matthee2017-SMHM-EAGLE}.

Stellar mass can be inferred observationally for large statistical samples, unlike the gas content of galaxies and haloes. However, given that stellar mass is only a small contribution to the baryon content of the Universe \citep{Fukugita1998,Driver2018}, it is imperative to explore how the abundance of different gas phases correlate with halo mass.
\hi\ is particularly interesting because it is the intermediate state in the baryon cycle. The \hi--halo mass scaling relation (HIHM) is likely to be much more complex than the stellar--halo mass relation because observations show that the correlation between \hi\ mass and stellar mass is characterised by a large scatter (e.g. \citealt{Catinella2010,Brown2015TheGalaxies,Brown2017ColdClusters,Catinella2018}). This is implied by the work of \citet{Chauhan2019}, who used galaxy formation simulations to show that the correlation between \hi\ mass and \hi\ velocity width - a tracer of a galaxy's dynamical mass - is complex, with variations of $>2$~dex in \hi\ mass at fixed velocity width. 

Several empirical studies have inferred limits on the form of the HIHM relation. \citet{Eckert2017} attempted to measure the ``cold'' baryon mass (stars plus ISM mass) vs. halo mass relation, for which they combined $21$~cm-derived \hi\ masses with empirical estimates of the gas mass in galaxies based on the correlation between the \hi\ mass and optical colours in galaxies with detected \hi. The difficulty with this approach is the unknown systematic effects in the application of the empirical estimation to a wider parameter space than probed by actual \hi\ detections \citep[see][]{Eckert2015-Resolve-baryon-estimates}. Other approaches use \hi-clustering measurements to infer an HIHM relation \citep{Padmanabhan2017ConstrainingHydrogen,Obuljen2019TheFromALFALFA}, as well as \hi\ spectral stacking, which has been used to calculate the mean \hi\ content of groups identified in optical redshift surveys \citep{Guo2020}. \hi~clustering provides an indirect way of measuring the HIHM relation because it relies on abundance matching to match the \hi\ with the respective halo that will be expected to host galaxies of the observed \hi\ mass. In contrast, \hi\ stacking provides a direct measurement of the {\it mean} \hi\ mass inside haloes of a given mass, typically using an estimate of the halo radius to choose the stacking area. However, it relies on group finders and halo mass estimates based on optical redshift surveys and so care must be taken because of the well known issue that optically selected and \hi-selected galaxies do not fully overlap, such that \hi-selected surveys typically miss the most massive, gas-poor galaxies (e.g. \citealp{deBlok1996-LSBvsHI,Schombert2001-LSBvsHI}). The HIHM relation is also expected to differ from the stellar--halo mass relation because, as previous work has shown, the distribution of \hi-selected galaxies depend not only on halo mass but also on the halo's formation history \citep{ Gao2005TheClustering, Guo2017ConstrainingClustering} and on halo spin parameter \citep{Maddox2015VariationMass, Obreschkow2016AngularDisks, Lutz2018TheGalaxies}.  

While these observational inferences provide highly valuable constraints on the {\it average} HIHM relation, they do not constrain the scatter. The HIHM relation has been investigated extensively using different theoretical models, including semi-analytic models of galaxy formation \citep{Kim2017TheGalaxies,Baugh2018-PMillennium,Spinelli2019_Marta} and hydrodynamical simulations \citep{Villaescusa-Navarro2018}, which have consistently shown that the HIHM relation is characterised by a large scatter (especially in the region \solarValue{12} $<$\subsuperscript{M}{vir}{} $<$ \solarValue{13}) - much larger than the stellar--halo mass relation, by $>0.5$~dex. However, the predicted scatter of the HIHM appears to be largely model-dependent and no observational constraints have been obtained yet. For instance, both \citet{Baugh2018-PMillennium} and \citet{Spinelli2019_Marta} attribute the scatter in the relation to feedback from Active Galactic Nuclei (AGN), which suppresses gas cooling in the halo, preventing further gas accretion onto the central galaxy. \citet{Spinelli2019_Marta} also find that the HIHM relation depends on the detailed assembly history of haloes, which agrees with inferences based on  \hi\ clustering studies in  \citet{Guo2017ConstrainingClustering}. \citet{Villaescusa-Navarro2018}, using the IllustrisTNG hydrodynamical simulations, also report a larger scatter in their HIHM relation at \subsuperscript{M}{vir}{} between \solarValue{12}--\solarValue{13}, compared to what is found for the stellar-halo mass relation in their simulation. 

The current paucity of observational constraints on the shape, scatter and evolution of the HIHM is likely to change in the coming decade, ultimately with the Square Kilometer Array (SKA; see \citealp{Abdalla2005-SKA-paper}), but also with its pathfinders (e.g. MeerKAT, see \citealp{Holwerda2011} and the Australian SKA Pathfinder, ASKAP; see \citealp{Duffy2012,Koribalski2020}). With these transformational instruments on the horizon, it is imperative that we use current galaxy formation models and simulations to explore the physics shaping the HIHM relation to offer predictions and aid the interpretation of these upcoming observations. This is the main motivation of this paper. 

Another important challenge is the fact that the SKA is expected to probe cosmological volumes much larger than those we currently use to study galaxy formation \citep{Power2015GalaxyArray}, even in the case of semi-analytic models of galaxy formation - whose accessible volumes are already $2$-$3$ orders of magnitude larger than what we can reliably do with hydrodynamical simulations. In the case of semi-analytic models, the typically used cosmological volumes are usually limited by the fact that we require enough resolution to accurately model the assembly and growth history of the haloes. The challenge is even greater if we focus on cosmological studies with the SKA, which require thousands of statistical realisations of the universe with trustworthy models describing how to populate haloes with \hi\ mass.
This demands a physically motivated way of populating DM-only simulations with \hi\ without the need of running computationally expensive physical galaxy formation models on them. This is an important second motivation for our work.

These motivations require an in-depth exploration of the astrophysical processes that shape the HIHM relation and the development of an analytical model for how to populate dark matter haloes with \hi. We aim to understand what physical parameters are responsible for how \hi\ populates haloes, and what drives the shape and scatter of the relation. For this, it is necessary to assess how the baryon physics included in galaxy formation simulations and halo formation history affect the HIHM relation across cosmic time. We explore which (other) halo properties affect the HIHM relation (e.g. spin, substructure mass fraction etc.). We do this by the use of the {\sc Shark} semi-analytic model of galaxy formation \citep{Lagos2018-Shark} and leverage its modularity and flexibility to test the effect of different physical models and parameters on the shape of the HIHM relation. We expect our numerical model showing how to populate DM haloes with \hi\ to be beneficial for designing \hi-stacking and \hi-intensity mapping experiments.

The structure of this paper is as follows. Section~$2$ summarises the relevant features of \shark. Section~$3$ validates our semi-analytic model against the local Universe \hi\ observations that capture the average HIHM relation. In Section~$4$, we delve into the properties responsible for the shape and scatter of the HIHM relation, and see how much impact these properties have. In Section~$5$, we present our physically motivated HIHM relation along with providing information on its evolution with redshift. We draw conclusions in Section~$6$. The Appendices show how the HIHM relation evolves with redshift and provide tabulated fits to populated halos with \hi\ mass.


\section{Modelling the \textsc{Hi} content of galaxies and  Haloes}
\label{sec:HI_content_haloes}
In this section, we describe the semi-analytical model that is used in the study, and which prescriptions are applied to calculate the \hi\  content of galaxies and haloes. The result of using these models are discussed in Section \ref{sec:Understanding-HI-halo}.

\subsection{The \shark\ semi-analytical model of galaxy formation}
\label{subsec:SAM}
We use the semi-analytical model of galaxy formation (SAM), \shark\ \citep{Lagos2018-Shark}. SAMs use halo merger trees, which are produced from a cosmological DM only $N$-body only simulation, and follow the formation and evolution of galaxies by solving a set of equations that describe all the physical processes that (we think) are relevant for the problem (see reviews by \citealt{Baugh2006, Somerville2015}).

\shark\footnote{https://github.com/ICRAR/shark/} is an open-source, flexible and highly modular SAM that models the key physical processes of galaxy formation and evolution. These include \begin{enumerate*}[label=(\roman*)]
    \item the collapse and merging of DM haloes;
    \item the accretion of gas onto haloes, which is governed by the DM accretion rate;
    \item the shock heating and radiative cooling of gas inside DM haloes, leading to the formation of galactic discs via conservation of specific angular momentum of the cooling gas;
    \item the formation of a multi-phase interstellar medium and subsequent star formation (SF) in galaxy discs;
    \item the suppression of gas cooling due to photo-ionisation;
    \item chemical enrichment of stars and gas;
    \item stellar feedback from evolving stellar populations;
    \item the growth of supermassive black holes (SMBH) via gas accretion and merging with other SMBHs;
    \item heating by AGN;
    \item galaxy mergers driven by dynamical friction within common DM haloes, which can trigger bursts of SF and the formation and/or growth of spheroids; and
    \item the collapse of globally unstable discs leading to bursts of SF and the creation and/or growth of bulges. 
\end{enumerate*} 
\shark\ also includes several different prescriptions for gas cooling, AGN feedback, stellar and photo-ionisation feedback, and SF. 

Using these models, \shark\ computes the exchange of mass, metals, and angular momentum between the key baryonic reservoirs in haloes and galaxies, which include hot and cold halo gas, the galactic stellar and gas discs and bulges, central black holes, as well as the ejected gas component that tracks the baryons that have been expelled from haloes. In Section~\ref{subsec: HI_SAM_calculation}, we describe in detail the modelling of star formation, AGN feedback, stellar feedback, reionisation, and gas stripping in satellite galaxies, all of which are relevant for the discussions in Sections \ref{sec:Understanding-HI-halo}, \ref{sec:Model-development} and \ref{sec:Discussion}. 

The models and parameters used in this study are the \shark\ defaults, as described in \citet{Lagos2018-Shark} and used in \citet{Chauhan2019} to study the \hi\ content of galaxies. These have been calibrated to reproduce the $z = 0,\ 1,\ \text{and}\ 2$ stellar mass functions; the $z = 0$ black hole-bulge mass relation; and the disc and bulge mass-size relations. This model also successfully reproduces a range of observational results that are independent of those used in the calibration process. These include the total neutral, atomic and molecular hydrogen-stellar mass scaling relations at $z=0$; the cosmic star formation rate (SFR) density evolution up to $z \approx 4$; the cosmic density evolution of the atomic and molecular hydrogen at $z \leq 2$ or higher in the case of the latter; the mass-metallicity relations for gas and stellar content; the contribution to the stellar mass by bulges; and the SFR--stellar mass relation in the local Universe. \citet{Davies2018} show that \shark\ reproduces the scatter around the main sequence of star formation in the SFR--stellar mass plane; \citet{Chauhan2019} show that \shark\ can reproduce the \hi~mass and velocity widths of galaxies observed in the ALFALFA survey; and  \citet{Amarantidis2019} show that the predicted AGN luminosity functions (LFs) agree well with observations in X-rays and radio wavelengths.

In addition, \citet{Lagos2019-SED} has shown that \shark\ can reproduce the panchromatic emission of galaxies throughout cosmic time; most notably, \shark\ reproduces the number counts from GALEX UV to the JCMT $850$ microns band, the redshift distribution of sub-millimetre galaxies, and the ALMA bands number counts \citep{Lagos2020}.  \citet{Bravo2020} show that \shark\ also reproduces reasonably well the optical colour distribution of galaxies across a wide range of stellar masses and redshift, as well as the fraction of passive galaxies as a function of stellar mass.

We use the \surfs\ suite of DM only $N$-body simulations for our study \citep{Elahi_SURFS}, which consist of $N$-body simulations of differing volumes, from $40$ to $210$ \subsuperscript{h}{}{-1}\ cMpc on a side, and particle numbers, from $\sim$130 million up to $\sim$8.5 billion particles. The simulations adopt the \lcdm~\textit{Planck} cosmology (Planck Collaboration XIII \citeyear{PlanckXIII}), which assumes total matter, baryon, and dark energy densities of $\Omega_{\rm m} = 0.3121$, $\Omega_{\rm b} = 0.0491$ and $\Omega_{\rm \Lambda } = 0.6751$, and a dimensionless Hubble parameter of $h = 0.6751$. 

For this analysis, we use the L40N512 and L210N1536 runs, referred to as micro-\surfs\ and medi-\surfs, respectively and whose properties are described in Table \ref{tab:sims}. By using two different resolution runs of different volumes, we can probe over 6 orders of magnitude in DM halo mass, thus giving us an optimal dynamic range for exploring the HIHM scaling relation.
We show the results of \shark\ using micro-\surfs\ at halo masses below \solarValue{11.2}, while medi-\surfs\ is used for higher halo masses. This transition mass is chosen as according to   \citet{Elahi_SURFS} at this mass haloes in medi-\surfs\ comprise $\ge 200$ particles, making them reliable for our calculation (because their merger trees will be sufficiently well resolved).
Merger trees and halo catalogues were constructed using the phase-space finder \velociraptor\ \citep{Elahi2019-Velociraptor, Canas2019-VELOCIRAPTOR} and the halo merger tree code \textsc{treefrog} \citep{Poulton_Treefrog2018,Elahi2019-TreeFrog}.

We define three types of galaxies in our analysis: \textit{centrals}, \textit{satellites} and \textit{orphans}. \shark\ uses the merger trees and subhalo catalogues as a skeleton, that is required to evolve our galaxies, and so we use this information to describe our galaxy types as well. In \shark, the central subhalo of every halo in the catalogue is defined as the most massive subhalo of every existing halo at $z=0$, and then subsequently making the main progenitor of those centrals as the centrals of their respective halo. Every subhalo/halo is connected to its progenitor(s) and descendant subhalo/halo, which is connected to the merger tree they belong to. Haloes point to their central and satellite subhaloes, with the subsequent subhaloes pointing to the list of galaxies they may contain. Following the subhalo and merger tree information, we define \textit{centrals} or \texttt{type=0} to be the central galaxy of the central subhalo. We only allow the central subhaloes to host the central galaxy, which in turn becomes the central galaxy of the hosthalo. The \textit{satellite} or \texttt{type = 1} galaxies are the central galaxies of the other existing subhaloes for that hosthalo (satellite subhaloes). The galaxies belonging to a subhalo that merges onto another one and is not the main progenitor become the \textit{orphan} or \texttt{type = 2} galaxies. A central subhalo in \shark\ can have only one central galaxy and any number of orphan galaxies, whereas the satellite subhalo can only have one \texttt{type = 1} galaxy. When a subhalo becomes a satellite subhalo, any orphan galaxies in that subhalo are transferred to the central subhalo.  

\begin{table}
\makeatletter
 \def\@textbottom{\vskip \z@ \@plus 1pt}
 \let\@texttop\relax
\makeatother

        \setlength\tabcolsep{2pt}
        \centering\footnotesize
        \caption{\surfs\ simulation parameters of the runs being used in this paper. We refer to L40N512 and L210N1536 as micro-\surfs\ and medi-\surfs, respectively.}
        \begin{tabular}{@{\extracolsep{\fill}}l|cccc|p{0.45\textwidth}}
                \hline
                \hline
                Name & Box size & Number of & Particle Mass & Softening Length\\
                & $L_{\rm box} [\rm cMpc/h]$ & Particles $N_{\rm p}$ & $m_{\rm p}$ [\rm \M/h]  & $\epsilon [\rm ckpc/h]$ \\
                 \hline
                L40N512     & $40$  & $512^3$   & $4.13\times10^7$ & 2.6  \\
                L210N1536   & $210$ & $1536^3$  & $2.21\times10^8$ & 4.5  \\
                \hline
        \end{tabular}
        \label{tab:sims}
        
\end{table}

\subsection{Halo properties as calculated in \shark}
\label{subsec: Halo_properties_SAM}

\shark\ assumes the masses of DM haloes
(\subsuperscript{M}{halo}{}) to be those calculated by \velociraptor. The virial mass is defined as $M_{\rm halo} \equiv M_{\rm 200} = 4\pi R^3_{\rm 200} \Delta\rho_{\rm crit}/3$, with $\rho_{\rm crit}$ being the critical density of the universe, with $M_{\rm 200}$ and $R_{\rm 200}$ being the mass and radius of the halo, respectively, when the density within the halo becomes $200$ times of the critical density of the universe. It is assumed that the mass profile of the halo follows an NFW profile \citep{Navarro1997-HierarichalClustering}. The halo concentration is estimated using the \citet{Duffy2008} relation between concentration, the halo's virial mass and redshift. The spin parameter of the haloes are drawn from a log-normal distribution of mean $0.03$ and width $0.5$. These parameters correspond to those measured in \surfs\ with the well-resolved haloes \citep{Elahi_SURFS}. 

\subsection{Modelling of key physical processes in \shark}
\label{subsec: HI_SAM_calculation}
As stated in Section \ref{subsec:SAM}, \shark\ is a modular SAM, and so the user can adopt a range of models for different physical processes. Although we use the default \shark\ model for the derivation of the HIHM scaling relation, we also want to understand what drives the shape of the HIHM relation, and so varying the models and parameters adopted in \shark\ is necessary. Here, we describe a subsample of the models and physical processes that are relevant for the HIHM relation.

We compare the \hi\ in haloes based on two different ISM gas-phase models, different AGN and stellar feedback efficiencies, and different ram pressure stripping considerations, as well as altering the photoionisation of \hi\ in haloes.

\subsubsection{Gas phases in the interstellar medium and star formation}
\label{subsubsec:SF_models}

In the default \shark\ model, hereafter referred to as \shark-ref, we use the prescription described in \citet{Blitz2006}, hereafter referred to as BR06, to compute the amount of atomic and molecular hydrogen (\hi and \hmol, respectively) in the gas disc and bulge of the galaxy. The gas, once it cools, is assumed to settle in an exponential disc of half-mass radius, \subsuperscript{r}{gas,disc}{}. In BR06 the ratio of the molecular to atomic hydrogen gas surface density in galaxies is a function of the local hydro-static pressure in the mid-plane of the disc, with a power-law index close to 1, 
\begin{equation}
    R_{\rm mol} \equiv \frac{\Sigma_{\rm H_2}}{\Sigma_{\rm HI}} = \left( \frac{P}{P_0} \right)^{\alpha_{\rm P}} ,
\end{equation} 

\noindent where \subsuperscript{P}{0}{} and $\alpha_{\rm P}$ are parameters measured in observations and have values $P_0/\kappa_{\rm B} = 1,500-40,000 \rm cm^{-3}\rm K$ and $\alpha_{\rm P} \approx 0.7-1$ \citep{Blitz2006,Leroy2008}. The hydrostatic pressure from the surface densities of gas and stars is calculated following \citet{Elmegreen1989}, 
\begin{equation}
    P = \frac{\pi}{2} G \Sigma_{\rm gas} \left (\Sigma_{\rm gas} + \frac{\sigma_{\rm gas}}{\sigma_{\star}} \Sigma_{\star} \right),
\end{equation}

\noindent where $\Sigma_{\rm gas}$ and $\Sigma_{\star}$ are the total gas (atomic, molecular and ionised) and stellar surface densities, respectively, and $\sigma_{\rm gas}$ and $\sigma_{\star}$ are the gas and stellar velocity dispersions. The stellar surface density is assumed to follow an exponential profile with a half-mass stellar radius of $r_{\rm \star,disc}$. We adopt $\sigma_{\rm gas} = 10$\kms and calculate $\sigma_{\star} = \sqrt{\pi G h_{\star} \Sigma_{\star}}$, where $h_{\star} = r_{\star}/7.3$ \citep{Kregel2002FlatteningGalaxies}, with $r_{\star}$ being the half-stellar mass radius. The \hi\ surface densities cannot extend to infinitely small surface densities because the UV background will ionise very low-density gas; thus a minimum threshold of $\Sigma_{\rm thresh} = 0.1 \rm M_{\odot}\,pc^{-2}$ is applied, following the results of the hydrodynamical simulations of \citet{Gnedin2012ONGALAXIES}. All the gas at lower densities is considered to be ionised.

In order to understand how the default \shark\ ISM prescription works against another available ISM model in \shark, we carry out another run using an alternative
prescription - in this case, \citet{Gnedin2014LINEGALAXIES}, hereafter referred to as GD14.
The GD14 model uses the dust-to-gas ratio, \subsuperscript{D}{MW}{}, and the local radiation field, \subsuperscript{U}{MW}{}, with respect to that of the solar neighbourhood, to estimate the ratio of \hi\ to \hmol\ in the gas disc. These two parameters are estimated as $D_{\rm MW} = Z_{\rm gas}/Z_{\odot}$ and {\bf $U_{\rm MW} = \Sigma_{\rm SFR}/\Sigma_{\rm MW}$}, where \subsuperscript{Z}{gas}{}\ is the metallicity of the ISM. The values $Z_{\odot} = 0.134$ \citep{Asplund2009} and $\Sigma_{\rm MW} = 2.5\, \rm M_{\odot} \rm yr^{-1}$ \citep{Bonatto2011ConstrainingClusters} are estimates from the solar neighbourhood. Hence, $D_{\rm MW}$ and $U_{\rm MW}$ are quantities that vary with galaxy properties.
Using the argument presented in \citet{Wolfire2003NeutralGalaxy}, where it is stated that the pressure balance between the warm and the cold neutral media can only be achieved if the density is larger than a minimum density, we can approximate the minimum density to be proportional to \subsuperscript{U}{MW}{}. Hence, assuming that the pressure equilibrium between warm/cold media is a necessity for the formation of ISM, then $U_{\rm MW}\ \text{will be proportional to}\ \rho_{\rm gas}$, with $\rho_{\rm gas}$ being the gas density. As galaxies show an almost constant $\sigma_{\rm gas}$, it can be assumed that the gas scale height is also close to constant, which allows us to replace $\rho_{\rm gas}$ by $\Sigma_{\rm gas}$ above. Based on \subsuperscript{D}{MW}{} and \subsuperscript{U}{MW}{} we calculate \subsuperscript{R}{mol}{} following \citet{Gnedin2014LINEGALAXIES}, 
\begin{equation}
    R_{\rm mol} = \left(\frac{\Sigma_{\rm gas}}{\Sigma_{\rm R = 1}} \right)^{\alpha_{\rm GD}},
\end{equation}
where \begin{equation}
    \alpha_{\rm GD} = 0.5 + \frac{1}{1 + \sqrt{U_{\rm MW}D^{2}_{\rm MW}/600}} ,
\end{equation}
\begin{equation}
    \Sigma_{\rm R = 1} = \frac{50 \rm M_{\odot} \rm pc^{-2}}{g} \frac{\sqrt{0.01 + U_{\rm MW}}}{1 + 0.69\sqrt{0.01 + U_{\rm MW}}} ,
\end{equation}
and  
\begin{equation}
    g = \sqrt{D^{2}_{\rm MW} + D^{2}_{\star}} .
\end{equation}
Here, $D_{\star} \approx 0.17$ for scales $> 500 \rm pc$.

Independent of how the \textsc{Hii}/\hi/\hmol\ is computed, our default star formation model assumes the SFR surface density to be proportional to the \hmol surface density. The SFR surface density is then calculated by assuming a constant depletion time for the molecular gas, following 
\begin{equation}
    \Sigma_{\rm SFR} = \nu_{\rm SF} f_{\rm mol} \Sigma_{\rm gas}.
    \label{eq:SFmodel}
\end{equation}
Here, $\nu_{\rm SF}$ is the inverse of the \hmol\ depletion timescale with $f_{\rm mol} = \Sigma_{\rm mol}/\Sigma_{\rm gas}$, where $\Sigma_{\rm mol}$ is the molecular gas surface density and $\Sigma_{\rm gas}$ is the total gas surface density; $\Sigma_{\rm SFR}$ is integrated over a radii range of $0-10$~\subsuperscript{r}{gas,disc}{}. Equation~\ref{eq:SFmodel} applies to both the BR06 and GD14 models. Note that two different values of $\nu_{\rm SF}$ are adopted in \shark. For star formation in disks, $\nu_{\rm SF}=1\,\rm Gyr^{-1}$, while for starbursts triggered by galaxy mergers and disk instabilities, $\nu_{\rm SF}=0.1\,\rm Gyr^{-1}$ \citep[these values are based on ][]{Sargent2014-MolecularGasRedshiftIndependence}. This is motivated by the bimodality observed in the $\Sigma_{\rm SFR}-\Sigma_{\rm mol}$ plane for normal star-forming galaxies and starbursts \citep{Genzel2010}.

\subsubsection{AGN feedback}
\label{subsubsec:AGN_models}

AGN feedback influences the amount of gas that cools and hence replenishes the ISM content of galaxies. The default AGN feedback model used in \shark\ is that of \citet{Croton2016}, hereafter referred to as Croton16. Croton16 assumes a Bondi-Hoyle \citep{Bondi1952} like accretion mode 
\begin{equation}
    \dot{M}_{\rm BH,hh} = 2.5 \pi G^2 \frac{m^2_{\rm BH}\rho_0}{c^3_{\rm s}}
\end{equation}
where \subsuperscript{c}{s}{}\ and $\rho_{\rm 0}$ are the sound speed and average density of the hot gas in the halo that accretes on to the SMBH, respectively, where \subsuperscript{c}{s}{} $\approx$ \subsuperscript{V}{vir}{} and  \subsuperscript{V}{vir}{} is the halo's virial velocity. $\dot{M}_{\rm BH,hh}$ is the accretion rate calculated for the hot-halo mode, as described below.
$\rho_{0}$ is calculated by equating the sound travel time across a shell of diameter twice the Bondi radius to the local cooling time. This is also termed the ``maximal cooling flow'' by \citet{Nulsen-Fabian2000}, which leads to 
\begin{equation}
    \dot{M}_{\rm BH,hh} = \kappa_{\rm agn} \frac{15}{16} \pi G \mu m_{\rm p} \frac{\kappa_{\rm B} T_{\rm vir}}{\Lambda} m_{\rm BH}.
    \label{eq:agn_feedback}
\end{equation}
$\kappa_{\rm agn}$ is a free parameter that was introduced in \citet{Croton2006TheGalaxies} to counteract the approximations used to derive the accretion rate. $\kappa_{\rm B}$ and $\Lambda$ are the Boltzmann constant and the cooling function that depends on \subsuperscript{T}{vir}{}\ and the hot gas metallicity. From Equation~\ref{eq:agn_feedback}, we can
estimate the BH luminosity (\subsuperscript{L}{BH}{}) in this accretion mode, which in turn is used to calculate the heating provided by the BH for the halo as shown,
\begin{equation}
    \dot{M}_{\rm heat} = \frac{L_{\rm BH}}{0.5V^2_{\rm vir}},
 \end{equation}
where $L_{\rm BH} = \eta \dot{M}_{\rm BH,hh} c^2$, with $\eta$ and $c$ being the luminosity efficiency (based on \citealt{Lagos-Padilla-Nelson2009}) and speed of light, respectively. 

To understand the effect of the AGN feedback, we vary the value of the free parameter $\kappa_{\rm agn}$ between $0$ (no AGN feedback) to $1$. Note that the default value in \shark\ is $0.002$.


\subsubsection{Stellar feedback}
\label{subsubsec:Stellar_Feedback}

The stellar feedback in \shark\ is separated into two main components: the outflow rate of the gas that escapes from the galaxy, $\dot{m}_{\rm outflow}$, and the ejection rate of the gas that escapes from the halo, $\dot{m}_{\rm ejected}$. \citet{Lagos2018-Shark} describes $\dot{m}_{\rm outflow} = \psi f(z, V_{\rm circ})$, where $\psi$ is the instantaneous SFR, $z$ is the redshift and \subsuperscript{V}{circ}{} is the maximum circular velocity of the galaxy, where the ejection rate is $>\ 0$ only when the total injected energy of the outflow is greater than the binding energy of the halo. The terminal wind velocity,  \subsuperscript{V}{w}{}, is based on the FIRE simulation suite \citep{Muratov2015},
\begin{equation}
    \frac{V_{\rm w}}{\rm km s^{-1}} = 1.9 \left(\frac{V_{\rm circ}}{\rm km s^{-1}} \right)^{1.1} .
\end{equation}
The terminal wind velocity is required to compute the excess energy that will be used to eject the gas out of the halo:
\begin{equation}
    E_{\rm excess} = \epsilon_{\rm halo} \frac{V^{2}_{\rm w}}{2}f(z, V_{\rm circ}),
\end{equation}
where $\epsilon_{\rm halo}$ is a free parameter. The net ejection rate can then be calculated as,
\begin{equation}
    \dot{m}_{\rm ejected} = \frac{E_{\rm excess}}{V_{\rm circ}^2/2} - \dot{m}_{\rm outflow}. 
    \label{eq:vcirc_wrt_ejected_gas}
\end{equation}
If $\dot{m}_{\rm ejected} < 0$ no ejection from the halo takes place and we limit $\dot{m}_{\rm outflow} = E_{\rm excess}/(V_{\rm circ}^2/2)$. 

In \shark-ref, we use the modelling presented in \citet{Lagos2013}, referred to as Lagos13, where they follow the evolution of the expansion of SNe driven bubbles from an early epoch of adiabatic expansion to the momentum-driven phase of expansion. They used this model to estimate $\dot{m}_{\rm outflow}$ and find,
\begin{equation}
    f = \epsilon_{\rm disc} \left(\frac{V_{\rm circ}}{v^{\prime}_{\rm hot}}\right)^{\beta},
    \label{eq:stellar_feedback}
\end{equation}
\begin{equation}
    v^{\prime}_{\rm hot} = v_{\rm hot}(1 + z)^{\rm z_{\rm P}}
\end{equation}

\shark-ref uses the default values as described in \citet{Lagos2018-Shark} with $\epsilon_{\rm disc} = 1$ and $z_{\rm P} =0.12$. We vary the value of $\beta$ from $0.5$ to $5$ in increments of $1$, with the default value in \shark-ref being $4.5$, to understand how stellar feedback influences the amount of \hi\ in haloes. For the no-stellar-feedback run, we set $\epsilon_{\rm disc} = 0$.


\subsubsection{Photoionsation feedback}
\label{subsubsec:Photoionisation_Feedback}

Photoionisation feedback refers to the
feedback arising from the ionising radiation background produced by the first generation of stars, galaxies and quasars during the epoch of reionisation. The large ionising radiation density affects small haloes, keeping the baryon temperature higher than the virial temperature, thus suppressing radiative cooling.  

\shark-ref follows the results of the one-dimensional collapse simulations of \citet{Sobacchi2013TheReionization}, which suggest that the effects of reionisation can be captured by allowing only those haloes that satisfy a redshift-dependent threshold velocity to be occupied. \shark-ref uses the the \textit{Sobacchi \& Mesinger parametric} form, as adapted by \citet{Kim2015}, which depends on the halo's \subsuperscript{V}{circ}{}\ based on the spherical collapse model of \citet{Cole1996} instead. This predicts $M_{\rm halo} \propto V^{3}_{\rm circ}$. Thus, haloes with circular velocities below $v_{\rm thresh}(z)$ are not allowed to cool their halo gas, where
\begin{equation}
    v_{\rm thresh}(z) = v_{\rm cut}(1 + z)^{\alpha_{v}} \left[1 - \left(\frac{1 + z}{1 + z_{\rm cut}}\right)^2 \right]^{2.5/3}. 
    \label{eq:photoionisation_feedback}
\end{equation}
Here, \subsuperscript{v}{cut}{}, \subsuperscript{z}{cut}{}\ and $\alpha_{\rm v}$ are free parameters that are constrained by the \citet{Sobacchi2013TheReionization} simulation. We use different \subsuperscript{v}{cut}{}\ values, ranging from $20$ \kms to $50$ \kms, to study the effect on the \hi\ content of the haloes. The default value in \shark-ref is $35$~\kms. We keep the other two parameters fixed to $z_{\rm cut} = 10$ and $\alpha_{\rm v} = -0.2 $, which are default in \shark-ref.

\subsubsection{Gas Stripping in Satellite galaxies}
\label{subsubsec:stripping}

Following the model of ``instantaneous ram-pressure stripping'' described in \citet{Lagos2014-stripping}, \shark\ assumes that as soon as galaxies become satellites, their halo gas is instantaneously stripped and transferred to the hot gas of the central galaxy, a process that is commonly referred to as ``strangulation". Thus, gas can only accrete onto the central galaxy in the halo and not onto satellite galaxies. Cold gas in the discs of galaxies is not stripped. 
\shark\ also allows us to switch off this process, in turn assuming that satellite galaxies can retain their hot halo gas, and hence their ISM can continue to be replenished for some time, until their halo gas reservoir is exhausted. We note that the quenching of satellites also happens in this case as satellite subhaloes, where satellite galaxies reside, are cut off from cosmological accretion, and hence their halo gas reservoir is not replenished. We test the effect of turning on and off the ``instantaneous ram pressure stripping'' on the overall \hi\ mass contained in haloes, with stripping `on' being used in \shark-ref.  

Regardless of whether or not the stripping is `on' or `off', the gas that is ejected from satellite galaxies due to stellar feedback is transferred to the ejected gas reservoir of the central galaxies, and hence that gas cannot be reincorporated into the hot halo gas of the satellites.


\section{Validation of the \shark\  model against local Universe \hi\  observations and previous models}
\label{sec:General_overview}

In this section, we describe how the total \hi\ in the haloes compares with available observations, with the aim of validating the model before we analyse in detail what drives the shape and scatter of the HIHM relation. In particular, we compare with the observed HIHM relation (Section \ref{subsec:comapring_HI_halo}) and \hi\ clustering (Section \ref{subsec:Correlation_function}).
We remind the reader that previous papers have shown that \shark-ref reproduces well the \hi\ mass function, \hi--stellar mass scaling relation \citep{Lagos2018-Shark}, \hi\ mass and velocity width distributions and the \hi\ mass--velocity width relation observed in ALFALFA \citep{Chauhan2019}.

\subsection{The local Universe HIHM relation}
\label{subsec:comapring_HI_halo}

\begin{figure}
\centering
\includegraphics[width=0.51\textwidth]{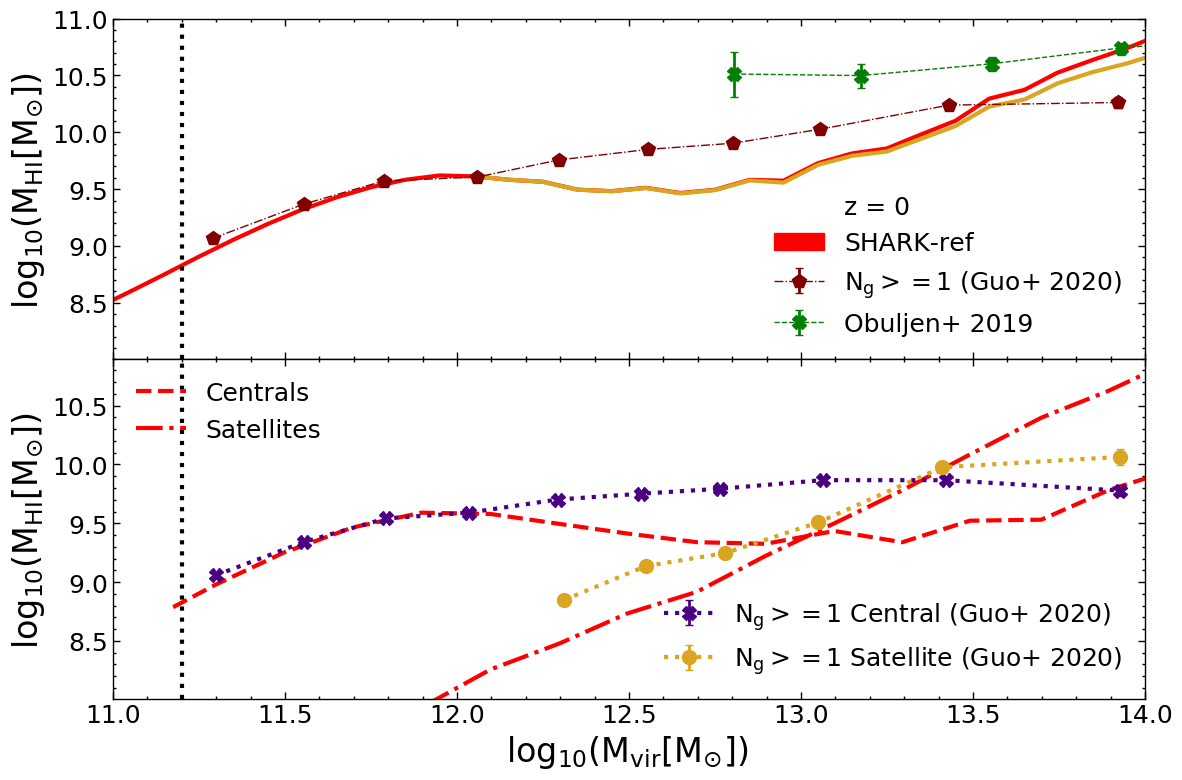}
\caption{The mean of the total \hi\ content in halos as a function of halo mass at $z=0$. In the upper-panel, the red line shows predictions from \shark-ref, with the vertical dashed line showing the convergence point between micro-\surfs\ and medi-\surfs. The yellow-line shows the \hi\ contained in subhaloes that are associated with the host-halo and are within one virial radius of the host halo. The symbols with error bars show the observed values of \hi\ shown in \citet{Guo2020} and \citet{Obuljen2019TheFromALFALFA}, as labelled. Note that \shark-ref predicts the \hi\ content of $N_{\rm g} \geq 1$ reasonably well until \subsuperscript{M}{vir}{} $\approx$ \solarValue{12}, with all the points agreeing with \shark-ref, at which point \shark-ref  starts to deviate from the \citet{Guo2020} and \citet{Obuljen2019TheFromALFALFA} points, either over-predicting or under-predicting the content at various points. The lower-panel shows the central and satellite \hi\ contribution from \citet{Guo2020} compared with \shark-ref. We see the centrals agreeing with \shark-ref until \subsuperscript{M}{vir}{} $\approx$ \solarValue{12}, but the satellite population agrees reasonably well with \shark-ref over the entire range. }
\label{fig:HI_showing_plot}
\end{figure}

In Figure~\ref{fig:HI_showing_plot}, we compare the results from \shark-ref with observations. We use the results shown in \citet{Guo2020}, where they calculate the \hi\ content of groups from the Sloan Digital Sky Survey DR7 Main Galaxy \citep[SDSS,][]{Lim2017-SDSS-group} sample by stacking the \hi\ spectra obtained from ALFALFA survey. SDSS is a major multi-spectral and spectroscopic redshift survey that covers over 35\% of the sky. We use data from the main SDSS galaxy survey, which is sensitive to $17.77$ r-band magnitude. The ALFALFA (Arecibo Legacy Fast ALFA) survey, on the other hand, is a blind \hi\ survey covering $6900$ \subsuperscript{\rm deg}{}{2} in the Northern Hemisphere, with $\sim 31,000$ direct \hi\ detections \citep{Giovanelli2005,Haynes2018TheCatalog} and going out to redshift $z = 0.06$. 

\citet{Guo2020} use the SDSS DR7 group catalogue to identify galaxies with available spectroscopic redshifts, which is about 98 per cent complete. The halo masses of these groups were calculated using the proxy of galaxy stellar mass, with the halo radius, \subsuperscript{r}{200}{}, estimated from the definition that the mean mass density within \subsuperscript{r}{200}{} is $200$ times the mean density of the universe at a given redshift. For stacking the \hi\ for these groups and galaxies, they use ALFALFA \texttt{IDL} \citep[see][]{Fabello2011}, which integrates over a square aperture and returns the \hi\ spectrum. They have used $2$\subsuperscript{r}{200}{} as the aperture for groups, with $200$ kpc being the apertures for centrals. They were able to extract $25,906$ group spectra and $25,868$ central spectra for their analysis. We present their final sample (with an occupancy number $N_{\rm g} \geq 1$), which includes all the haloes with $1$ or more galaxies in it, and compare with \shark-ref.  

We also use the data from \citet{Obuljen2019TheFromALFALFA}, who estimate the \hi\ masses in dark matter haloes by directly integrating the \hi\ mass functions over the available range of \hi\ masses. \citet{Obuljen2019TheFromALFALFA} model the abundance and clustering of neutral hydrogen through a halo-model based approach, where they parametrise the HIHM relation
as a power law with an exponential mass cut-off (see Equation 6 in \citealp{Obuljen2019TheFromALFALFA}). In contrast to \citet{Guo2020}, \citet{Obuljen2019TheFromALFALFA} do not directly measure the \hi\ content of haloes, but instead use empirical relations to derive it. There is clearly some tension between these two approaches because they appear to be more than $2$-sigma away from each other at $M_{\rm halo} > 10^{13}$\M. Some of this may be due to the SDSS group catalogue not sampling the high halo mass end with enough statistics, as well as the \citet{Obuljen2019TheFromALFALFA} model not correctly capturing the \hi\ mass in the massive haloes (where the \hi\ content of galaxies is generally undetected by  ALFALFA).

In the upper-panel of Figure~\ref{fig:HI_showing_plot} compares \shark-ref with observations. We calculate the error on the mean \hi\ content of \shark-ref haloes via bootstrapping. The error is too small to be noticeable in the plot shown here. The observational data plotted are taken from \citet{Guo2020} and  \citet{Obuljen2019TheFromALFALFA}. It can be seen that \shark-ref is consistent with the \hi\ mass content of groups until $M_{\rm vir} < 10^{12}$\M. For the \hi-stacking points with $M_{\rm vir} > 10^{12}$\M, \shark-ref consistently under-predicts \hi\ in haloes, while it over-predicts it for $M_{\rm vir} > 10^{13.2}$\M. The inferred relation of \citet{Obuljen2019TheFromALFALFA} seems to be flatter than our predictions, which results in the model under- (over-) predicting the \hi\ content of haloes at $M_{\rm halo} <(>) 10^{13.8}$\M. 

In the lower-panel of Figure~\ref{fig:HI_showing_plot}, we compare the \hi~contribution from the satellite and central populations to the \hi~content of haloes at $z=0$. We also show the \hi-stacking results for the contribution of \hi\ from centrals and satellites as presented in \citep{Guo2020}. The error-bars for centrals (from the observational data) are the values presented in \citep{Guo2020}. We estimate the errors, $\Delta$, for the satellites from those reported for the total \hi\  and central galaxy contributions as  $\Delta_{\rm sat} = \sqrt{\Delta_{\rm total}^{2} + \Delta_{\rm central}^2}$, with $\Delta_{\rm total}$ and $\Delta_{\rm central}$ being the errors calculated for the total \hi\ content of the halo and centrals, respectively. We find that the observed centrals data are consistent with the \shark-ref predictions until $M_{\rm vir} < 10^{12}$\M, and thereafter \shark-ref under-predicts the \hi\ contained in the centrals. The satellites data, in contrast, are in better agreement with \shark-ref predictions.

Note that the relation derived in Figure~\ref{fig:HI_showing_plot} has not taken into account limitations that are inherent in observational surveys. \citet{Bravo2020}, using a \shark-derived lightcone to produce an analogue of the Galaxy and Mass Assembly (GAMA) survey \citep[e.g.][]{Robotham2011GalaxyG3Cv1}, showed that assigning galaxies to groups and classifying them as centrals and satellites in the same way as is done in observations has an important impact on how we understand  satellite/central galaxy quenching \citep[also see][]{Stevens2017PhysicalSage}. This is because $\sim$15\% of satellites/centrals are wrongly classified as such (according to the intrinsic definition provided by the halo/subhalo catalogue). In this work, we compare directly \velociraptor\ groups to the stacking results of \citet{Guo2020} without considering the effects shown in \citet{Bravo2020}. Because the SDSS group catalogue used by \citet{Guo2020} is expected to have an even higher contamination than the GAMA groups analysed by \citet{Bravo2020} (see \citealt{Robotham2011GalaxyG3Cv1} for details), we expect this to play an even greater role in our comparison. In future work, we will make a detailed comparison with observations by mimicking the \hi\ stacking procedure, with the aim of quantifying the systematic effects above.
As is shown in \citet{Chauhan2019}, accounting for observational limitations and producing mock-catalogues for comparison is essential when comparing simulations with observational data. 

After comparing with the observations, we compare \shark\ against other SAMs, such as GALFORM \citep{Cole_2000} and GAEA  \citep[GAlaxy Evolution and Assembly,][]{Xie2017_Marta}. \citet{Baugh2018-PMillennium} analysed the HIHM relation in a recalibrated GALFORM variant, using the Planck Millennium $N$-body simulation, which is the latest addition to the ``Millennium'' series of simulations of structure formation. For reference, Planck Millennium has a DM particle mass of $2.12 \times 10^9$\M~and a box of length $542.6$ \subsuperscript{h}{}{-1} cMpc \citep{Baugh2018-PMillennium}. 

GAEA on the other hand was run on the Millennium I \citep{Springel2005} and Millennium II simulations \citep{Boylan-Kolchin2009}, whose DM particle masses are $1.7 \times 10^{10}$\M and $1.4 \times  10^8$\M, respectively, in boxes of length of $500$ and $100$ \subsuperscript{h}{}{-1} cMpc, respectively. We also compare to the HIHM relation derived from the hydrodynamical simulation Illustris-TNG100  \citep{Nelson2018,Pillepich2018}, which is publicly available \citep{Nelson2019}.  This simulation has a box size of $75$\subsuperscript{h}{}{-1}~cMpc and a DM particle mass of $7.5 \times 10^6$\,\M.  The \hi\ content of Illustris-TNG100 galaxies was calculated in post-processing, following the `inherent' method outlined in \citet{Stevens2019AtomicSurveys}, using the \citet{Gnedin2014LINEGALAXIES} prescription.  We exclusively sum the \hi\ masses of Illustris-TNG100 \emph{galaxies} within $R_{\rm vir}$ to calculate a halo's total \hi\ mass.  In other words, we intentionally exclude any \hi\ contribution from the CGM.  This makes the results from Illustris-TNG directly comparable to SAMs, which do not include \hi\ in the CGM by design.
We also compare with the semi-empirical HIHM relation described in \citet{Padmanabhan2017Constraints12Gyr}, which was derived at $z \sim 0$ by abundance matching dark matter haloes with \hi-selected galaxies. They use the \hi-mass function from HIPASS \citep{Meyer2004} and ALFALFA \citep{Martin2012THESURVEY} along with the \citet{Sheth-Tormen2002} dark matter halo-mass function to match the \hi-selected galaxies to dark matter haloes. They assume that each dark matter halo hosts one \hi\ galaxy with its \hi\ mass is proportional to the host dark matter halo mass. By construction, this means that the most massive \hi\ galaxies inhabit the most massive haloes. 

\begin{figure}
  \includegraphics[width=\linewidth]{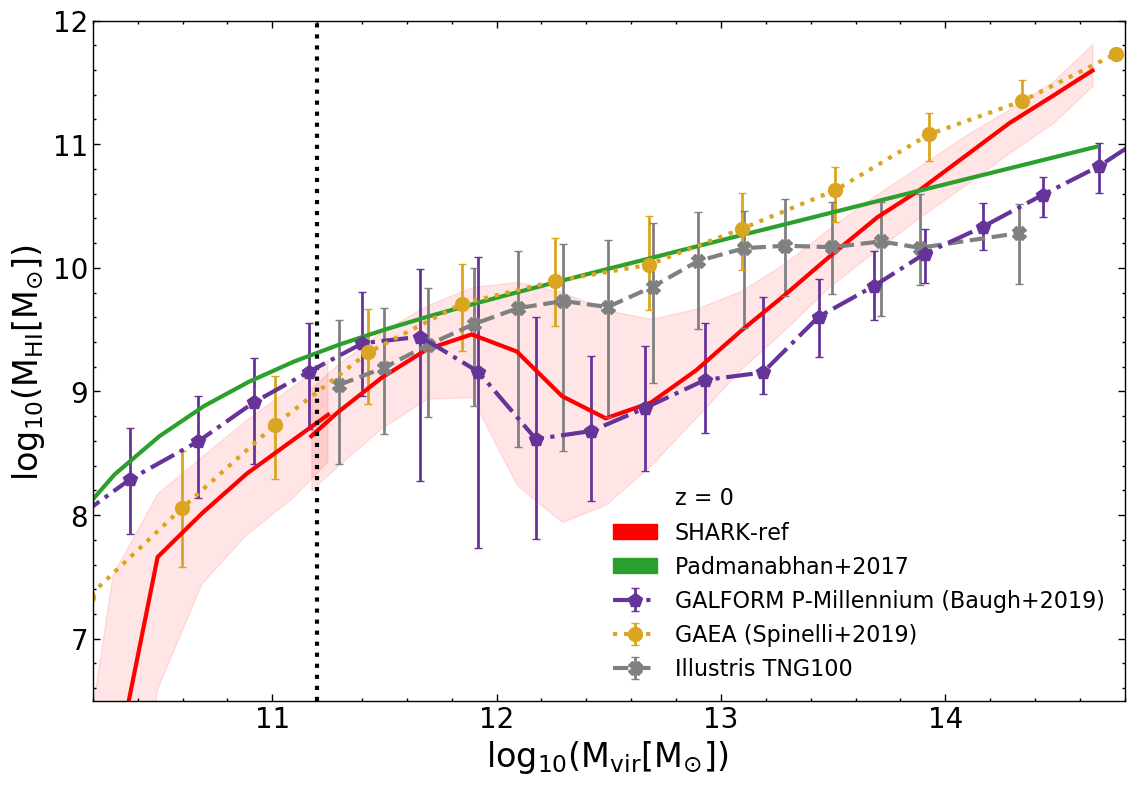}
\caption{The median and the \subsuperscript{16}{}{th}--\subsuperscript{84}{}{th} percentile range of \hi\ content of haloes as a function of halo mass by GALFORM \citep{Baugh2018-PMillennium}, GAEA \citep{Spinelli2019_Marta}, TNG100 \citep{Stevens2019AtomicSurveys} and \shark\ \citep{Lagos2018-Shark}. The dip in the median \hi\  mass occurs at lower halo masses for GALFORM than for \shark-ref, though for GAEA and TNG100 we do not see a prominent dip at all. This is an effect of different AGN feedback and SF models implemented by the different SAMs presented here, as the strength of the AGN feedback affects the position and shape of the drop. As for TNG100, the dip and \hi\ value depends on how it has been calculated, as for the current comparison, the CGM contributions to the \hi\ in the haloes has been removed from the TNG100 to make it more comparable with the SAMs presented. The purple-dotted line with the errorbars are the \hi\ values as shown in \citet{Baugh2018-PMillennium}, with the errorbars showing the \subsuperscript{10}{}{th}-\subsuperscript{90}{}{th} percentile range of the distribution, whereas the yellow-dotted line represents the values obtained from GAEA with the errorbars showing the \subsuperscript{16}{}{th}--\subsuperscript{84}{}{th} percentile range of the distribution. The grey dashed line represents TNG100, with error-bars showing the \subsuperscript{16}{}{th}--\subsuperscript{84}{}{th} percentile range of the distribution. The solid green line represents the \hi--halo scaling relation developed by \citet{Padmanabhan2017Constraints12Gyr}. The red solid line is the prediction from \shark-ref with the shaded region representing the \subsuperscript{16}{}{th}--\subsuperscript{84}{}{th} percentile range of the distribution.}

\label{fig:HI_showing_plot_SAM}
 \end{figure}

In Figure~\ref{fig:HI_showing_plot_SAM}, we plot the median of the total \hi\ content as a function of halo mass for the SAMs, \shark, GALFORM \citep{Baugh2018-PMillennium} and GAEA \citep{Spinelli2019_Marta}; the hydrodynamical simulation Illustris-TNG100; and the empirical relation by \citet{Padmanabhan2017Constraints12Gyr}.

Both GALFORM and \shark\ predict qualitatively similar curves, which display a prominent dip in the median \hi\ mass of halos at intermediate masses. The exact mass at which the dip happens differs between the models, with GALFORM predicting this to take place at $M_{\rm halo}\approx 10^{12}\rm \, M_{\odot}$, while for \shark\ this happens at $M_{\rm halo}\approx 10^{12.5}\rm \, M_{\odot}$. At lower (higher) halo masses, GALFORM predicts a higher (lower) median \hi\ mass than \shark. GAEA on the other hand, displays a very weak dip in the median \hi\ mass with halo mass. 

The \citet{Padmanabhan2017Constraints12Gyr} semi-empirical relation by construction shows a monotonically increasing \hi\ mass vs. halo mass.
This behaviour is qualitatively very different to the SAMs shown here, particularly \shark\ and GALFORM. We show in Section \ref{subsec:Understanding_shape} that the non-monotonic relation between the \hi\ and halo mass is due to the modelling of AGN feedback. The difference in the sharpness of the dips seen in \shark\ and GALFORM is due to the AGN feedback modelling used in the SAMs. As mentioned in Section~\ref{subsubsec:AGN_models}, \shark\ uses the \citet{Croton2016} model for AGN feedback, where the BH heating is estimated based on the luminosity of the BH, which is then used to adjust the cooling rate to respond to the heating. The heating radius is then estimated based on the radius within which the energy injected by the AGN equals that of the halo gas internal to that radius that would be lost if the gas were to cool. Whereas when looking at the AGN feedback in GALFORM, which is based on the \citet{Bower2006} model, AGNs are assumed to quench gas cooling only if the available AGN power is comparable to the cooling luminosity. the latter makes the AGN heating a binary mode, resulting in a sharper transition in GALFORM. GAEA produces massive galaxies that are less quenched than observations suggest at stellar masses $>10^{10}\rm \, M_{\odot}$ (see for example Figure $3$ in \citealt{Xie2020}), which may be an indication that their AGN feedback is not efficient enough. Illustris-TNG100, on the other hand, displays a mild dip at around $M_{\rm halo}\approx 10^{12.5}\rm \, M_{\odot}$, but much weaker than that displayed in \shark\ and GALFORM. This dip goes away when we include the CGM \hi\ contribution in the total \hi\ mass of the halos (not shown here), strongly suggesting that the CGM makes up a non-negligible amount of the \hi\ in groups. Unlike SAMs, Illustris-TNG100 predicts a flat median \hi\ mass at $M_{\rm halo}\gtrsim 10^{13}\rm \, M_{\odot}$. We caution that the definition of $M_{\rm halo}$ is not the same in all these simulations, but differences in definitions are much smaller ($\lesssim 0.2$~dex) than the differences seen here in the position of the \hi\ mass dip. The abrupt drop in the \hi\ abundance of haloes at \subsuperscript{M}{vir}{} $\lesssim$ \solarValue{10.4} is caused by the strength of the UV background being sufficient to keep the gas in those low-mass haloes ionised (see \ref{subsubsec:Photoionisation_Effect} for more details).

 When looking at the scatter around the median HIHM relation for all simulations, we find that all galaxy formation simulations shown here (\shark, GALFORM, GAEA and Illustris-TNG100) agree in that the scatter is maximal at $M_{\rm halo}\approx 10^{12}-10^{13}\rm \, M_{\odot}$, although the exact mass at which this occurs, and the magnitude of the scatter, varies from simulation to simulation. \shark, GALFORM, and Illustris-TNG100 produce a similarly large scatter ($\approx 1-1.5$~dex) at around the position where the dip in \hi\ mass takes place, while GAEA predicts a much smaller scatter of $\approx 0.3$~dex. This shows that observational constraints on the scatter of the HIHM relation are essential if we are to judge the success of the models.


\subsection{The \hi\ Correlation function}
\label{subsec:Correlation_function}

\begin{figure}
\centering

\includegraphics[width=\linewidth]{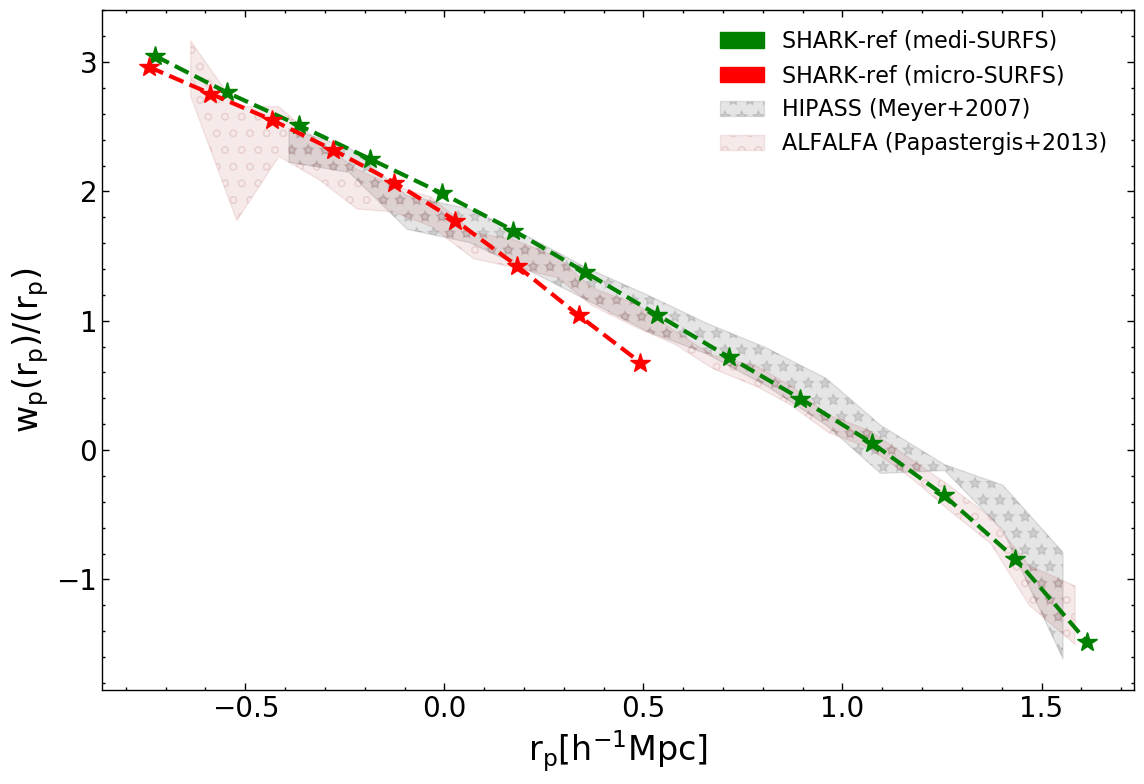}  

\caption{The projected two-point correlation function of the \shark-ref model for micro-SURFS (red-dashed line) and medi-SURFS (green-dashed line) compared with the observations of \citet{Meyer2007TheGalaxies} (grey-shaded region with stars) and \citet{Papastergis2013THEPROPERTIES} (brown-shaded region with circles), for the HIPASS and ALFALFA 40\% surveys, respectively. There is good agreement between the predictions and observations within the errorbars. For micro-\surfs, the predictions deviate at $r_{\rm P}\gtrsim 1 \, \rm h^{-1}\, Mpc$ due to the small size of the simulated box.}
\label{fig:correlation}
\end{figure}

The correlation function is defined as the excess clustering of a target distribution of galaxies over a random distribution, and thus is a measure of the spatial distribution of galaxies. It encodes information about both the underlying cosmology and the physics of galaxy formation, and its form is subject to how galaxies are selected (e.g. optically selected or \hi\ selected).

We use the $z=0$ medi- and micro-\surfs\ boxes to measure the projected two-point correlation function (2PCF) of galaxies with \hi\ masses $>$ \solarValue{8} for medi-\surfs\ and \hi\ masses $>$ \solarValue{7} for micro-\surfs. We employ the CorrFunc\footnote{https://github.com/manodeep/Corrfunc} \citep{CorrFunc} python routine developed to compute correlation functions and other clustering statistics for simulated and observed galaxies, as follows:
\begin{equation}
\frac{w_{\rm p}(r_{\rm p})}{r_{\rm p}} = \frac{2}{r_{\rm p}} \int^{\pi_{max}}_{0} \xi(r_{\rm p}, \pi) d\pi.    
\end{equation}
Here, we have measured the correlation function as a two-dimensional histogram, $\xi(r_{\rm p}, \pi)$, with the count of galaxy pairs as a function of both projected separation (\subsuperscript{r}{p}{}) and line-of-sight separation ($\pi$). By integrating $\xi(r_{\rm p}, \pi)$ over $\pi$, we can account for the effect of peculiar velocities. The $\pi_{\rm max}$ values adopted for our micro- and medi-\surfs\ boxes are $10$ \subsuperscript{h}{}{-1} cMpc and $30$ \subsuperscript{h}{}{-1} cMpc, respectively. Different $\pi_{\rm max}$ values are used to incorporate the different box sizes of micro- and medi-\surfs. These values reproduce the observational measurements of \citet{Papastergis2013THEPROPERTIES} and \citet{Meyer2007TheGalaxies}, with medi-\surfs\ using the same $\pi_{\rm max}$ values as were used in the observations. As for micro-\surfs, we opted for a lower $\pi_{\rm max}$ value because of the relatively small volume of the simulation box, which impacts the strength of clustering \citep{Power2006-boxsize-impact}.

In Figure~\ref{fig:correlation}, we reproduce the clustering measurements using the criteria used by \citet{Papastergis2013THEPROPERTIES} and \citet{Meyer2007TheGalaxies}, and show the predicted 2PCF of \hi\ selected galaxies in \shark\ in both simulated SURFS boxes, micro-\surfs\  and medi-\surfs. We also show the observational measurements of \citet{Meyer2007TheGalaxies} using HIPASS and \citet{Papastergis2013THEPROPERTIES} using ALFALFA. Both these observational measurements apply a volume correction and hence are comparable to the 2PCF obtained from the simulated box, which by construction is volume-limited. \citet{Meyer2007TheGalaxies} adopted a higher mass threshold of $M_{\rm HI} \simeq$ \solarValue{9} for their analysis, whereas \citet{Papastergis2013THEPROPERTIES} utilise the entire $40$\% ALFALFA data sample, with the \hi\ masses limiting to $M_{\rm HI} > $\solarValue{7.5}. Despite also using different \subsuperscript{M}{HI}{}\ limits for our different resolution boxes, we find agreement between them, although the micro-\surfs\ predictions start deviating at about  $r_{\rm p} \gtrsim 1$ \subsuperscript{h}{}{-1} Mpc as a result of the small volume of micro-\surfs.    

HIPASS and ALFALFA have different volumes and depth, and hence they are expected to trace different \hi\ mass distributions. This can, in principle, lead to different clustering signals if the 2PCF is \hi-mass dependent. \citet{Papastergis2013THEPROPERTIES} and \citet{Meyer2007TheGalaxies} tested this dependence and found that the clustering amplitude were largely insensitive to the \hi\ mass (see however \citealt{Guo2017ConstrainingClustering} for a different conclusion). \citet{Crain2017} also tested the \hi-clustering dependancy on the \hi\ mass of the galaxies in EAGLE hydrodynamical simulation \citep{Schaye2015-EAGLE-reference,Crain2015-EAGLE-reference} by looking at the clustering measurements of galaxies belonging to the same stellar bin, and found that \hi-poor galaxies seem to be more clustered. We tested this in our simulated boxes and found that the clustering amplitude was independent of the \hi\ mass selection (not shown here). This is also the reason why micro- and medi-\surfs\ agree well in Figure~\ref{fig:correlation} despite having different \hi\ mass lower limits.


\section{The physical drivers of the HIHM relation}
\label{sec:Understanding-HI-halo}

In this section, we explore the physical processes that drive the shape and the scatter of the HIHM relation.
In what follows, we compute a halo's 
\hi\ mass by summing over the \hi\ masses of all galaxies embedded in that halo. Note that \shark\ does not model the atomic content of the intra-halo gas and hence our measurement only reflects the total \hi\ content in the ISM of galaxies that belong to the same group.

\begin{figure}
  \includegraphics[width=\linewidth]{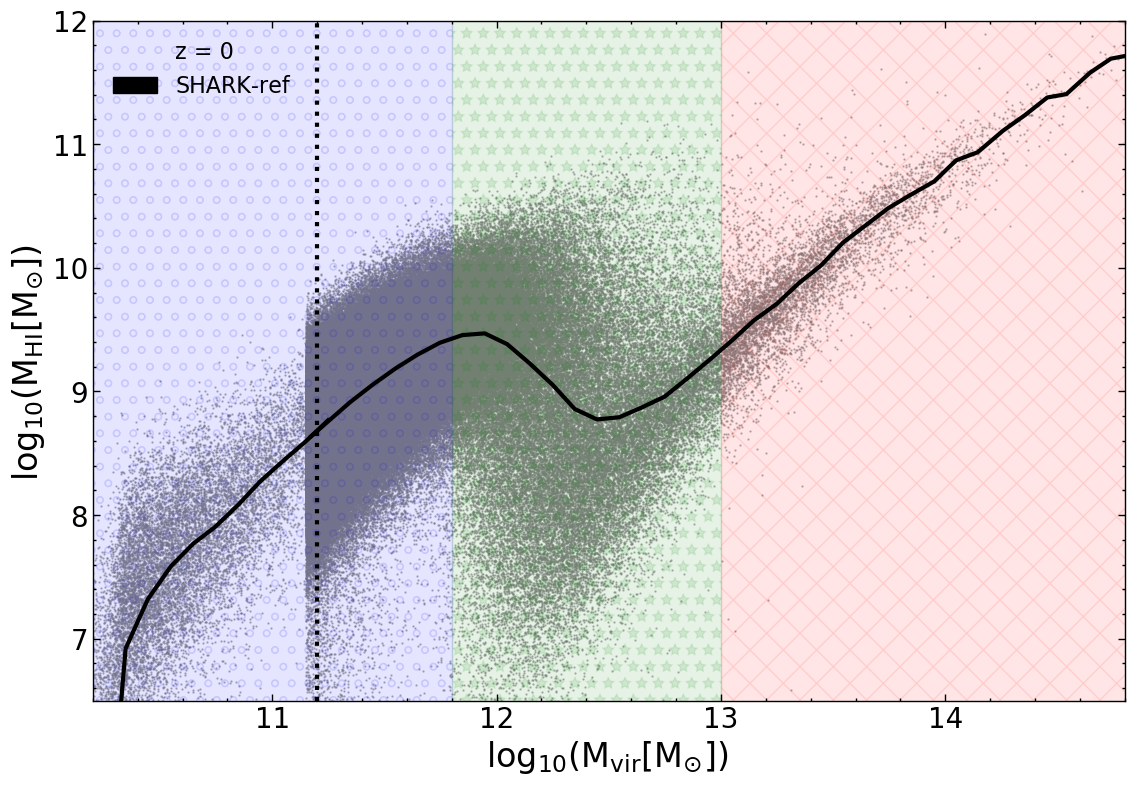}
\caption{The HIHM relation in \shark\ at $z=0$. Each point  is an individual halo, while the line shows the median of the relation. The three regions used to study the HIHM relation are shown with different shaded styles. The vertical dotted line represents the converging point of the two resolution boxes we are using - micro-\surfs\ and medi-\surfs.}
\label{fig:HI_transition}
 \end{figure}

In order to better understand the physical drivers of the HIHM relation, we divide the relation into three regions, as shown in Figure~\ref{fig:HI_transition}:
 
 \begin{enumerate}[label=\alph*)]
    \item \textbf{Low-mass Region}: includes haloes with $M_{\rm vir} < 10^{11.8}$\M. In this region the \hi\ mass monotonically increases with halo mass. We show in Section~\ref{subsec:Understanding_shape} that here the majority of the \hi\ content is in the central galaxy, with satellites contributing little to nothing, as many of these centrals are isolated (i.e. have no satellites).
    
    \item \textbf{Transition Region}: includes haloes with $10^{11.8}M_{\odot} \leq M_{\rm vir} < 10^{13}M_{\odot}$. Here the \hi\ content of haloes displays a non-monotonic dependence on halo mass. In this region some haloes have most of their \hi\ content in the central galaxy, while others are dominated by their satellites. As a result, this is the region of largest scatter. 
    
    \item \textbf{High-mass Region}: includes haloes with $M_{\rm vir} > 10^{13}$ \M. In this region, the \hi\ mass returns to a monotonically increasing relation with the halo mass. Here, the majority of \hi\  is contained in the satellite population. 
\end{enumerate} 

\subsection{Understanding the shape of the HIHM relation}
\label{subsec:Understanding_shape}

In order to unveil the physical drivers behind the shape of the HIHM relation, we leverage on the flexibility and modularity of \shark\ to explore different models and parameters for any one physical process. In this section we show how the HIHM relation is affected by these variations and break down the analysis into the effect of different physical processes.


\subsubsection{AGN feedback effect}
\label{subsubsec:AGN_feedback_Effect}

\begin{figure}
  \includegraphics[width=\linewidth]{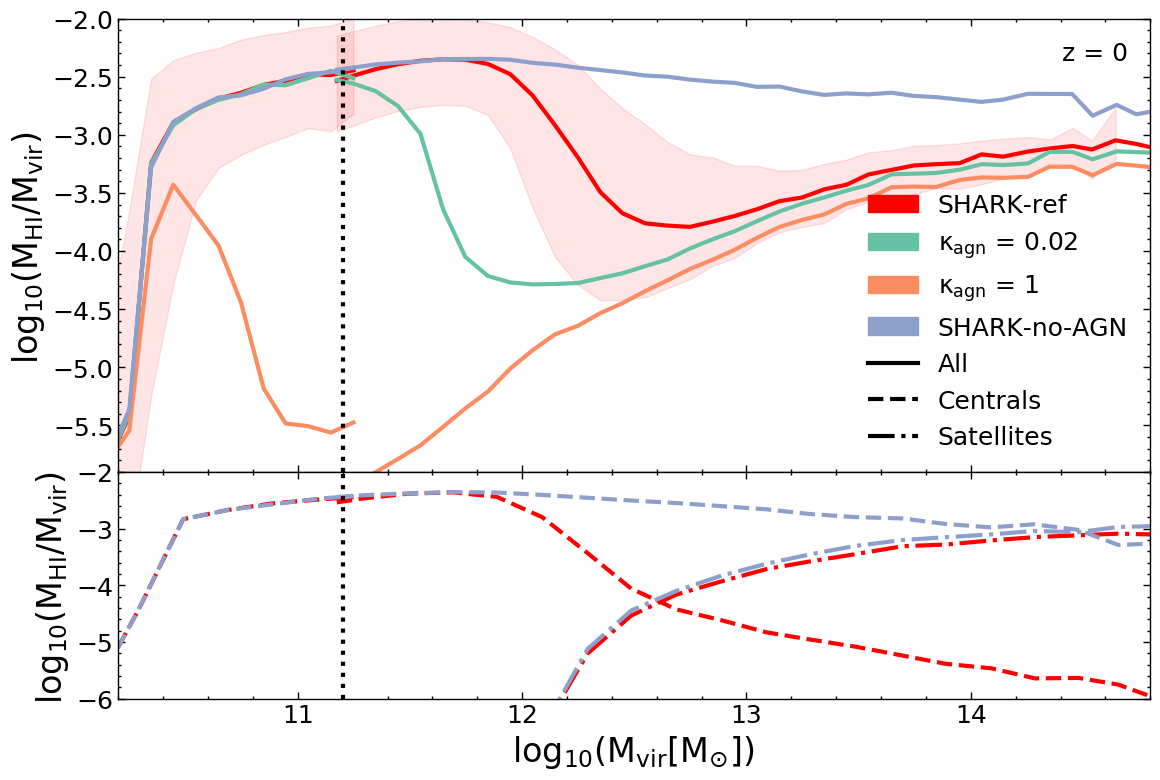}
\caption{Median \subsuperscript{M}{HI}{}/\subsuperscript{M}{vir}{} ratio as a function of \subsuperscript{M}{vir}{}. The shaded region represents the one-sigma scatter on the median \subsuperscript{M}{HI}{}/\subsuperscript{M}{vir}{} relation for our default (\shark-ref) model, and other lines representing different strengths of the feedback (as labelled). $\kappa_{\rm agn}$ is the free parameter that regulates the AGN feedback efficiency (see Equation \ref{eq:agn_feedback}); the higher the value, the stronger the feedback. It should be noted that as AGN feedback becomes more efficient, the knee of the relation shifts towards smaller virial masses, making AGN feedback efficiency a major contributor to the shape of the HIHM scaling relation. The vertical dotted line represents the shift from micro-\surfs\ (dashed-dotted lines) to medi-\surfs\ (solid lines). \textit{Lower Panel}: The median \hi\ contribution from central and satellite galaxies to the total \hi\ of the halo. For clarity we show this for the \shark-ref and \shark-no-AGN runs only. The centrals, which are major contributors to the HIHM relation at the transition region,  are significantly affected by changes in the AGN feedback efficiency.}

\label{fig:AGN-models}
 \end{figure}

As previously stated in Section \ref{subsubsec:AGN_models}, we vary the free parameter $\kappa_{\rm agn}$ (Equation \ref{eq:agn_feedback}), which controls the strength of AGN feedback. In Figure~\ref{fig:AGN-models}, we show how this efficiency affects the overall median \hi\ content of the halo at $z=0$. The different colours represent different values of $\kappa_{\rm agn}$, with the shaded region representing the \subsuperscript{16}{}{th}--\subsuperscript{84}{}{th} percentile range of the \shark-ref model. We remind the reader that the vertical line demarcates the transition from the high resolution, small volume micro-\surfs\ box used at low halo masses, to the moderate resolution, large volume, medi-\surfs\ box, used at high halo masses. This demarcation style is used throughout the figures in this paper, and has the purpose of increasing the dynamical range explored. \citet{Lagos2018-Shark} analysed the convergence between these two boxes and found that the stellar mass function was very well converged down to $10^8\,\rm M_{\odot}$ in medi-\surfs, while the \hi\ mass function was converged at $10^{8.5}\rm \, M_{\odot}$. We therefore adopt a transition between the boxes that roughly corresponds to these masses.


We find the \subsuperscript{M}{HI}{}/\subsuperscript{M}{vir}{} ratio increases as \subsuperscript{M}{vir}{} increases, reaching a peak value and then rapidly dropping  to a minimum (except for the \shark-no-AGN run) to then gradually rise again. This drop corresponds to our \textit{transition region} (for \shark-ref), and is mostly influenced by the strength of the AGN feedback. As we move from $\kappa_{\rm agn} = 0.002\  \text{(the default in \shark-ref)}, 0.02$ and $1$, the drop shifts from \subsuperscript{M}{vir}{} $=$ \solarValue{12} to \solarValue{11.2} to \solarValue{10.6}, respectively. As for the case of \shark-no-AGN feedback, $\kappa_{\rm agn} = 0$, we see that the ratio reaches a peak and then gradually decreases with the halo mass and this peak corresponds to the peak achieved by \shark-ref model. This is because shock heating of the accreted gas onto haloes plays a role in slowing down the cooling in the more massive haloes and hence the replenishment of the ISM of central galaxies, producing the mild decrease in \hi-to-halo mass ratio. 
It should be noted that despite the drop becoming steeper and taking place at lower halo masses with increasing AGN feedback efficiency, the \hi\ contained in the haloes gradually rises up to similar values at the cluster regime ($M_{\rm vir} > 10^{14.3}\rm \, M_{\odot}$), which is a consequence of satellites dominating this regime. As for the smallest haloes, there is not much difference in their \hi\ content as AGN feedback does not play a role here. 

More efficient AGN feedback has the consequence of steepening the drop in the \hi-to-halo mass ratio in the \textit{transition region}, as this shifts to lower halo masses. This is driven by the fact that as AGN feedback becomes more efficient, gas cooling becomes extremely inefficient, hampering the replenishment of the ISM of central galaxies.

A related consequence is that satellite galaxies become more prominent \hi\ reservoirs of the halo at lower halo masses as the AGN feedback efficiency increases, which can be seen in the lower panel of Figure~\ref{fig:AGN-models}. The lower panel shows the central and satellite \hi\ contributions for \shark-ref and \shark-no-AGN runs. We find that for the \shark-no-AGN run, the centrals remain the primary \hi\ reservoir of haloes as massive as $M_{\rm vir}\approx$ \solarValue{14.6}, thereafter satellites become dominant. On the other hand, in the \shark-ref run we find satellites start to become major \hi\ contributors at much lower halo masses, $M_{\rm vir}\approx$ \solarValue{12.5}.


\subsubsection{Stellar feedback effect}
\label{subsubsec:Stellar_feedback_Effect}

Stellar feedback in \shark\ is a two step process: gas is first expelled from the galaxy, and then from the halo depending on the excess energy of the outflow compared to the bounding energy (discussed in detail in Section \ref{subsubsec:Stellar_Feedback}). 

In the first step, the outflow rate from the galaxy depends on the maximum circular velocity of the galaxy to the power $- \beta_{\rm disc}$. For reference, an energy conserved outflow should have a $\beta_{\rm disc}=2$, while a momentum-conserved outflow has $\beta_{\rm disc}=1$. \citet{Lagos2013} found that once outflows are followed throughout their evolution in the interstellar medium from the adiabatic expansion to the snow-plough phase, $\beta_{\rm disc}$ can take higher values, and in fact, \shark-ref adopts $\beta_{\rm disc}=4.5$.
Here, we vary the value of $\beta_{\rm disc}$ to examine the effect this has on the \hi\ content of haloes.

\begin{figure}
  \includegraphics[width=\linewidth]{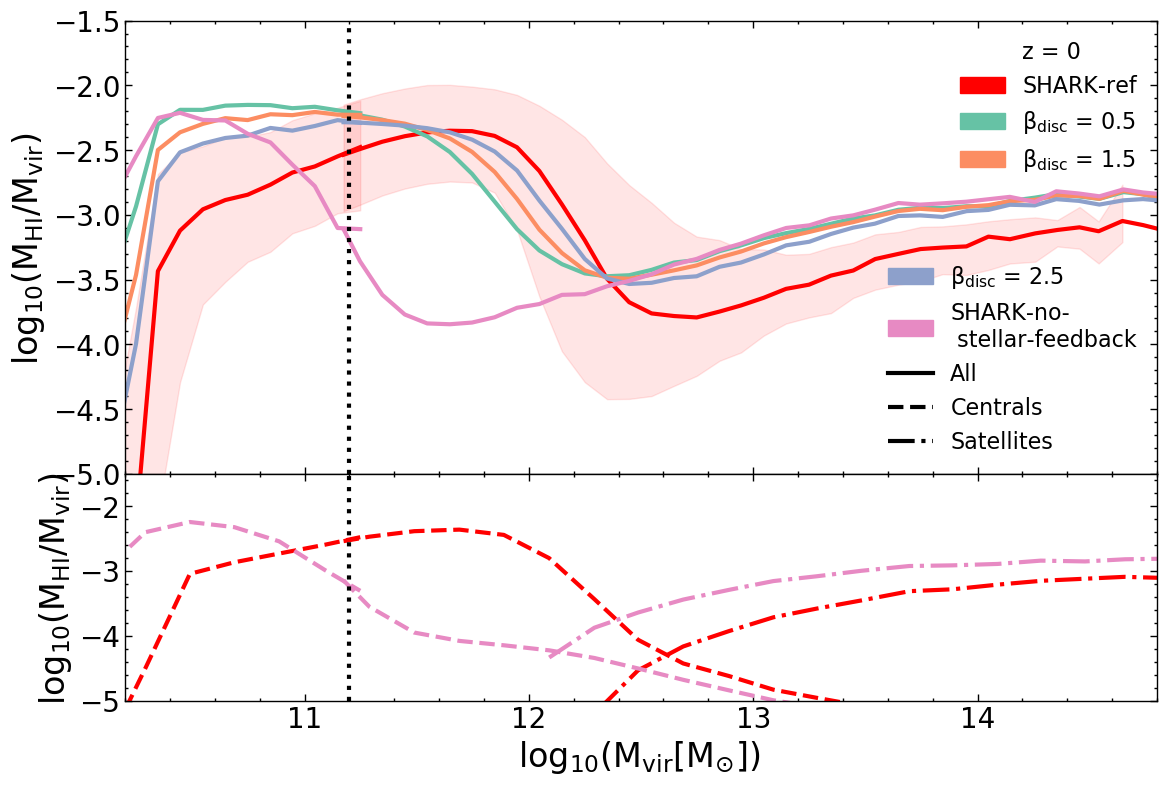}
\caption{As in Figure~\ref{fig:AGN-models}, but for different $\beta_{\rm disc}$, which represents the power-law exponent in the circular velocity dependence of the mass loading due to stellar feedback (see Equation \ref{eq:stellar_feedback}). Although we see an effect of this parameter over the whole mass range, it is more prominent at low masses, with a weaker stellar feedback being associated to a higher \hi-to-halo mass ratio. The impact of stellar feedback is anyway weaker than that of AGN. \textit{Lower Panel}: The  median \hi\ contribution from centrals and satellites to the total \hi\ of the halo. For clarity we show the \shark-ref and \shark-no-stellar-feedback runs only. In the latter, centrals see a decrease of their \hi\ content at very low halo masses compared to \shark-ref.}

\label{fig:Stellar-Feedback-models}
 \end{figure}

In Figure~\ref{fig:Stellar-Feedback-models}, we present the effect of varying $\beta_{\rm disc}$ on the \subsuperscript{M}{HI}{}/\subsuperscript{M}{vir}{}-\subsuperscript{M}{vir}{} relation at $z=0$. We change the value of $\beta_{\rm disc}$ from $0.5$ to $5$. The way $\beta_{\rm disc}$ affects the \hi\ content of haloes is different at different halo masses. The \hi\ content of haloes below the virial mass of \solarValue{11.2} is affected the most, with higher $\beta_{\rm disc}$ values inducing a smaller amount of \hi\ in the halo. A similar trend is seen in haloes above the mass \subsuperscript{M}{vir}{} $>$ \solarValue{12.6}.


These trends are caused by a higher value of $\beta_{\rm disc}$ driving higher outflow rates, and hence depleting the ISM of both centrals and satellites alike. In the transition region we see that a higher $\beta_{\rm disc}$ value is associated to {\it higher} \hi-to-halo mass ratios. This at first appears counter-intuitive as more outflows should lead to a lower \hi\ content. However, this can be reconciled by the fact that what drives this trend is the transition from \hi\ being dominated by the central galaxy to the satellites moving towards lower halo masses as $\beta_{\rm disc}$ increases.

One interesting aspect of having no stellar feedback (\shark-no-stellar-feedback), is seen in Figure~\ref{fig:Stellar-Feedback-models}. The \subsuperscript{M}{HI}{}/\subsuperscript{M}{vir}{} ratio is very similar to the $\kappa_{\rm AGN} = 10$ run, i.e. very high AGN feedback efficiency (Section~\ref{subsubsec:AGN_feedback_Effect}), for the \hi\ content of the entire halo. 
With stellar feedback off, we end up with more elliptical galaxies at lower halo masses which is indicative of the galactic disc being unstable and unable to  sustain itself. This leads to galaxies being bulge-dominated at $M_{\rm stellar}\gtrsim 10^{8.5}\,\rm M_{\odot}$ compared to $M_{\rm stellar}\gtrsim 10^{10}\,\rm M_{\odot}$ in \shark-ref. Because the BH mass scales with the bulge mass in \shark, AGN feedback can now be effective in galaxies of much lower stellar masses compared to \shark-ref. In short, AGN feedback becomes overly efficient in the absence of stellar feedback across the whole stellar mass range. A similar effect was noticed in the EAGLE hydrodynamical simulations \citep[see][]{Wright2020-feedback-effects-hydrosim}, where AGN feedback becomes much more efficient when there is no stellar feedback present. We also vary other parameters related to stellar feedback. In particular we tested varying $\epsilon_{\rm disc} = 1,\ 3,\ 5,\ 7\ \text{and}\ 10$. We find that the effect of changing the $\epsilon_{\rm disc}$ has a similar effect on the \subsuperscript{M}{HI}{}/\subsuperscript{M}{vir}{}-\subsuperscript{M}{vir}{} relation as varying $\beta_{\rm disc}$. 

In the lower panel of Figure~\ref{fig:Stellar-Feedback-models}, the \hi\ contribution of central and satellites is shown for the \shark-ref and \shark-no-stellar-feedback runs. The \hi\ content of centrals decreases rapidly for the \shark-no-stellar-feedback and starts at a lower halo mass of \subsuperscript{M}{vir}{} $\approx$ \solarValue{10.4}, whereas for the \shark-ref centrals, the \hi\ content starts decreasing at \subsuperscript{M}{vir}{} $\approx$ \solarValue{12}. We also find that the \hi\ content of satellites in the \shark-no-stellar-feedback run is more significant than in the \shark-ref run relative to the total, with the satellites becoming a major \hi\ contributors at lower halo masses.  

Despite stellar feedback having a clear effect on the HIHM relation, it appears like AGN feedback has a more dramatic effect on the shape of the HIHM relation. This makes sense as stellar feedback hardly quenches galaxies but instead plays a role in the self-regulation of star formation. AGN, on the contrary, is very efficient at quenching galaxies above a stellar mass threshold, that in \shark-ref happens roughly at $M_{\rm stellar}\approx 10^{10.5}\rm \, M_{\odot}$.

\subsubsection{The effect from other physical mechanisms}
\label{subsubsec:moved_processes_summary}

In addition to stellar and AGN feedback, we explore other physical mechanisms in \shark, which we present in Appendix \ref{appendix:understanding-shape}. These include photoionisation feedback, ISM modelling and environmental effects. These other mechanisms have a lesser effect on the HIHM relation compared to AGN and stellar feedback. Here, we provide short description of the main conclusions. 

As stated in Section \ref{subsubsec:Photoionisation_Feedback}, we vary the value of $v_{\rm cut}$, which directly affects the circular velocity ($v_{\rm thresh}$) of the haloes under which the halo gas is not allowed to cool down and thus remains ionised (see Equation \ref{eq:photoionisation_feedback}). We find that changing $v_{\rm cut}$ does not have any effect on the drop seen in the \textit{transition region}, which remains at the \subsuperscript{M}{vir}{} $\sim$ \solarValue{12} mass scale for all the runs with varying $v_{\rm cut}$. Though, a lower photoionisation feedback efficiency does result in higher \hi\ content for haloes of  \subsuperscript{M}{vir}{} $>$ \solarValue{12.4}. This is caused by the fact that with lower photoionisation feedback smaller haloes are allowed to retain their \hi\ content, and when they become satellites, their \hi\ contribution to the total \hi\ of a halo increases (see Figure \ref{fig:Photoionisation-models}). See Appendix \ref{subsubsec:Photoionisation_Effect} for more details.

We also tested the effect of using different models for the molecular-to-atomic gas partition on the total \hi\ content of the halo. We compared the BR06 (the default model of choice) and GD14 models for gas partition in the ISM (see Section \ref{subsubsec:SF_models}). We find that the \textit{transition region} for the model adopting the GD14 prescription occurs at lower halo masses, \subsuperscript{M}{vir}{} $\approx$ \solarValue{11.5} compared to \subsuperscript{M}{vir}{} $\approx$ \solarValue{12} for BR06. We find that this is due to the interplay between AGN feedback and the ISM model, as bigger BHs are produced in the GD14 run compared to BR06 at fixed halo mass in the transition region, again highlighting the complex interplay between the physical processes modelled in \shark. We also find that using GD14 results in higher \hi\ content for low- and high-mass haloes as opposed to BR06 (see Figure \ref{fig:SF-models}). The latter is due to the fact that the centrals of low-mass haloes are more \hi-rich in GD14 than BR06, which boosts the \hi\ content of those, but also of high-mass haloes as they become satellites. We delve deeper into the ISM model effect in Appendix \ref{subsubsec:SF_model_effect}.

Finally, we test the ram-pressure stripping effect on the \hi\ content of  haloes, by switching between the stripping mode `on' and `off' (see Section \ref{subsubsec:stripping}). We find that the total amount of \hi\ in either model is approximately the same, though the stripping `off' model leads to a slightly lower \hi\ in the transition region (see Figure \ref{fig:stripping_models}). More details on this effect are given in Appendix \ref{subsubsec:Stripping_effect}. 

One major inference made through these tests was that despite the variations above, the shape of the HIHM relation essentially remained the same.

\subsubsection{Summary}

In conclusion, we find that several physical processes affect the shape of the \subsuperscript{M}{HI}{}/\subsuperscript{M}{vir}{}--\subsuperscript{M}{vir}{} relation and therefore we cannot isolate a single process that is the sole contributor for this. We can, nonetheless, rank different processes by their apparent effect. By doing this we find that AGN feedback appears to have the strongest effect as the transition region changes shape dramatically with varying AGN feedback efficiency, and moreover, the existence of a transition region (regardless of its shape) seems to be solely determined by AGN feedback. We expect the exact way of modelling AGN feedback to also have an effect (though this is not tested explicitly here). Other physical processes, such as stellar feedback, the ISM modelling and photoionisation feedback have a noticeable effect on the shape of the relation but qualitatively the relation continues to clearly have three distinct regions.

\subsection{Physical drivers behind the scatter of the HIHM relation}
\label{subsec:Understanding-scatter}

The shape of the HIHM is only half the story. To fully characterise the HIHM scaling relation, we also need to understand the underlying scatter and its physical drivers. This is necessary for the purpose of Section~$5$, in which we aim to develop a numerical way of populating DM-only simulations with \hi. For the latter, it is then important to explore how the scatter correlates with different halo properties which are accessible in these simulations. With this in mind, we explore how the scatter of the HIHM relation related to halo properties such as the halo mass assembly history, the halo's spin parameter, etc, in the following sections.
Here, we focus on the \shark-ref model only.


\subsubsection{Spin parameter effect}
\label{subsubsec:spin_parameter_effect}

An intrinsic halo property that has recently been discussed in length in the literature in connection to the \hi\ content of galaxies is the spin parameter. The spin parameter of a halo is normally quantified as follows \citep{Peebles1969}, 
\begin{equation}
    \lambda = \frac{J\sqrt{|E|}}{G\, M^{5/2}},
\end{equation}
where $J$ is the magnitude of the angular momentum vector of the particles within the virial radius, $M$ is the virial mass, $E$ is the total energy of the system and $G$ is the gravitational constant. \citet{Maddox2015VariationMass} and \citet{Obreschkow2016AngularDisks} have suggested based on ALFALFA and THINGS \citep{Walter2008-THINGS} observations that the angular momentum of a galaxy regulates its \hi\ mass and the atomic-to-baryon mass fraction; the idea being that a galaxy with high angular momentum can support a larger \hi\ disc, thus sustaining more \hi\ mass as well, compared to a lower angular momentum disc, which is subject to more instabilities. Empirically this has been observed as a correlation between the angular momentum, \hi\ content and physical extent \citep{Lutz2018TheGalaxies}. Angular momentum in haloes scales steeply with mass, dependence that is removed when focusing instead on the spin parameter. Hence, for our purpose - studying what drives the scatter of \hi\ content in haloes at fixed halo mass - the halo spin is a more natural property to focus on than angular momentum.

\begin{figure}
\centering
    \includegraphics[width=\linewidth]{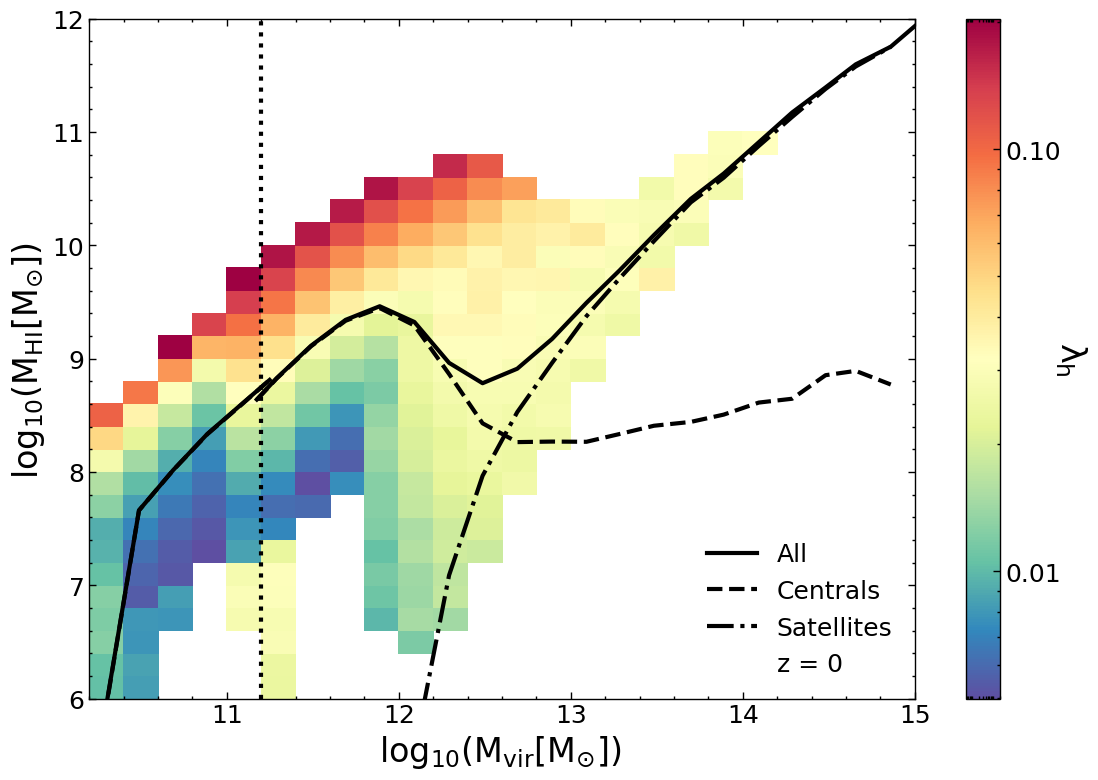}
\caption{As in Figure~\ref{fig:age_50} but here bins are coloured by the median halo's spin parameter, as labelled in the colour bar. There is a strong correlation between the \hi\ mass and the spin parameter at fixed halo mass for haloes with $M_{\rm vir} < 10^{12}\,\rm M_{\odot}$. Haloes with higher spin parameters are \hi-richer than their counterparts. This trend becomes less prominent at the transition regions and completely disappears in the high mass region.}
\label{fig:spin parameter}
\end{figure}

Figure~\ref{fig:spin parameter} shows the \subsuperscript{M}{HI}{}--\subsuperscript{M}{vir}{} relation with bins in this space this time coloured by the median spin parameter of haloes. 
The halo's spin parameter is very strongly correlated with the scatter in the HIHM relation at \subsuperscript{M}{vir}{} $<$ \solarValue{12}, with higher spin parameters being associated to more \hi-rich haloes. The \hi\ content in haloes at the low-mass region is primarily contributed by the central galaxy. Hence, the relation between the \hi\ mass and spin parameter for haloes is pretty much a reflection of the relation between the \hi\ content and angular momentum of the central galaxy. 

We would like to caution our readers that we use the halo spin parameter as opposed to the spin of the galaxies and these can be very different. The cited observations have no access to the halo spin. The strong correlation seen in  Figure~\ref{fig:spin parameter} could be exaggerated due to the simplistic model assumptions. For instance, \shark\ assumes that the halo gas has the same specific angular momentum as the halo's DM, with the specific angular momentum of the gas being conserved as it cools. \shark\ also assumes the specific angular momentum of the galaxy's components and halo to be aligned.


As we move towards the transition and high-mass regions, this correlation is no longer observed. This is because in these regions we see the emergence of the satellite population as the main contributors of \hi\ in haloes and hence the relation between \hi\ mass and angular momentum of the central galaxy is no longer relevant. Satellite galaxies on the other hand, have angular momenta which is largely uncorrelated with the host-halo's spin. Satellite galaxies in \shark\ have a specific angular momentum that is inherited from their hosthalo last time they were centrals. Due to the stochastic nature of the halo spin parameter, by $z=0$ satellite galaxies have stellar spins, and therefore \hi\ masses, that are uncorrelated with the central galaxy spin.

We also study the evolution of the \hi--halo mass--$\lambda$ relation towards high redshift, up to $z=2$ (see Appendix~\ref{appendix:Redshift-trends}). We find that the trend remains prominent throughout the whole redshift range. We also find evidence of the \textit{transition region} shrinking in dynamic range due to the systematic effect of AGN feedback efficiency decreasing as we move to higher redshifts.


\subsubsection{Substructure mass effect}
\label{subsubsec:mvir_ratio_effect}

\begin{figure}
\centering
\includegraphics[width=\linewidth]{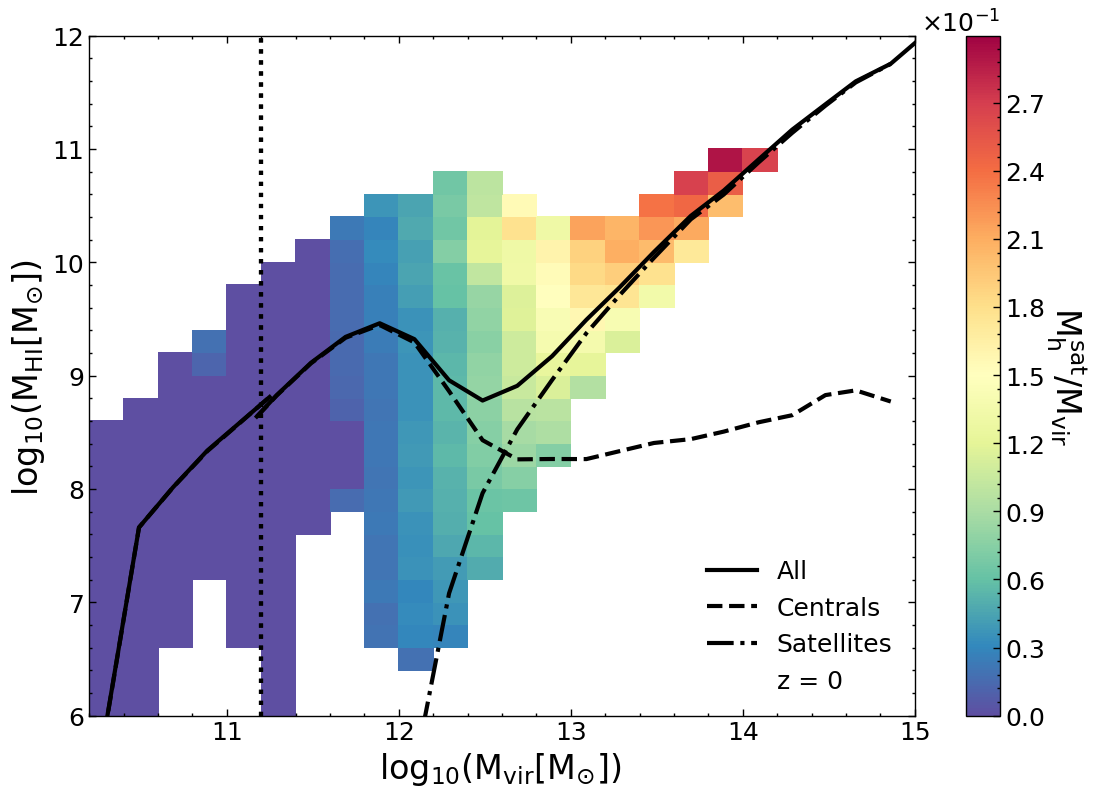}
\caption{As in Figures~\ref{fig:age_50} and \ref{fig:spin parameter}, but here bins are coloured by the median ratio between the total mass in subhaloes (\subsuperscript{M}{h}{sat}) to the total halo mass (\subsuperscript{M}{vir}{}), as labelled in the colour bar. At $M_{\rm vir} > 10^{12}$ \M, a correlation emerges with higher \fracmvir\ associated to a higher \hi\ content at fixed halo mass.}
\label{fig:fraction_satellite}
\end{figure}

As stated in previous sections, satellite galaxies are the primary source of \hi\ in haloes in the high-mass region. Hence, we expect the amount of substructure to be a good predictor of the scatter in the HIHM relation at high halo masses. To explore this idea, Figure~\ref{fig:fraction_satellite} shows the HIHM relation with bins now coloured by the fraction of mass in a halo that is contained in subhaloes, \fracmvir. Note that here we use subhalo and halo masses of the \velociraptor\ catalogues of the micro-\surfs\ and medi-\surfs.

We note that already at the transition region the effect of substructure on the \hi\ content of haloes is visible, but certainly becomes clearer in the high mass region, in a way that haloes with higher \fracmvir\ also have more \hi. This is largely due to the larger number of satellites a halo with a higher \fracmvir\ has compared to one with a lower \fracmvir\ at fixed halo mass. The fact that the trend is weaker in the transition region than at \subsuperscript{M}{vir}{} $>$ \solarValue{12.5} is due to the fact that many of those haloes have very few or no satellites. The clear correlation we obtain between the \hi\ mass and \fracmvir\ at high halo masses makes it a good candidate to be used to predict the \hi\ content of massive haloes.


We explore the evolution of the \hi--halo mass relation dependence on \fracmvir\ over the redshift range $0\le z\le 2$  in Appendix~\ref{appendix:Redshift-trends}, and find the trend to remain prominent and continue to be the main parameter that correlates with the scatter of the \hi--halo mass relation at the high halo mass end ($\gtrsim 10^{13}\, \rm M_{\odot}$).

\subsubsection{Other Halo Properties}
\label{subsubsec:other-halo-properties}
In addition to the halo parameters analysed here, we also explored the halo concentration and the effect of formation age (redshift at which the halo has assembled $50$\% of its present mass) of the halo on the \hi\ content.
We found no correlation between the \hi\ content of haloes and its concentration. This is due to the fact that \shark\ adopts the concentration model of \citet{Duffy2008}, which only depends on halo mass and time. Hence, naturally, at fixed halo mass, we obtain no dependence of the \hi\ content on concentration.

It has been speculated in previous studies that the formation age of haloes, hereafter referred to as \agefifty, is correlated to their \hi\ content \citep[see][]{Guo2017ConstrainingClustering, Spinelli2019_Marta}. When testing the effect of \agefifty\ with \shark, we find that a slight trend is noticeable in the \textit{transition region}, with younger haloes having more \hi\ than their counterparts of the same mass (see Figure \ref{fig:age_50}). We discuss more on the effects of formation age on the \hi\ content in Appendix \ref{appendix:understanding-the-scatter}, and its relation to AGN feedback in  Section~\ref{subsubsec:all_other_effect}.


\subsubsection{Baryon physics effects}
\label{subsubsec:all_other_effect}

As stated previously (see Section \ref{subsubsec:AGN_feedback_Effect}), the dip in the \hi--halo scaling relation (at \subsuperscript{M}{vir}{} $\approx$ \solarValue{12}) is caused by AGN feedback, which becomes prominent at these masses. AGN feedback is also responsible for the flaring of the scatter in the transition region, which increases from about $0.5$~dex in the low-mass region to almost $1.2$~dex at the transition region. As pointed out above, the halo spin parameter and \fracmvir\ are promising second variables to reduce the scatter at the low- and high-mass end regions, respectively. 

For the transition region, however, a combination of these two parameters is required, as in this region we get both types of haloes, those that have their \hi\ content mostly in their central, and those that have most of their \hi\ in satellite galaxies. But even when including both parameters, we still cannot reduce the residual scatter to below $0.9$~dex (discussed in detail in Section \ref{sec:Model-development}). 

\begin{figure}
\centering
\includegraphics[width=\linewidth]{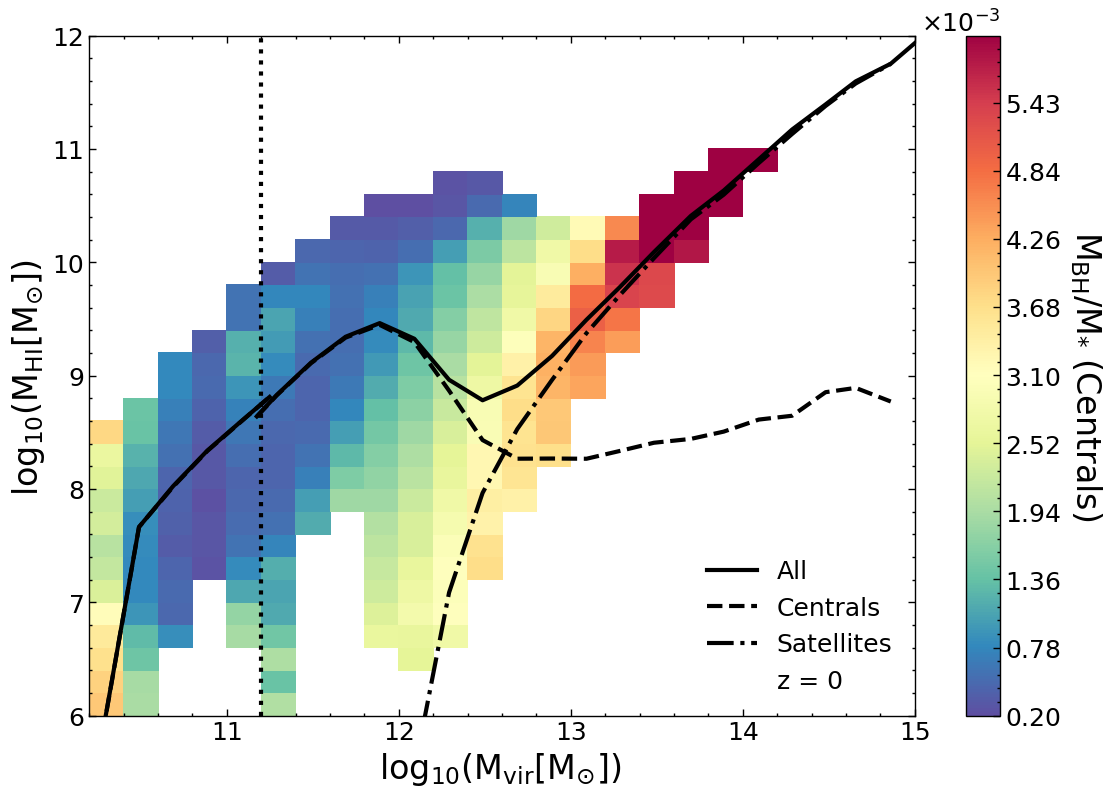} 
\caption{As in Figs~\ref{fig:age_50}, \ref{fig:spin parameter} and \ref{fig:fraction_satellite}), but bins here are coloured by the median of the fraction of BH mass (\subsuperscript{M}{BH}{}) to the central galaxy's stellar mass ($M_{\star}$) as labelled in the colour bar. A clear trend emerges in the transition region of more \hi\ residing in haloes whose central has a low-mass BH relative to its stellar mass.}
\label{fig:BH-mstar}
\end{figure}

This is to be expected, as the exact effect of AGN feedback cannot be trivially predicted from halo properties only but instead we require insight into the BH mass and cooling luminosity. To better illustrate the effect of AGN feedback at the transition region, Figure~\ref{fig:BH-mstar} shows the HIHM relation with bins coloured by the median ratio between the BH mass, \subsuperscript{M}{BH}{}, and stellar mass of the central galaxy. We find a stronger correlation between the halo \hi\ mass with ${M_{\rm BH}}/{M_{\star}}$ at fixed halo mass than that seen with \agefifty, halo spin parameter and \fracmvir. Haloes with low-mass BHs relative to the stellar mass of the central tend to have more \hi\ mass compared to haloes with more massive BHs. However, there still is a causal relation between the AGN feedback efficiency and \agefifty. We find that at fixed halo mass, more massive BHs inhabit older haloes, and hence more powerful AGN feedback is possible in older haloes. We do, however, find the correlation between the scatter of the HIHM relation at fixed halo mass to be stronger with the BH mass than with \agefifty.


Despite the significance of the BH mass in reducing the residual scatter of the HIHM relation, we do not use it in Section~$5$ to build up our numerical model for how to populate haloes with \hi. This is because we are interested in a model that can be applied to large-scale DM-only simulations. This analysis, however, serves to remind the reader that the complexity of baryon effects cannot be fully described with halo properties alone.


\subsubsection{The \hi\ content of subhaloes}
\label{subsubsec:subhalo_effect}

In this section, we discuss how the \hi\ mass inside the subhaloes is related to subhalo properties. 

\begin{figure}
\centering
    \includegraphics[width=\linewidth]{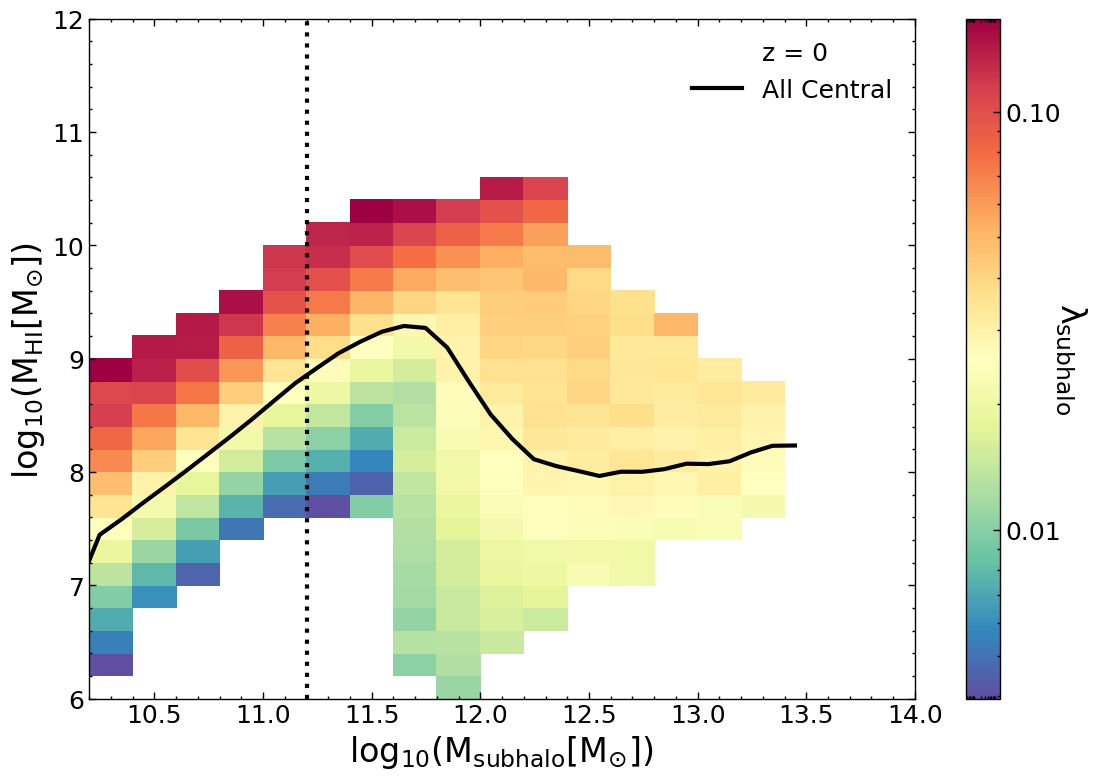}
\caption{The \hi\ content of central galaxies as a function of their subhalo mass at $z=0$. Bins are coloured by the median subhalo's spin parameter, as labelled in the colour bar. The solid black line shows the median \hi\ mass as a function of the mass of the subhalo, \subsuperscript{M}{subhalo}{}. Subhaloes with higher spin parameters are \hi-richer than their counterparts up to \subsuperscript{M}{subhalo}{} $\sim$ \solarValue{12}, after which the trend is almost completely lost at the transition and high mass regions.}
\label{fig:spin parameter-subhalo}
\end{figure}

Section~\ref{subsubsec:spin_parameter_effect} showed that there is a strong correlation between the HIHM scatter and the spin parameter of the halo at fixed halo mass  in the low-mass region. A possible interpretation of  Figure~\ref{fig:spin parameter} is that the weakening of the correlation at $M_{\rm halo}>10^{11.8}\,\rm M_{\odot}$ is due to the contribution of satellite galaxies becoming significant, and their subhalo's spin being uncorrelated to the host halo's spin. In this scenario, it is possible that the \hi\ content of the underlying subhalo population is well correlated with the subhalo's spin parameter instead. To test this idea, we plot the \subsuperscript{M}{HI}{}--\subsuperscript{M}{subhalo}{} relation for the central subhaloes in Figure~\ref{fig:spin parameter-subhalo}, colouring by the spin of the central subhalo. Here, we only include galaxies {\tt type=0} (centrals). We remind the reader that galaxies {\tt type=0} are centrals of the central subhalo in a halo, while galaxies {\tt type=1} are centrals of satellite subhalos.  

The solid line shows the median \hi\ content of the central subhalo as a function of the subhalo mass, at $z=0$. The dotted vertical line demarcates the micro- to medi-\surfs\ subhalo population transition. The central subhalo spin parameter is strongly correlated with the scatter in the \subsuperscript{M}{HI}{subhalo}--\subsuperscript{M}{subhalo}{} at $M_{\rm subhalo} < $ \solarValue{11.5}, after which the correlation becomes much weaker, similar to the behaviour we obtained for the total halo mass. On the other hand, we find that satellite subhaloes\footnote{We only use {\tt type=1} satellites as they are associated to a satellite subhalo. Galaxies {\tt type=2} are not included here as their host subhalo has been lost.} do not show a correlation between the \hi\ mass and the satellite subhalo's spin at fixed subhalo mass. This shows that the weakening of the correlation between the HIHM and halo's spin parameter is not driven by the effect of satellite galaxies, and instead central subhalos display the same behaviour.


\begin{figure}
\centering
    \includegraphics[width=\linewidth]{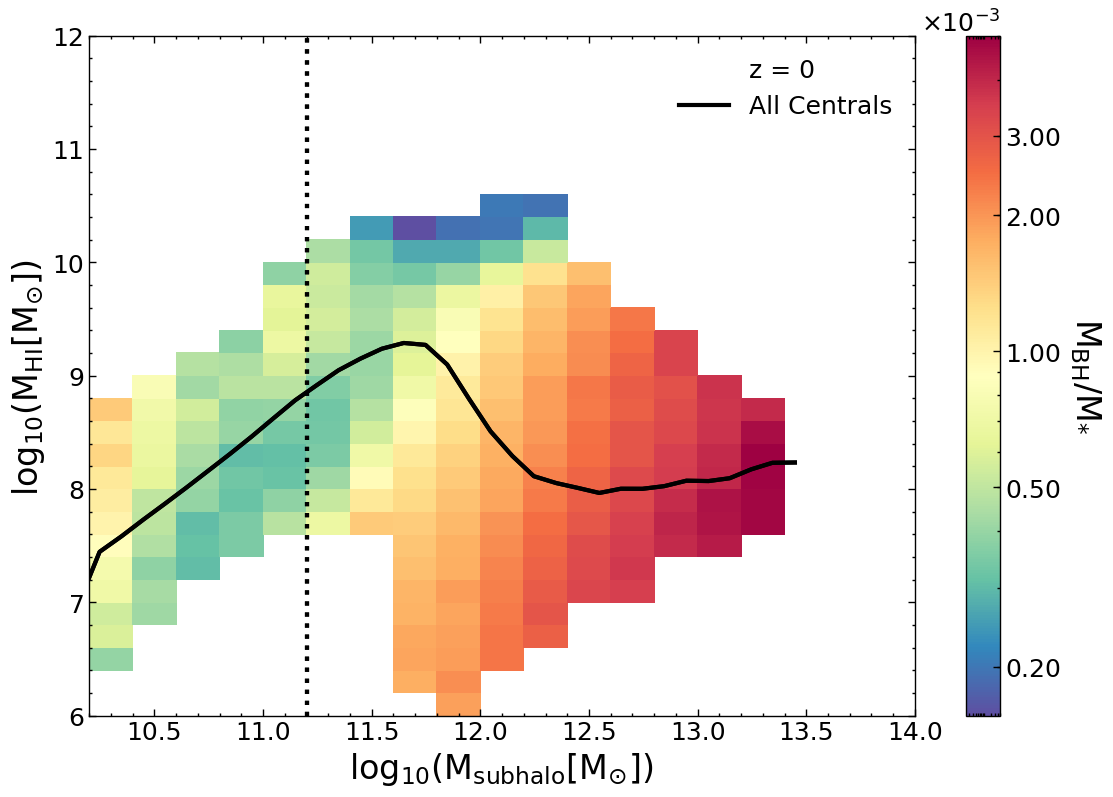}
\caption{Similar to Figure~\ref{fig:spin parameter-subhalo}, but here the bins are coloured by the median ratio between of BH and the stellar mass of the central galaxy of central subhaloes, as labelled in the colour bar at $z=0$. The solid line show the median \hi\ in central subhaloes. A clear trend emerges at $M_{\rm subhalo} \gtrsim $ \solarValue{11.4}, where we find that the subhaloes with higher BH-to-stellar mass ratio of the galaxy, lesser the \hi\ abundance.}
\label{fig:BH-star-subhalo}
\end{figure}

Figure~\ref{fig:BH-star-subhalo} explores the effect of AGN feedback in erasing the spin parameter dependency in the transition region at the subhalo level. We plot the \subsuperscript{M}{HI}{}--\subsuperscript{M}{subhalo}{} relation explicitly for central subhaloes, colouring the bins by the median $M_{\rm BH}$/$M_{\star}$ ratio, where $M_{\rm BH}$ and $M_{\star}$ are the BH and galaxy stellar masses, respectively, of the central galaxy of the central subhalo, at $z=0$.  
We find that the AGN does not show a strong correlation with the scatter of the HIHM relation for subhaloes for $M_{\rm subhalo} \leq$ \solarValue{11.5}, but at higher subhalo masses a clear correlation emerges. This shows that the weakening of the $\lambda_{\rm subhalo}$--\hi\ mass correlation at fixed subhalo mass in Figure~\ref{fig:spin parameter-subhalo}  in the transition region is driven by the effect of AGN feedback. We also find a similar, albeit weaker trend in  satellite subhaloes, meaning that AGN feedback is also playing a role in reducing the \hi\ content of massive satellites {\tt type=1}. This is similar to what we saw for the entire haloes: the significant increase in the scatter of the \hi\ mass-subhalo mass relation is driven by AGN feedback.


\begin{figure}
\centering
    \includegraphics[width=\linewidth]{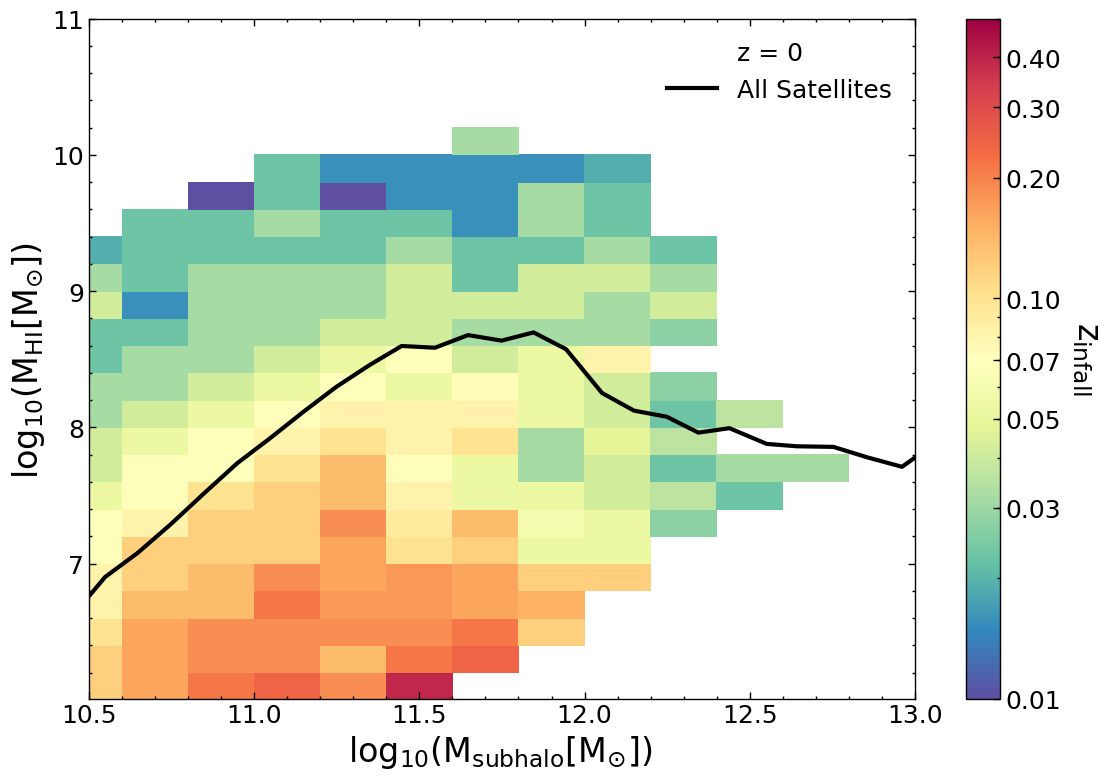}
\caption{The HIHM relation of the satellite subhaloes (type = 1) in \shark-ref at $z=0$, with each bin coloured by the median \subsuperscript{z}{infall}{} of the subhalo. The solid line represents the median \hi\ mass of all the satellite subhalos, irrespective to their \subsuperscript{z}{infall}{} as a function of \subsuperscript{M}{subhalo}{}. A clear trend is seen between the \hi\ of the satellite subhaloes and their \subsuperscript{z}{infall}{}, with later the \subsuperscript{z}{infall}{}, more \hi-rich is the satellite subhalo.}
\label{fig:zinfall-subhalo}
\end{figure}

In order to understand the \hi\ in satellite subhaloes and their lack of correlation with the subhalo's spin parameter, we explore the correlation between the \hi\ mass of the satellite subhalo and the redshift at which the subhalo became a satellite subhalo,  \subsuperscript{z}{infall}{}. In Figure~\ref{fig:zinfall-subhalo}, we plot the \subsuperscript{M}{HI}{}--\subsuperscript{M}{subhalo}{} relation for  satellite subhaloes, colouring by the median \subsuperscript{z}{infall}{} of the satellite subhaloes in each bin at $z=0$. For this figure we limit ourselves to using medi-\surfs\ only, as there are not enough satellite subhaloes in micro-\surfs\ for a statistical study at $M_{\rm subhalo}<10^{11}\,\rm M_{\odot}$. A clear trend emerges, where we see later infalling subhaloes being \hi-richer than earlier infallers. 

We remind the reader that here we are only including {\tt type=1} satellites, as these quantities are not well defined for {\tt type=2} satellites. This trend is expected as in \shark\ we implement instantaneous stripping of the hot halo of subhalos that become satellites, leaving the ISM to exhaust itself by continuing star formation. This process of stripping plus starvation is the cause for the loss of correlation with the subhalo's spin.


\section{Developing a numerical model to populate dark matter haloes with \hi}
\label{sec:Model-development}

The relation between \hi\ and the underlying distribution of DM will be explored in significant detail over the coming years thanks to the advent of the SKA and its pathfinders. Hence, it becomes imperative that physical galaxy formation models explore the ways in which \hi\ and DM trace each other in advance of these experiments. Most atomic hydrogen is expected to reside in dense systems in or around galaxies, where \hi\ is shielded from ionising UV photons \citep{Spinelli2019_Marta}. Understanding this distribution and evolution opens up new avenues for cosmology and galaxy evolution. A significant challenge in \hi\ cosmology applications is the requirement to produce thousands of mock observations to measure the statistical uncertainties in parameter determinations. The only plausible way of doing this is by approximate $N$-body, dark-matter only simulations (see \citealt{Howlett2015a} for an example in the optical and \citealt{Howlett2015b} for an example of fast methods to produce $N$-body halo catalogues). Having a physical way of populating these simulations with \hi\ is a crucial step.

As discussed previously, both the functional form and scatter of this relation can be described in terms of non-baryonic halo properties. This presents a unique advantage and the possibility to apply the phenomenological behaviour in which \hi\ traces DM haloes we described above to large simulations. In this section we present a numerical method to populate DM haloes with \hi\ based on \shark-ref. We perform exhaustive fits to the relations analysed in Section~\ref{sec:Understanding-HI-halo} in the same three halo mass regimes presented there.

We develop our numerical model in the redshift range $0\le z \le 2$, as \shark\ predictions for the cosmic density of \hi\ starts to deviate significantly from the observations at higher redshifts (see \citealt{Lagos2018-Shark,Hu2019}). \citet{Lagos2018-Shark} argue that the reason for this discrepancy is the fact that \shark\ models only the \hi\ content in the ISM of galaxies, while it does not explicitly model the \hi\ content in the circumgalactic medium. Hydrodynamical simulations, e.g. \citet{VandeVoort2012, Diemer2019}, show that at $z\gtrsim 2$ the majority of \hi\ resides in the circumgalactic medium. 

We caution the reader that the fits presented here are for one physical model of galaxy formation (\shark-ref), though we do expect different models to behave differently (see Figure~\ref{fig:HI_showing_plot_SAM}). Hence, this should not be taken as a unique way of populating haloes in DM-only simulations with \hi, but {\it a} way of doing it that reflects a physical model that matches a variety of observational constraints.

\subsection{The total \textsc{Hi}--halo mass scaling relation}
\label{subsec:HI-halo_scaling}

To develop our numerical model for how to populate haloes with \hi\ (here \hi\ being the total \hi\ content of a halo), we perform a fit to our simulation in two parts. We first fit the shape of the relation, $f_{M_{\rm HI}}(M_{\rm vir},z)$, which depends solely on halo mass and redshift, and then a perturbation component, $\delta_{M_{\rm HI}}$, which scales with halo properties other than mass,
\begin{equation}
    \log_{10}(M_{\rm HI}) = f_{M_{\rm HI}}(M_{\rm vir},z) + \delta_{M_{\rm HI}}.
    \label{eq:2-part-equation}
\end{equation}
The median HIHM relation of \shark\ is fitted with a polynomial function $f_{M_{\rm HI}}(M_{\rm vir},z)$, with the fit done in bins of $0.1$~dex of halo mass. We use different polynomial fits for different regions, which will be expanded upon later in this section. Our polynomial fit for the median can formally be written as
\begin{equation}
    f_{\rm M_{\rm HI}}(M_{\rm vir},z) = \sum_{i=0}^{n}\ a_{i}(z) \left(\log_{10}(M_{\rm vir})\right)^{i}.
    \label{eq:spline_fit}
\end{equation}
The value of $n$ differs between halo mass regions: $n = 2,\ 5,\ \text{and}\ 1$ respectively for the low-mass, transition, and high-mass regions. These were found upon iterating with different dimensions and finding the minimum $n$ that provides a reasonable fit. 

After fitting the median, we use the {\sc R} \textsc{hyper-fit} package of \citet{Robotham-Hyperfit} to fit a plane to the residual scatter ($\delta_{\rm M_{HI}}$) around the HIHM relation. \textsc{hyper-fit} derives a general likelihood function that is maximised to recover the best-fitting model describing a set of $D$-dimensional data points with a ($D - 1$)-dimensional plane, with some intrinsic scatter. The secondary parameters involved in fitting the residual scatter vary according to regions. Sections~\ref{subsec:lowmass}, \ref{subsect:middlemass} and \ref{subsec:highmass} provide details of these fits for the low-mass, transition, and high-mass regions, respectively. We report the vertical scatter around the best fit plane provided by {\sc hyper-fit} and use that to quantify the goodness of the fit.


\subsubsection{H\,{\textsc i}--halo scaling relation: Low-mass region}\label{subsec:lowmass}

For the \textit{low-mass region}, we use a quadratic ($n=2$) polynomial to fit the median \hi--halo relation. A quadratic is needed to incorporate the slight downturn seen at the end of the low-mass region (around $M_{\rm vir} \! \simeq \! 10^{11.8}\,{\rm M}_{\odot}$). We find that the best-fitting coefficients of the median relation change with redshift. This redshift dependence can itself be fitted well with polynomials, as follows: 

\begin{equation}
    \begin{aligned}
    & a^{\rm low}_{\rm 0} = -101.322 + 15.853\,z, \\ & a^{\rm low}_{\rm 1} = 17.982 - 2.757\,z - 1.9808\,z^2, \\
    & a^{\rm low}_{\rm 2} = -0.7725 + 0.2759\,z,
    \end{aligned}
    \label{eq:low_mvir_coefficients}
\end{equation}
where \subsuperscript{a}{0-2}{low} are the coefficients for the polynomial fit of Equation \ref{eq:spline_fit} for the low-mass region, and $z$ is redshift. 

We have shown in Section \ref{subsubsec:spin_parameter_effect} that for fixed $M_{\rm vir}$ in the low-mass region, the halo spin parameter is strongly correlated with the amount of \hi\ contained in the halo. We therefore use that as our sole property to constrain the scatter in this region.  When fitted, we find
\begin{equation}
    \delta^{\rm low}_{M_{\rm HI}}(\lambda_{\rm h}) =
    \begin{aligned}
    & 1.433 \left(\log_{10}(\lambda_{\rm h})\right) + 2.124.
    \end{aligned}
    \label{eq:low_vir_fit}
\end{equation}
Here, $\lambda_{\rm h}$ is the halo spin parameter. We get a vertical scatter of $\sigma = 0.19$~dex around our relation when we fit the residual scatter of the HIHM relation with $\lambda_{\rm h}$ using \textsc{hyper-fit}.
By residual scatter we refer to the residual left after subtracting the fitted $f_{M_{\rm HI}}(M_{\rm vir},z)$ from the intrinsic \shark-ref \subsuperscript{M}{HI}{} values.
We find that the residual scatter-$\lambda_{\rm h}$ fit for the low-mass region does not change significantly over the redshift range $0\le z\le 2$ and hence, the above equation is at least valid for $z\!\lesssim\!2$, which is the tested regime.


\subsubsection{H\,{\textsc i}--halo scaling relation: Transition region}\label{subsect:middlemass}

The fitting is the hardest at the \textit{transition region}, as this region is dominated by AGN feedback, and the inherent scatter cannot be defined solely on halo properties. When we focus solely on halo properties, it is seen in Figures~\ref{fig:spin parameter} and \ref{fig:fraction_satellite} that both halo spin parameter and \fracmvir\ play a role in defining the scatter of the transition region. 

When fitting the median relation, $f_{M_{\rm HI}}(M_{\rm vir},z)$, we use a quintic ($n=5$) polynomial fit for our model, in order to incorporate the squiggle seen in the region from \subsuperscript{M}{vir}{}$=$ \solarValue{11.8}--\solarValue{13}. The coefficients for this fit have been tabulated in Table~\ref{tab:transition-parameters}, as the parameters of the fit change with redshift in ways that are not easy to parametrise.

We note that although the halo spin parameter becomes less important in the \textit{transition region}, haloes with a higher spin systematically retain more \hi\ up to \subsuperscript{M}{vir}{} $\simeq$ \solarValue{12.5}. In this region (\subsuperscript{M}{vir}{} $<$ \solarValue{12.5}), the \hi\ is still prominently contained in the central galaxies of these haloes, even though we see the beginning of the emergence of satellite population. At \subsuperscript{M}{vir}{} $\gtrsim$ \solarValue{12.5}, satellites become the dominant \hi\ reservoirs of the halo and the host halo's spin parameter is not a meaningful property to define the \hi\ content of satellite subhaloes. When we lose the spin parameter dependence, the vertical scatter around the best fit plane in the transition region is captured almost entirely by \fracmvir.

We find the HIHM relation's scatter to be reasonably well captured by
\begin{multline}
    \delta^{\rm TR}_{\rm M_{\rm HI}}\left(\lambda_{h},\frac{M^{\rm sat}_{\rm h}}{M_{\rm vir}}\right) = 
    b_{\rm frac}(z) {\rm log}_{10}\left(\frac{M^{\rm sat}_{\rm h}}{M_{\rm vir}}\right) \\ 
    +   b_{\lambda}(z) {\rm log}_{10}(\lambda_{\rm h}) + b_{\rm 0}(z),
\end{multline}
\begin{equation}
    \begin{aligned}
        & \text{with}\\
    & b_{\rm frac}(z) = 0.25\,e^{-z} + 0.2192,\\
    & b_{\lambda}(z)  = 2.77\,e^{-z} + 0.7854,\\
    & b_{\rm 0}(z) = 4.56\,e^{-z} + 1.4041.
    \end{aligned}
    \label{eq:transition_fit}
\end{equation}

We find that the scatter around the transition region changes considerably with redshift. This is due to both the AGN and subhalo populations being markedly different at earlier epochs. Despite our finding in Section~\ref{subsubsec:other-halo-properties} and Appendix~\ref{appendix:understanding-the-scatter}, that there is a slight correlation between the \hi\ content of haloes and their \agefifty, we did not find \agefifty~to be useful at reducing the vertical scatter compared to $\lambda_{\rm h}$ and \fracmvir.

When we fit the residuals using \textsc{hyper-fit}, we obtain a vertical scatter of $\sigma = 0.91$~dex around the plane at $z=0$. Although this is much larger than the $0.19$~dex we achieve in the low-mass region, we find the vertical scatter decreasing to $\sigma = 0.27$~dex at $z=2$, making it highly redshift dependent. This is due to the fact that as we move to higher redshift, the AGN influence decreases and so does the scatter dependence on it, thus making it easier to fit the relation with spin parameter and \fracmvir\ at those redshifts. 

Another aspect which was discussed in Section~\ref{subsubsec:all_other_effect} is that the scatter in this region can be better described with baryon properties; for example, using the BH-to-stellar mass ratio of the central galaxy instead of \fracmvir\ brings the \textsc{hyper-fit} vertical scatter down to $\sigma = 0.8$~dex at $z=0$. But as the goal of this analysis is to define the HIHM scaling relation solely on the basis of halo properties, baryons are not included.

\subsubsection{H\,{\textsc i}--halo scaling relation: High-mass region}\label{subsec:highmass}

In the \textit{high-mass region} the dependence of \hi\ mass on the spin parameter or \agefifty\ becomes negligible. This is because haloes' \hi\ content is almost entirely contained in satellite galaxies. Thus, in this region we see that at fixed halo mass the \hi\ content is primarily correlated with the number of substructures present in that particular halo. In this region, we find that a linear function (i.e. polynomial fit of $n=1$) is sufficient to describe the dependence of the median \hi\ mass on halo mass.

The coefficients of this linear fit vary as a function of redshift as follows:
\begin{equation}
    \begin{aligned}
    & a_{\rm 0}^{\rm high} = -8.9448 + 8.7511\,z - 5.153\,z^2 + 0.891\,z^3,\\
    & a_{\rm 1}^{\rm high} = 1.3918 - 0.4618\,z + 0.1756\,z^2. 
    \end{aligned}
\end{equation}
The scatter in the HIHM relation is then fitted as
\begin{equation}
    \delta^{\rm high}_{\rm M_{\rm HI}}\left(\frac{M^{\rm sat}_{\rm h}}{M_{\rm vir}}\right) = 
    \begin{aligned}
    & b_{\rm frac}(z) \log_{10}\left(\frac{M^{\rm sat}_{\rm h}}{M_{\rm vir}}\right) + b_{\rm 0}(z), 
    \end{aligned}
    \label{eq:high_vir_fit}
\end{equation}
with
\begin{equation}
    \begin{aligned}
    & b_{\rm frac}(z) = 0.498\, e^{-z} + 0.11,\\
    & b_{\rm 0}(z) = 0.669\, e^{-z} + 0.1734.
    \end{aligned} \label{scatterhighmas}
\end{equation}
The vertical scatter from \textsc{hyper-fit} for this fit comes out to be $\sigma = 0.3$~dex at $z=0$, making it a good fit for the residuals. As we move from $z=0$ to $z=2$, we find  $\sigma$ changing from $0.3$~dex to $0.23$~dex, which is a weak change and could be driven by the decreasing number of haloes in the high-mass region at higher redshifts.


\subsection{Assessing the effectiveness of the numerical model for the  \textsc{Hi}--halo mass scaling relation}

Figure~\ref{fig:model-fit-all} compares the actual \hi\ content of \shark-ref haloes at $z=0$ with that from our fitted HIHM scaling relation as applied to the same underlying halo population (see Equations \ref{eq:spline_fit} to \ref{eq:high_vir_fit}). 

We can see that our numerical model produces a comparable relation to that of the intrinsic model for the low-mass and high-mass regions, highlighting the fits approximately capture the correct amount of scatter. However, for the \textit{transition region}, the \subsuperscript{16}{}{th} and \subsuperscript{84}{}{th} percentiles of our fit are higher than in the \shark-ref model, though when comparing with the cumulative \hi\ in the simulation boxes (see Section~\ref{subsec:cumulative_HI}), this might not make a huge inconsistency as the percentage of \hi\ contributed from this region is small.

This numerical model represents significant progress over previous work, which focused only on the median \hi\ content of haloes, without considering the scatter around the relation. This is important as \hi-selected surveys will always be preferentially biased towards the more gas-rich systems rather than the typical at fixed halo mass. To properly capture this effect in mock observations it is crucial to have an understanding on how much scatter the underlying relation displays. Our numerical model offers exactly this and hence we expect it will prove useful for future \hi\ surveys planning.

\subsection{H\textsc{i} evolution with redshift}
\label{subsec:appendix-section-5}

It has been shown in previous sections (see Section \ref{subsec:HI-halo_scaling}), that the coefficients of $f_{\rm M_{\rm HI}}$ are redshift-dependent. We find that as we move to higher redshifts, AGN feedback is less efficient at preventing the halo gas from cooling. Also, haloes at higher redshifts have not had enough time to assemble all their mass, leading to a lesser number of substructures. Both of these factors  significantly contribute to the evolution in the scatter at the \textit{transition region}. By $z=2$, the transition is barely visible. We explore this in detail in Appendix~\ref{subsec:redshift_dependence}.

\begin{figure}
  \includegraphics[width=\linewidth]{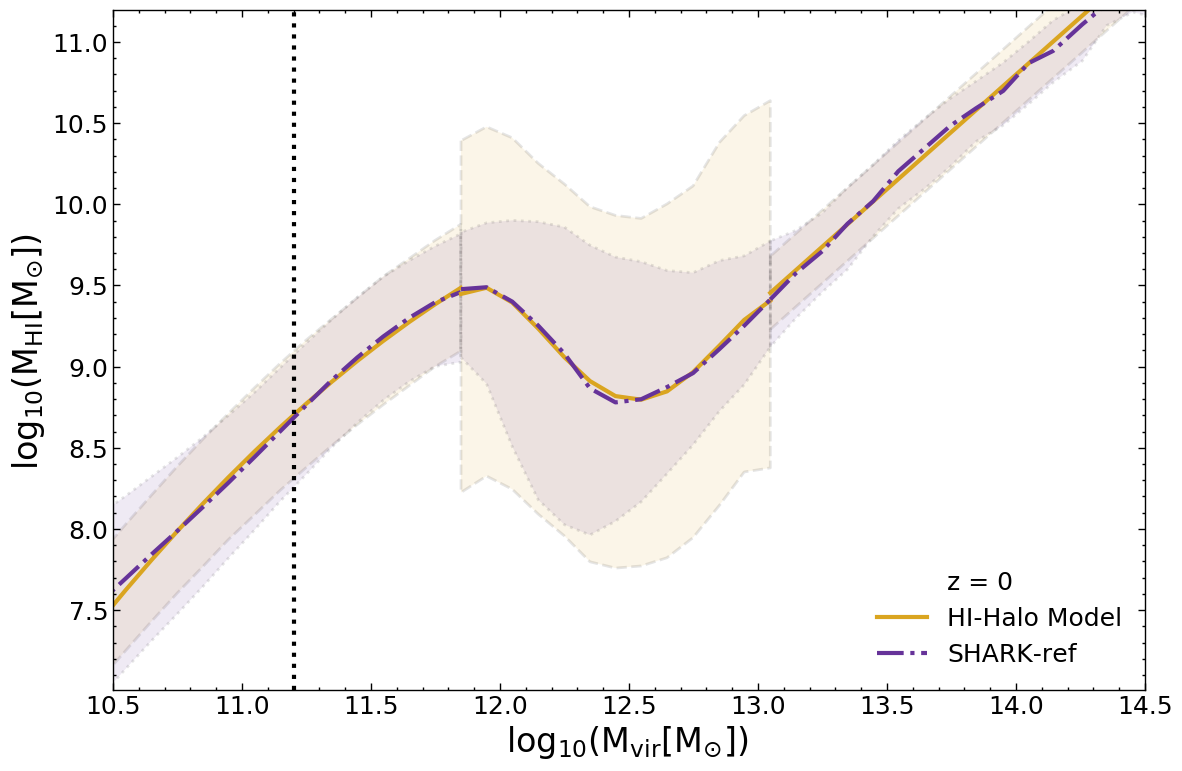}
\caption{Overall \hi\ content in a halo as a function of \subsuperscript{M}{vir}{} as predicted by our scaling relation (see Section \ref{sec:Model-development}), which was fitted to the  output of \shark-ref compared here. The purple (dot-dashed) and yellow (solid) line represent our the median relations for \shark-ref and our model, respectively. The shaded region of corresponding colour around each relation shows the \subsuperscript{16}{}{th}--\subsuperscript{84}{}{th} percentile range. Our scaling relation stays close to the values predicted by the SAM, but we see a slightly higher scatter in the transition region}

\label{fig:model-fit-all}
 \end{figure}

We have seen in Figure \ref{fig:model-fit-all} that the transition region is characterised by  a large scatter that is difficult to fully account for with halo properties alone.
It it therefore informative to know how much \hi\ in \shark-ref resides in the transition region, to quantify the impact inaccurate estimates of the scatter can have on studies that focus on unveiling the total \hi\ content of the Universe.
We find at $z=0$, about $25$\% of the total \hi\ resides in the transition region and $60$\% in the low-mass region. At $z=2$, we find that almost $80$\% of the \hi\ in the \shark-ref resides in the low-mass region. We explore the evolution of the cumulative \hi\ in \shark-ref in Appendix \ref{subsec:cumulative_HI}.


\section{Conclusions}
\label{sec:Discussion}

Understanding the evolution of \hi\ throughout cosmic time provides key insights into cosmology and galaxy evolution. Unlike the stellar--halo mass relation, the HIHM relation is not necessarily monotonic and is likely to be characterised by a large scatter (given the large scatter in the \hi--stellar mass relation; \citealt{Catinella2018}). In this paper, we have used the state-of-the-art semi-analytic galaxy formation model \shark, with the aim of understanding the physical processes behind the shape, scatter and evolution of the \hi--halo mass relation at $0\le z\le 2$.

We compared the \hi--halo mass relation and the \hi\ clustering of \shark\ with available observations. These observations were not used as part of the tuning of the free parameters of \shark, and can hence be considered predictions. We find the predicted \hi\ clustering in \shark\ to be in excellent agreement with the observations. However, when comparing with observational inferences of the \hi--halo mass relation, coming mostly from \hi\ stacking of groups, we found that \shark\ reproduces well the \hi\ abundance in halos of masses $<10^{12}\,\rm M_{\odot}$ and $>10^{13.3}\,\rm M_{\odot}$, but in the range $10^{12}-10^{13.3}\,\rm M_{\odot}$, \shark\ under-predicts the abundance of \hi\ in halos. In an upcoming paper (Chauhan et al. in preparation), we show that these discrepancies are largely due to the uncertainty in group definition around that halo mass (that in current spectroscopic surveys have a small occupancy).

We then explored the effect of different physical processes in the shape of the HIHM relation, and what properties of halos are the best secondary parameter that correlates with the scatter in the HIHM relation. Our key results can be summarised as follows,
\begin{itemize}
   
   \item  The HIHM relation is characterised by three mass regions that display distinct behaviours. At $z=0$, we find that the total \hi\ content of haloes with \subsuperscript{M}{vir}{} $<10^{11.8}$\M, aka low-mass region, increases monotonically with the halo mass. In halos of masses $10^{11.8}\,{\rm M}_{\odot}<M_{\rm vir}<10^{13}\,\rm M_{\odot}$, aka the transition zone, the total \hi\ content of haloes peaks at \subsuperscript{M}{vir}{} $=$ \solarValue{12} and then declines with increasing halo mass. For halos of masses $M_{\rm vir}>10^{13}\,\rm M_{\odot}$, aka the high-mass region, the total \hi\ content of haloes starts to increase again with increasing halo mass. The scatter around the HIHM varies significantly in the three mass regions, being $\sim 0.5$~dex, $\sim 1.2$~dex and $\sim 0.4$~dex in the low-mass, transition and high-mass regions, respectively. 
    
   \item We find the contribution to the total \hi\ mass of the halo to be dominated by central galaxies for haloes of \subsuperscript{M}{vir}{} $<$ \solarValue{12.5}.  At higher halo masses,  satellite galaxies are the dominant contributor. The bump seen in the HIHM relation in the transition zone is caused by central galaxies, while the total \hi\ mass contributed by satellites scales monotonically with halo mass. The latter is what produces the increasing \hi\ mass with increasing halo mass in the high-mass zone.

   \item The peak of the HIHM relation in the transition region and the halo mass at which this peaks happens are largely determined by the AGN feedback efficiency, with stellar feedback, photoionisation feedback and ISM modelling  playing a lesser role. The dip in the HIHM relation is caused by the suppression of gas cooling in these haloes due to the influence of AGN feedback. At lower halo masses, AGN does not play an important role.
   
   \item We isolate the main secondary parameter responsible for the scatter of the HIHM relation. In the low-mass region, the scatter at fixed halo mass is highly correlated with the spin parameter of the halo, whereas for the high-mass zone, the scatter is correlated with the fractional contribution from substructure to the total halo mass, \fracmvir. As for the transition zone, we find the scatter to be highly dependent on the black hole-to-stellar mass ratio of the central galaxy, reflecting the importance of AGN feedback in this region. However, when we explored halo properties only, we find that a combination of halo's spin and \fracmvir\ is relatively successful at characterising the scatter of the HIHM relation in the transition zone. Once these secondary dependencies are included, the vertical scatter of the 2-dimensional plane (between the median-subtracted halo mass \hi~mass and the secondary parameter) at $z=0$ is significantly tighter than the HIHM relation, with values of $\sim$0.19~dex, $\approx 0.91$~dex and $0.3$~dex, in the low-mass, transition and high-mass zones, respectively. 
    
   \item As we move to higher redshifts, the transition zone starts to shrink, as AGN feedback becomes less efficient. The vertical scatter in the 3-dimensional plane over the transition zone decreases significantly with redshift, from $\sigma = 0.91$~dex at $z=0$ to $\sigma = 0.27$~dex at $z=2$. The latter values for the scatter already consider the dependency on spin and \fracmvir. In the low and high mass regions the decrease in the scatter is not as significant as in the transition zone, with the low-mass region hardly seeing a decrease in the vertical scatter (remaining at $\sim 0.19$~dex once the halo spin is considered) and the high mass region sees a decrease from $\sim 0.3$~dex at $z=0$ to $\sim 0.23$~dex at $z=2$, once \fracmvir\ is considered. By $z=2$, the HIHM relation is monotonic over the whole halo mass range.
\end{itemize}

Finally, we use the lessons learned to develop a numerical model to populate halos in DM-only simulations with \hi, depending on their halo mass, spin parameter, \fracmvir\ and redshift. Obvious applications of this numerical model include \hi\ intensity mapping,  \hi\ stacking and modelling of \hi\ clustering. This study also opens up avenues for exploring the role of different halo properties in the HIHM relation. With the upcoming SKA and its Pathfinders, we will be able to explore the role of halo properties in the HIHM relation observationally, providing better constraints and deeper insight into the HIHM relation.


\section*{Acknowledgements}

We would like to thank Marta Spinelli, Hong Guo, Jian Fu, Kristine Spekkens, Cullan Howlett, Mat\'ias Bravo, St\'ephane Courteau and Rob Crain for their constructive comments and useful discussions.
We also thank Aaron Robotham, Rodrigo Tobar and Pascal Elahi for their contribution towards \surfs\ and \shark, and Mark Boulton for his IT help.
GC is funded by the MERAC Foundation, through the Postdoctoral Research Award of CL, and the University of Western Australia.
Parts of this research were carried out by the ARC Centre of Excellence for All Sky Astrophysics in 3 Dimensions (ASTRO 3D), through project number CE170100013. CL is funded by ASTRO 3D. ARHS acknowledges receipt of the Jim Buckee Fellowship at ICRAR-UWA. This work was supported by resources provided by the Pawsey Supercomputing Centre with funding from the Australian Government and the Government of Western Australia. 

\section*{Data Availability}
The data that support the findings of this study are available upon request from the corresponding author, GC. The \surfs\ simulations used in this work can be freely accessed from
\url{https://tinyurl.com/y4pvra87} (micro-\surfs) and \url{https://tinyurl.com/y6ql46d4} (medi-\surfs).



\bibliographystyle{mnras}
\bibliography{Correct_bibliography}



\appendix

\section{Understanding the Shape - cont.}
\label{appendix:understanding-shape}

In this section, we explore a bit more on the impact using different models and parameters for a physical process have on the shape of the HIHM relation, which we have only briefly discussed in the main paper.


\subsection{Photoionisation effect}
\label{subsubsec:Photoionisation_Effect}

We discuss the implementation of photoionisation feedback in \shark\ in Section~\ref{subsubsec:Photoionisation_Feedback}, and briefly touched on the effect changing its parameters has on the HIHM relation in Section~\ref{subsubsec:moved_processes_summary}. Here we show the effect of photoionisation feedback on the overall \hi\ content of the haloes at $z=0$. We vary the value of $v_{\rm cut}$, which from Equation \ref{eq:photoionisation_feedback} directly affects the circular velocity ($v_{\rm thresh}$) of haloes under which the halo gas is not allowed to cool down.

\begin{figure}
  \includegraphics[width=\linewidth]{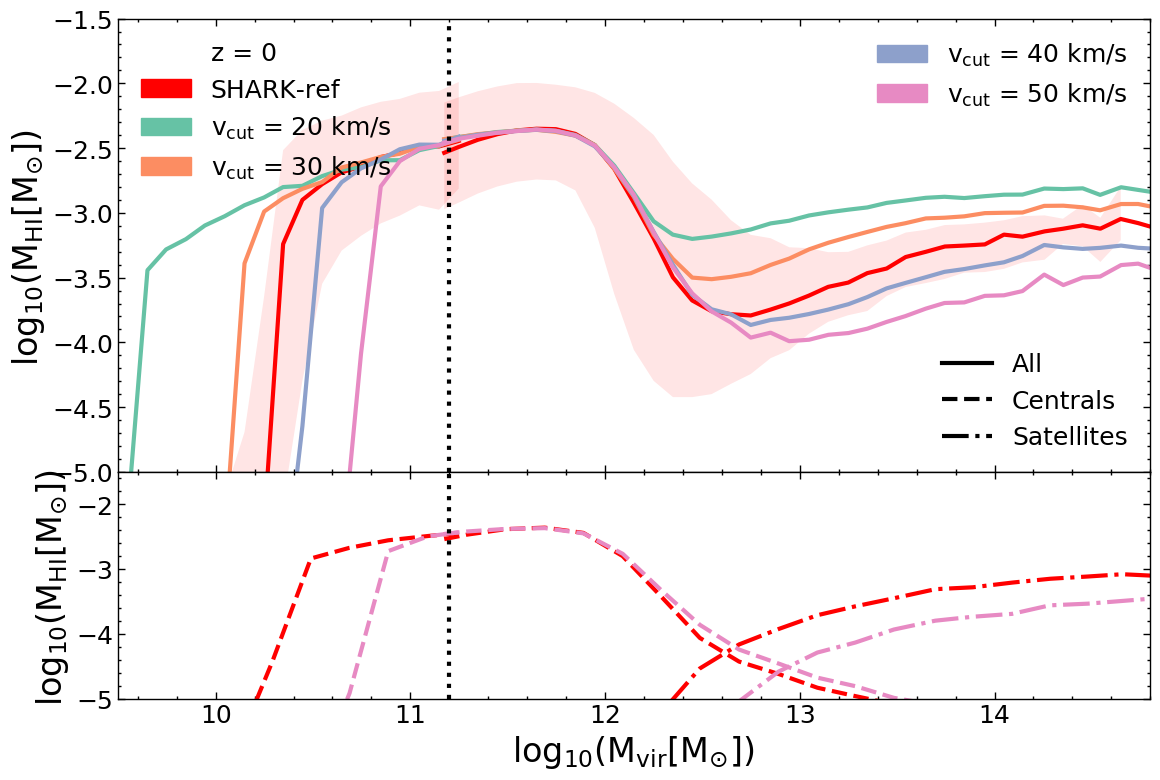}
\caption{As in Figure~\ref{fig:AGN-models} but for different values of \subsuperscript{v}{cut}{}, which represents the virial velocity threshold under which the gas in haloes is assumed to be kept ionised by the UV background, and is hence not allowed to cool down and replenish the interstellar medium of the central galaxy (see Equation \ref{eq:photoionisation_feedback}).
Different colour lines represent different \subsuperscript{v}{cut}{} values, with red representing the default \shark-ref model and the shaded region being the \subsuperscript{16}{}{th}--\subsuperscript{84}{}{th} percentile range. Photoionisation heating does not affect the knee of the HIHM relation, though it does affect the amount of \hi\ contained in haloes in the low and high mass regions. \textit{Lower Panel}: The median \hi\ contribution from centrals and satellites to the total \hi\ of the halo. For clarity we only show \shark-ref and the $v_{\rm cut}=50$ \kms variation. Unlike previous figures, changing the photoionisation feedback leads to a change in the \hi\ content of satellites, with higher the feedback the lesser the amount of \hi\ in satellites.
}

\label{fig:Photoionisation-models}
 \end{figure}

The effect of varying $v_{\rm thresh}$ is presented in Figure~\ref{fig:Photoionisation-models}, where haloes below a certain mass (which correspond to the circular velocity $v_{\rm thresh}$) do not have \hi\ in them, as the halo gas is kept ionised. As expected, increasing $v_{\rm cut}$ has the effect of shifting the steep decline of the \hi\ fraction-halo mass relation to higher halo masses. Though, changing $v_{\rm cut}$ value does not have any effect on the \textit{transition region} - the drop essentially remains at the same \subsuperscript{M}{vir}{} value (\solarValue{12}) for all the variations. We find that photoionsation feedback becomes more prominent for the \hi\ content of haloes after \subsuperscript{M}{vir}{} $>$ \solarValue{12.4}, with a smaller $v_{\rm cut}$ driving a higher \hi\ content in haloes. This effect is due to smaller haloes being allowed to cool down their halo gas under smaller $v_{\rm cut}$ values, increasing their \hi\ content. These centrals of low-mass haloes can then become satellites of larger haloes and contribute to the total \hi\ content of that halo.


In the lower panel of Figure~\ref{fig:Photoionisation-models}, we compare the \hi\ contributions from satellites and centrals for the \shark-ref and $v_{\rm cut} = 50$\kms  runs. We find that the central \hi\ contribution remains almost unchanged in both the runs, except for the halo mass below which the \hi\ content sharply decreases, which is at \subsuperscript{M}{vir}{} $\approx$ \solarValue{10.4} and \subsuperscript{M}{vir}{} $\approx$ \solarValue{11} for \shark-ref and $v_{\rm cut} = 50$\kms\ runs, respectively. On the contrary, the contribution from satellite galaxies are different in these runs, with \shark-ref having higher \hi\ content in satellites than the other extreme run. The latter is due to the galaxies that become satellites being more \hi-rich with smaller $v_{\rm cut}$ values.
 
\subsection{Interstellar medium model effect}
\label{subsubsec:SF_model_effect}

In \shark, stars form from molecular gas, and different models are implemented for how to split the ISM into ionised, atomic and molecular gas phases. Here, we compare two models for the molecular-to-atomic gas partition, specifically the BR06 (the default model of choice) and the GD14 model. In both cases, stars are formed from the molecular gas with a fixed efficiency (see Equation~\ref{eq:SFmodel}). A brief overview of the effect of changing the ISM model had been given in Section~\ref{subsubsec:moved_processes_summary}, here we delve into more details.

\begin{figure}
  \includegraphics[width=\linewidth]{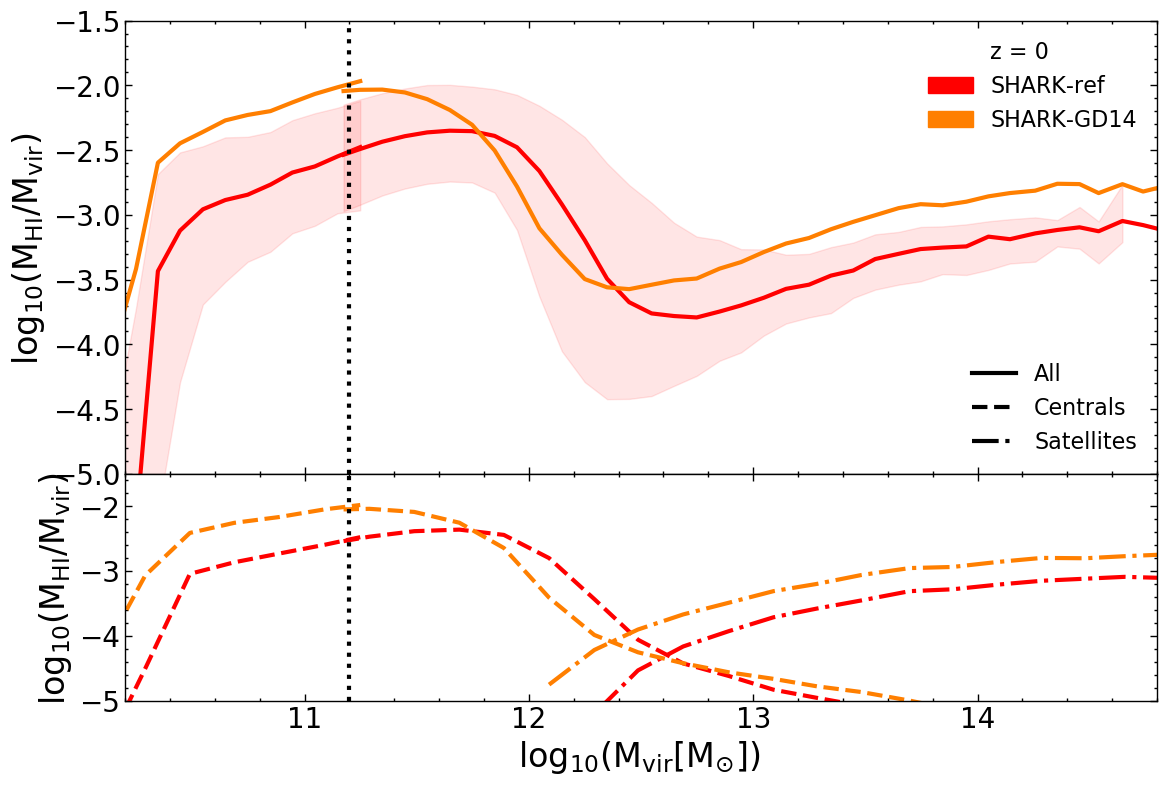}
\caption{As in Figure~\ref{fig:AGN-models} but for two variations of the molecular-to-atomic interstellar gas partition in \shark. The models being compared are the default \shark\  model as shown in \citet{Lagos2018-Shark}, which incorporates the \citet{Blitz2006} prescription (\shark-ref) to split atomic and molecular gas in the interstellar medium of galaxies, with a variant adopting the \citet{Gnedin2014LINEGALAXIES} atomic-to-molecular transition prescription (\shark-GD14). In both variants, stars from the molecular gas with the same efficiency. The top panel shows the entire \hi\  fraction whereas the bottom panel shows the central and satellite contributions.}

\label{fig:SF-models}
 \end{figure}

Figure~\ref{fig:SF-models} shows \subsuperscript{M}{HI}{}/\subsuperscript{M}{vir}{} as a function of \subsuperscript{M}{vir}{} for different \hmol-to-\hi\ partition models that are implemented in \shark, with the top-panel showing the total \subsuperscript{M}{HI}{}/\subsuperscript{M}{vir}{} ratio, and the bottom-panel showing the centrals and satellite contributions at $z = 0$. When using the GD14 prescription, the overall \hi\ content of haloes is higher than when adopting the BR06 prescription, except at halo masses between $10^{12}\,\rm M_{\odot}$ and $10^{12.7}\,\rm M_{\odot}$. The \textit{transition region} for the model adopting the GD14 prescription is at a lower halo masses, \subsuperscript{M}{vir}{} $\approx$ \solarValue{11.5} against \subsuperscript{M}{vir}{} $\approx$ \solarValue{12} for BR06. In this transition region, BR06 predicts a slightly higher abundance of \hi. However, at lower and higher halo masses, GD14 results in higher \hi\ content. The fact that centrals of low mass haloes are more \hi-rich in GD14 than BR06 is the cause for the higher \hi\ abundance at high halo masses, as many of the low mass centrals become satellites as time progresses.

When we compare the \hi\ contributions of centrals and satellites to the overall \hi\ of the halo (bottom panel in Figure~\ref{fig:SF-models}), we find that the \hi\ contribution of centrals in GD14 is higher than BR06 for haloes \subsuperscript{M}{vir}{} $<$ \solarValue{12}, while at higher masses there is virtually no difference. This happens due to the fact that \hi-\hmol\ partition in GD14 depends on the gas metallicity (among other parameters). This is not the case for BR06, which is a purely pressure-based model. \shark-ref predicts low metallicities for low-mass galaxies, which in turn makes the \hi\ value for low mass haloes to be higher in GD14, as the \hi\ in low-mass haloes is dominated by the centrals. 
As for the satellite contribution, we see that GD14 consistently predicts more \hi\ than BR06 throughout all virial masses, again due to the gas metallicity effect. 
The fact that the transition region happens at lower halo masses in GD14 than in \shark-ref is, however, unrelated to the SF law. We find that BH masses are slightly bigger at intermediate mass galaxies (around the break of the stellar mass function) in GD14, causing AGN feedback to be more efficient than in \shark-ref in those galaxies. As seen before, more efficient AGN feedback shifts the transition region to lower halo masses, which is effectively what happens in the GD14 run. This again highlights the complex interplay between the different baryon physics in models such as \shark.

  
 \subsection{Gas stripping effect}
 \label{subsubsec:Stripping_effect}

The last effect we want to test is the environmental effect, which we do by comparing the effect turning `off' ram-pressure stripping has on the overall \hi\ content of the haloes (see Section \ref{subsubsec:stripping}).

\begin{figure}
  \includegraphics[width=\linewidth]{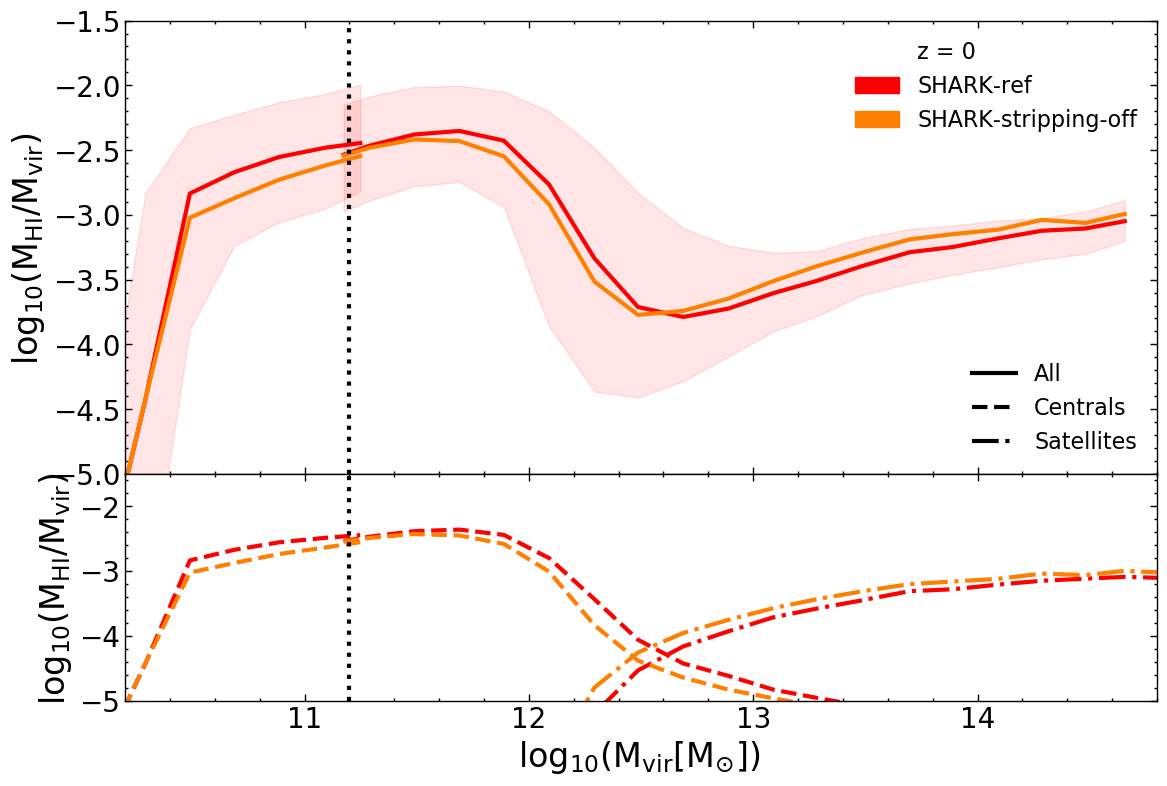}
  \caption{Similar to Figure~\ref{fig:SF-models}, but for the default model (red) vs a model with no gas stripping (yellow).
The top panel shows the total \hi\ fraction, whereas the bottom panel shows the central and satellite contributions. Differences between the two models are clear when we decompose the \hi\ contribution between centrals and satellites, but these difference compensate each other so that the total \hi\ in haloes is barely affected. The top panel shows the entire \hi\  fraction whereas the bottom panel shows the central and satellite contributions.}
\label{fig:stripping_models}
 \end{figure}

In Figure~\ref{fig:stripping_models} (top panel), we compare  \subsuperscript{M}{HI}{}/\subsuperscript{M}{vir}{}--\subsuperscript{M}{vir}{} relation for stripping mode `on' and `off' as a function of \subsuperscript{M}{vir}{}. We find that the total amount of \hi\ in either model is approximately the same, though stripping `off' tends to lead to a slightly lower \hi\ in the transition region and higher \hi\ in the high mass region. 

When looking at the central-satellite galaxies contribution to the total \hi\ mass of the halo (bottom panel), we find centrals to reduce their \hi\ content when stripping is off, while satellites become more important. This happens because when stripping is off, satellites are able to hold on their hot haloes for longer, which means that the hot halo of the central is now less massive than in the run with stripping. This leads to central galaxies accreting less gas (due to the smaller overall reservoir of gas), while satellite can continue to accrete gas for longer. Clearly these two competing effects compensate relatively well as to lead to small differences in the total \hi\ content of halos at $10^{12}\,\rm M_{\odot}<M_{\rm vir}<10^{13.5}\,\rm M_{\odot}$.

\section{Formation Age Effect}
\label{appendix:understanding-the-scatter}

Here we discuss the effect halo formation age has on the scatter in the HIHM relation. 
This was Briefly discussed in Section~\ref{subsubsec:other-halo-properties}.  

We define formation age (\agefifty) as the redshift at which the halo accreted $50$\% of its present mass. It has been speculated that \agefifty\ is correlated to the amount of \hi\ contained in a halo. \citet{Guo2017ConstrainingClustering} found from their clustering measurements of ALFALFA galaxies, that a way of describing the clustering bias dependence on scale was to assume \hi-rich galaxies to live in preferentially young haloes. Under this assumption, they developed a subhalo abundance matching model (SHAM) which was used to derive a strong correlation between the \hi\ content of the haloes and its \agefifty. A suitable explanation for this effect is the fact that young haloes would be expected to contain \hi-rich galaxies, as they had not had enough time to lose their cold gas via `ram-pressure stripping' or other environmental effect. \citet{Spinelli2019_Marta}, using the semi-analytic model of galaxy formation GAEA, found that in low-mass haloes there was no difference between young and old haloes in terms of their \hi\ content; but as the \subsuperscript{M}{vir}{} increased, a segregation appeared between young and old haloes, with the former being more \hi-rich in agreement with \citet{Guo2017ConstrainingClustering} inferences.

\begin{figure}
\includegraphics[width=\linewidth]{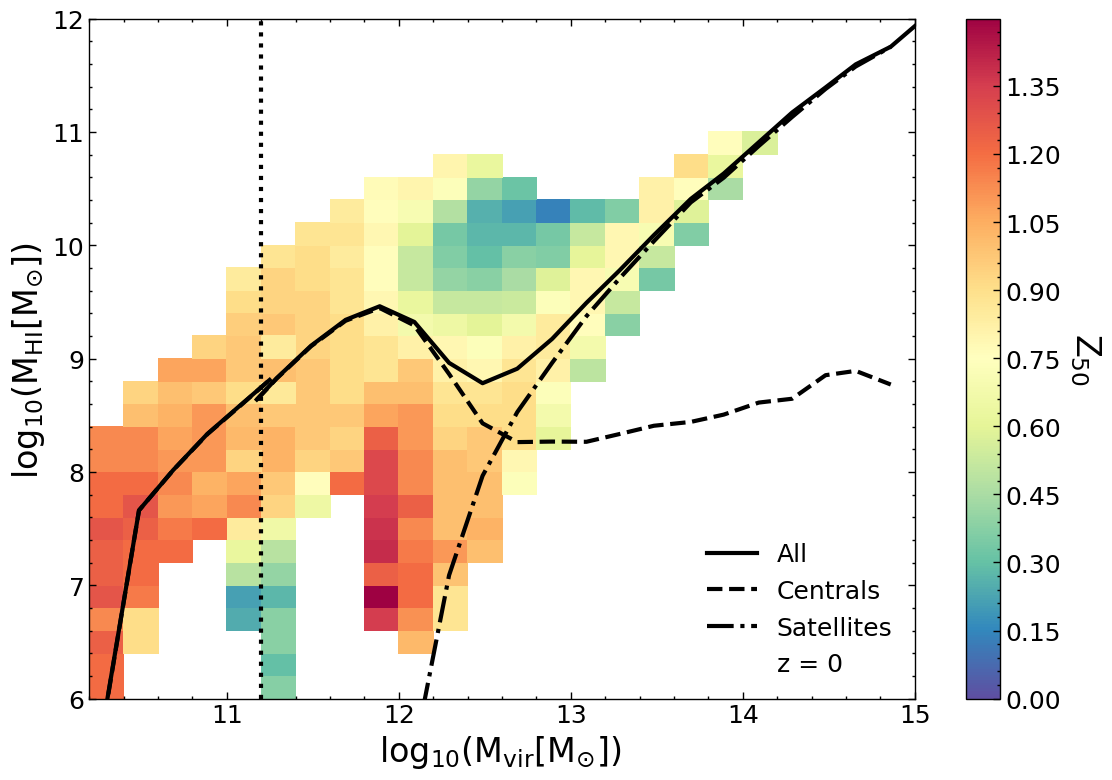} 
\caption{The HIHM relation of haloes in \shark-ref at $z=0$, with each bin being coloured by the median \agefifty\ of the haloes in that bin, as labelled in the colour bar. The solid line represents the median \hi\ mass of the halo as a function of \subsuperscript{M}{vir}{},  while the dashed and dashed-dotted lines represent the central and satellite galaxies contributions, respectively. The vertical dotted line shows the transition from micro-\surfs\ to medi-\surfs\ at lower and higher halo masses, respectively. A slight trend with \agefifty\ is seen at the transition region so that younger haloes tend to be more \hi-rich. This trend reverses though at higher halo masses.}
\label{fig:age_50}
\end{figure}

\begin{figure}
\includegraphics[width=\linewidth]{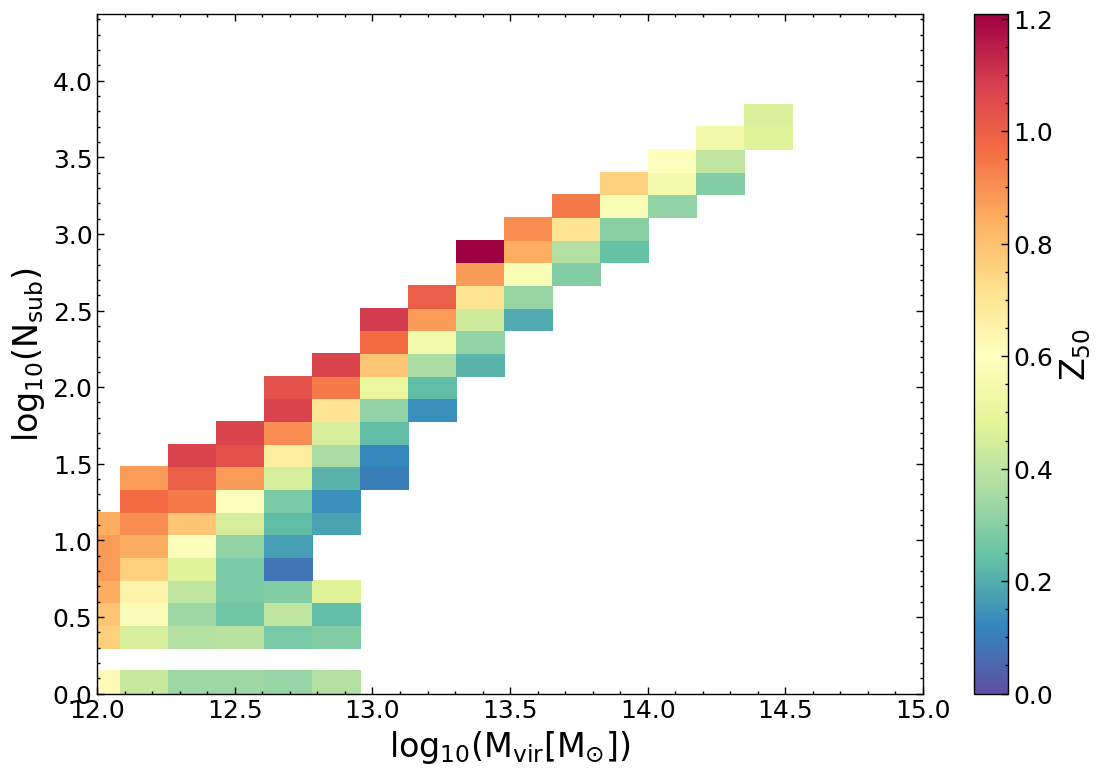} 
\caption{The number of subhaloes in a halo as a function of \subsuperscript{M}{vir}{}, with bins coloured according to the median \agefifty\ of that bin. Older haloes have more substructure than their younger counterparts at fixed halo mass. }
\label{fig:age_50_substructure}
\end{figure}

We test the effect of \agefifty\ here. Figure~\ref{fig:age_50} shows the HIHM relation at $z=0$ colouring each bin by the median formation age of haloes in that bin. We find that in \shark, \agefifty\ does not show a significant trend in the low-mass region, though a slight trend is noticeable in the transition region. We see that younger haloes (closer to $z=0$) tend to have more \hi\ than their counterparts of the same mass. We think the trend emerges here because it is in this region that satellites start to become a more prominent reservoir of \hi\ compared to the central galaxy. We see a slight opposite trend in the high-mass region, where later forming haloes tend to be \hi\ {\it poorer} which contradicts the conclusion in \citet{Guo2017ConstrainingClustering}. This is due to older haloes having on average more substructure and therefore more satellite galaxies at fixed halo mass \citep[see][]{Wechsler2018-galaxyDMhaloes, Croton2007HaloClustering}, which contribute to the total \hi\ content of the halo. We discuss this in more detail in Section~\ref{subsubsec:mvir_ratio_effect}. Figure~\ref{fig:age_50_substructure} shows the relation between the halo mass, \agefifty\ and the number of substructure per halo. We find that haloes formed earlier have more substructure as compared to their younger counterparts in the same mass bin.

Appendix~\ref{appendix:Redshift-trends}  presents the redshift evolution of the \hi--halo mass--\agefifty\ relation up to $z=2$. We find that the trend we see at $z=0$ holds at high redshift with the main difference being the expected lack of massive halos.


\section{Developing Numerical Model to populate dark matter haloes with HI - cont.}
\label{appendix:model-development}

After the brief overview given in Section~\ref{subsec:appendix-section-5}, here we explore a bit more on the evolution of the HIHM relation through different redshifts.

\subsection{Redshift Dependence}
\label{subsec:redshift_dependence}

As noted in Section~\ref{subsec:HI-halo_scaling}, the coefficients for $f_{M_{\rm HI}}(M_{\rm vir})$ are dependent on redshift. We find that as we move towards higher redshifts, the \textit{transition region} shrinks, with the noticeable bump (around \subsuperscript{M}{vir}{} $\simeq$ \solarValue{12}) becoming flatter (see Appendix~\ref{appendix:Redshift-trends}). By the time we reach $z=2$, the \hi--halo scaling relation becomes a monotonically increasing function of \subsuperscript{M}{vir}{}. One of the key reasons behind this outcome is that for higher redshifts AGN feedback is less efficient than at $z=0$ and therefore by $z=2$ AGN feedback does not play a significant role at keeping the halo gas hot and preventing gas cooling and accretion onto galaxies. In addition, as the haloes have not had enough time to assemble all of their mass, they do not have enough substructures yet to contribute to increasing the scatter in the transition region. In short, we find that there are no distinctive regions at high redshifts, i.e. the \emph{transition region} effectively disappears. 

In the low-mass region we find that, while the shape of the median HIHM relation carries a redshift dependence, the fits to the residuals ($\delta_{M_{\rm HI}}$, which captures the scatter) do not. That is to say, for example, the influence that halo spin has on the total \hi~in a halo of fixed virial mass is the same at all epochs.

\begin{figure}
  \includegraphics[width=\linewidth]{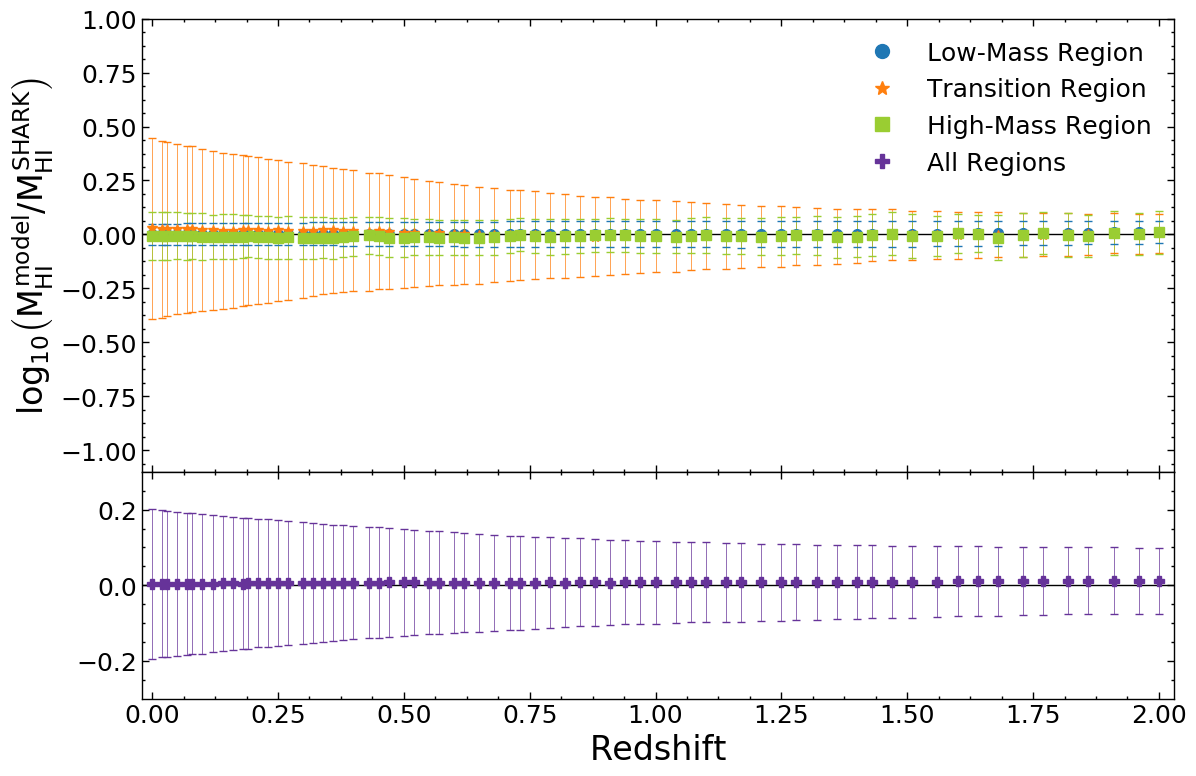}
\caption{The ratio between the true \hi\ masses of haloes in \shark-ref and the derived masses from Equations \ref{eq:low_mvir_coefficients} to \ref{scatterhighmas}, i.e. the \hi-mass residuals, as a function of redshift. Symbols with errorbars show the median and \subsuperscript{16}{}{th}--\subsuperscript{84}{}{th} percentile range. This is presented for all haloes in the simulation (lower panel), and for each halo mass region separately (top panel), as labelled. For reference, the horizontal lines show equality.}

\label{fig:redshift_HI_model}
 \end{figure}

In Figure~\ref{fig:redshift_HI_model}, we compare the \hi\ mass calculated by our model with the intrinsic \hi\ output from \shark-ref, at each snapshot out to $z=2$, to assess the performance of our numerical model. 
We show this for the individual halo mass regions as well as the total halo population (however, by sheer number, the low-mass region dominates the latter). It can be seen that as we move from low to high redshifts, the median of the residuals stays around 0, with small deviations of $\lesssim 0.02$~dex. We also find that the \subsuperscript{16}{}{th}--\subsuperscript{84}{}{th} percentile range decreases as we move to higher redshift. This shows that our numerical model is able to successfully capture the dependence of \hi\ mass on halo properties, within certain limits.


In Appendix~\ref{appendix:Redshift-trends}, we show the how the HIHM relation changes at higher redshifts. We find that the scatter around the median relation significantly changes for the transition region, as we move to higher redshifts, and this can be encapsulated in Equations~\ref{eq:low-appendix} to \ref{eq:high-appendix}. 


\subsection{Cumulative \textsc{Hi}}
\label{subsec:cumulative_HI}

\begin{figure}
  \includegraphics[width=\linewidth]{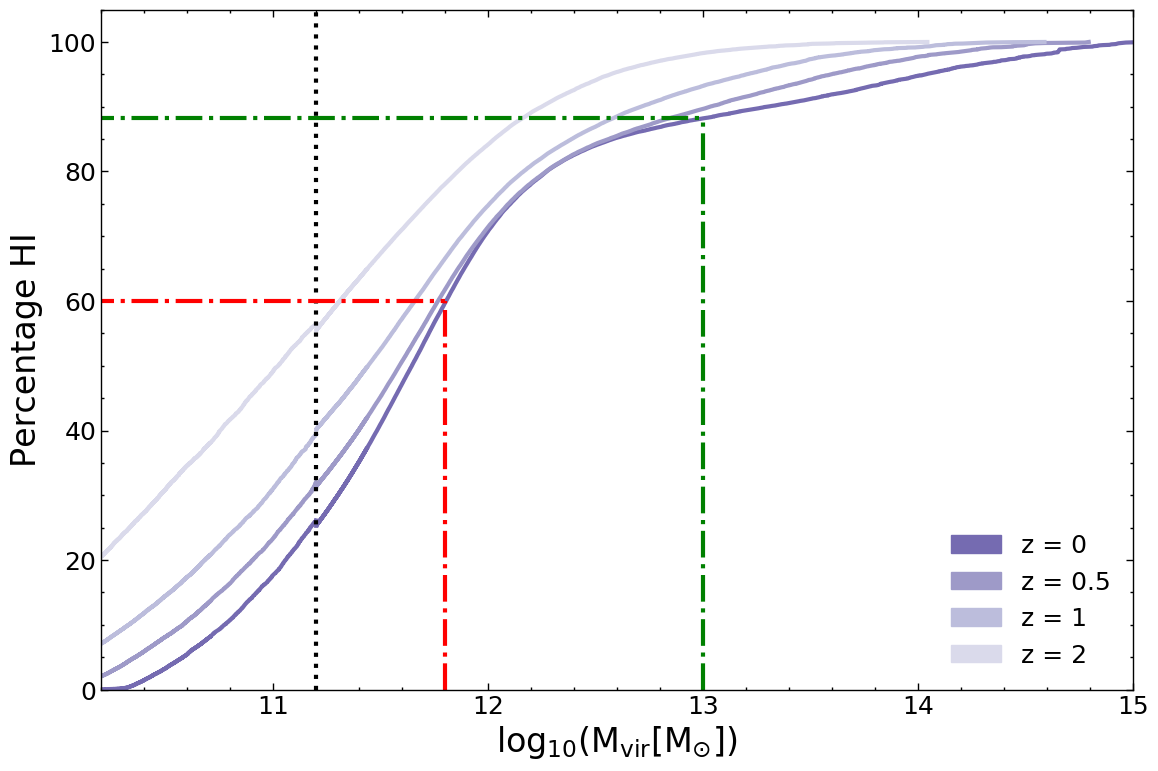}
\caption{The cumulative fraction of cosmic \hi\ mass contained in haloes as a function of virial mass at four different redshifts, as labelled. At $z=0$, $\sim 60$\% of the \hi\ is contained in haloes with $M_{\rm vir} < 10^{12}$\M, with about $\sim 25$\% lying in the transition region of $10^{12} M_{\odot} \leq M_{\rm vir} < 10^{13} M_{\odot}$ and the rest in haloes with $M_{\rm vir} > 10^{13}$\M. For reference, these halo mass thresholds are shown with dot-dashed lines. At higher redshift, the contribution from the lower-mass region becomes even greater.}

\label{fig:cumulative-HI}
 \end{figure}

As seen in Figure~\ref{fig:model-fit-all}, the scatter is well constrained for the low- and high-mass regions, by invoking secondary parameters (the halo's spin parameter and \fracmvir, respectively) but at the transition region we find this to be more difficult. It is therefore informative to ask how much of the total \hi\ in \shark-ref resides in the transition region. In Figure~\ref{fig:cumulative-HI}, we plot the cumulative \hi\ mass as a function of halo mass in \shark-ref. We find that at $z=0$ $\sim 60$\% of the \hi\ is contained in haloes with $M_{\rm vir} < 10^{11.8}$\M, with about $\sim 25$\% lying in the transition region of $10^{11.8}\,{\rm M}_{\odot} \leq M_{\rm vir} < 10^{13}\,{\rm M}_{\odot}$. The rest, $\sim 15$\%, is in haloes with masses $M_{\rm vir} > 10^{13}$\,\M.

As we move to higher redshifts, we find that the low-mass region becomes more important, with contributions that increase from $60$\% at $z=0$ to $80$\% at $z=2$. 
 
This shows that, even if our numerical model is less reliable around the transition region, the majority of \hi\ lies in regions that are very well modelled by our numerical method. This is particularly important in, for example, \hi\ stacking or intensity mapping experiments, when the relevant quantity is the aggregated \hi\ mass at a given redshift.


\section{Redshift dependence of the HIHM relation}
\label{appendix:Redshift-trends}

\begin{figure}
\centering
\subfloat{
    \includegraphics[width=0.5\textwidth]{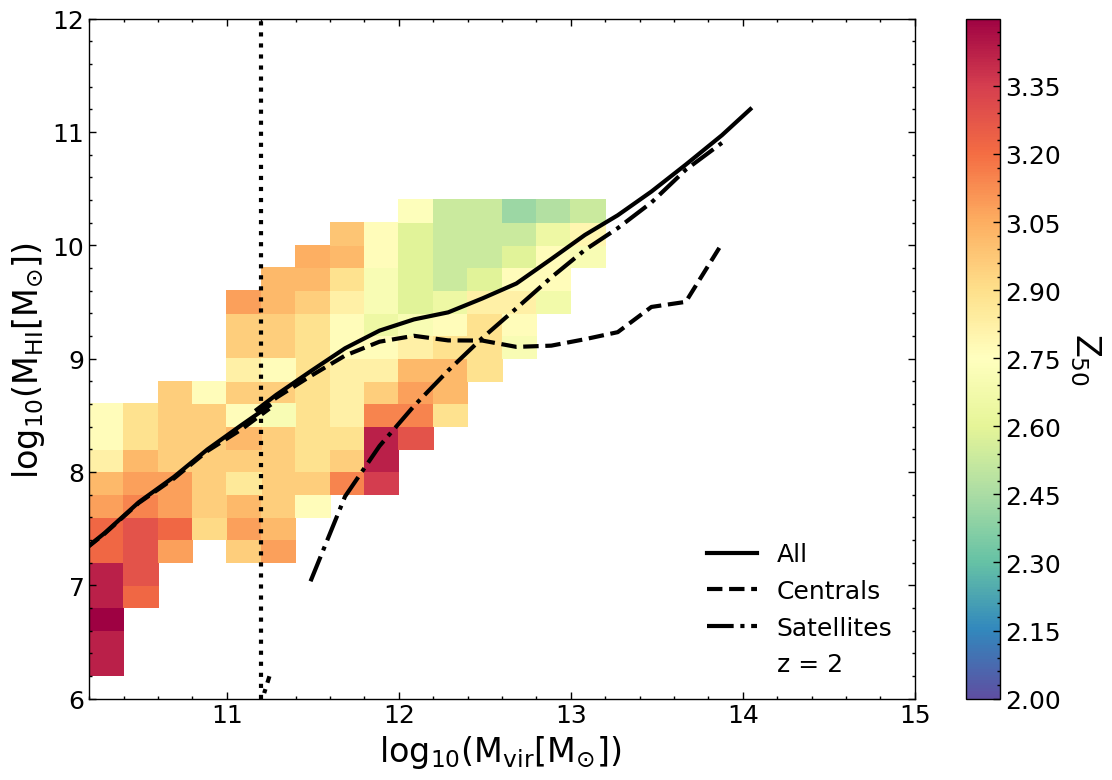} }\\[-3.1ex]
\subfloat{
    \includegraphics[width=0.5\textwidth]{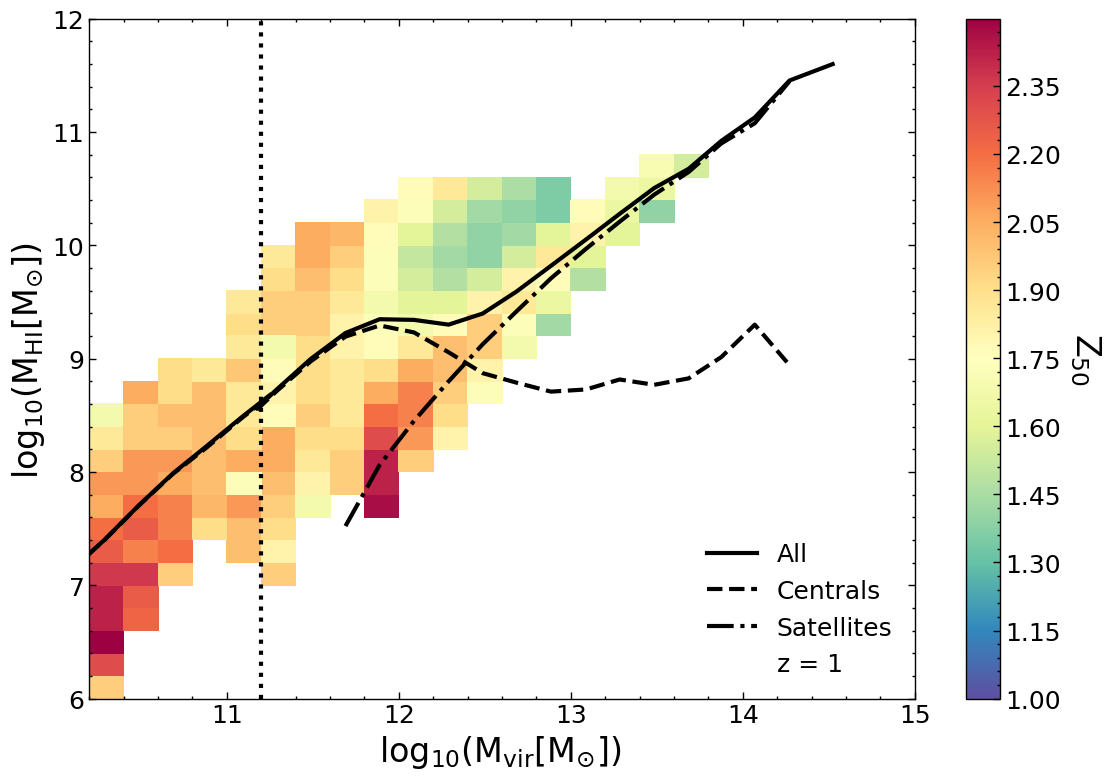} }\\[-3.1ex] 
\subfloat{
    \includegraphics[width=0.5\textwidth]{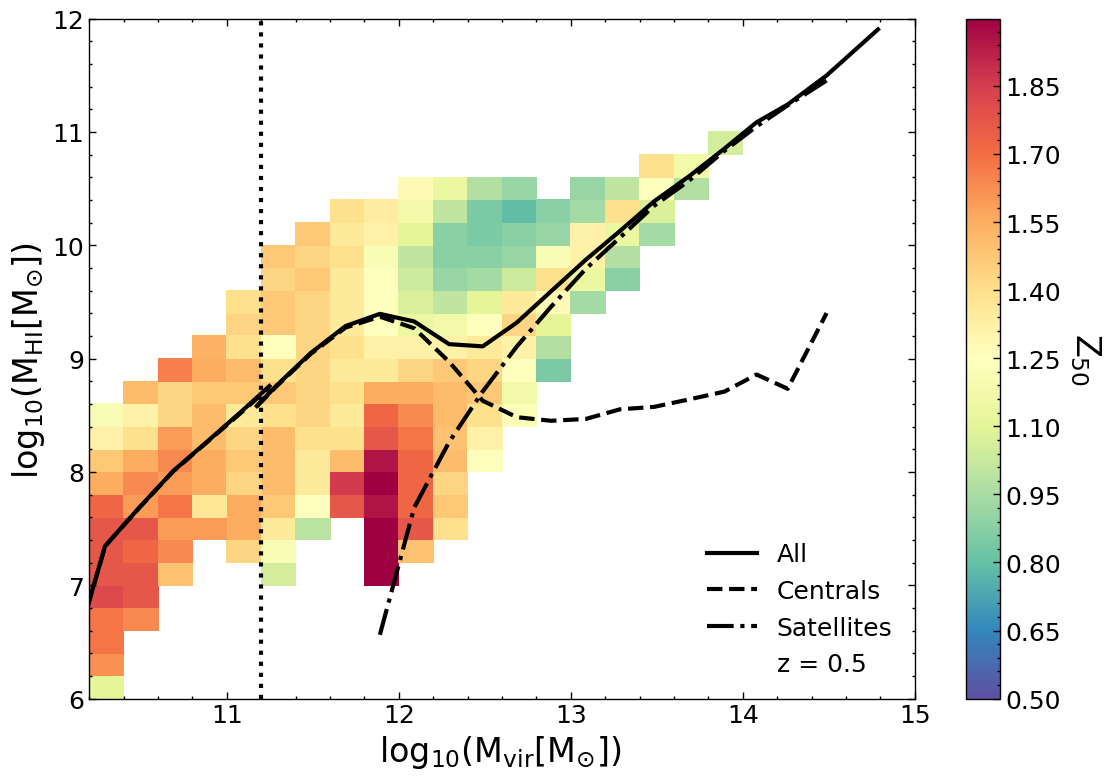}} 
\caption{The HIHM relation of haloes in \shark-ref at $z=0.5,\ 1$ and $2$, with each bin being coloured by the median \agefifty\ of the haloes in that bin, as labelled in the colour bar. The solid line represents the median \hi\ mass of the halo as a function of \subsuperscript{M}{vir}{},  while the dashed and dashed-dotted lines represent the central and satellite galaxies contributions, respectively. The vertical dotted line shows the transition from micro-\surfs\ to medi-\surfs\ at lower and higher halo masses, respectively. Though not much can be seen, there is a slight trend with the younger formed haloes being more \hi\  rich than the older ones.}
\label{fig:age_50-appendix}
\end{figure}

\begin{figure}
\centering
\subfloat{
    \includegraphics[width=0.5\textwidth]{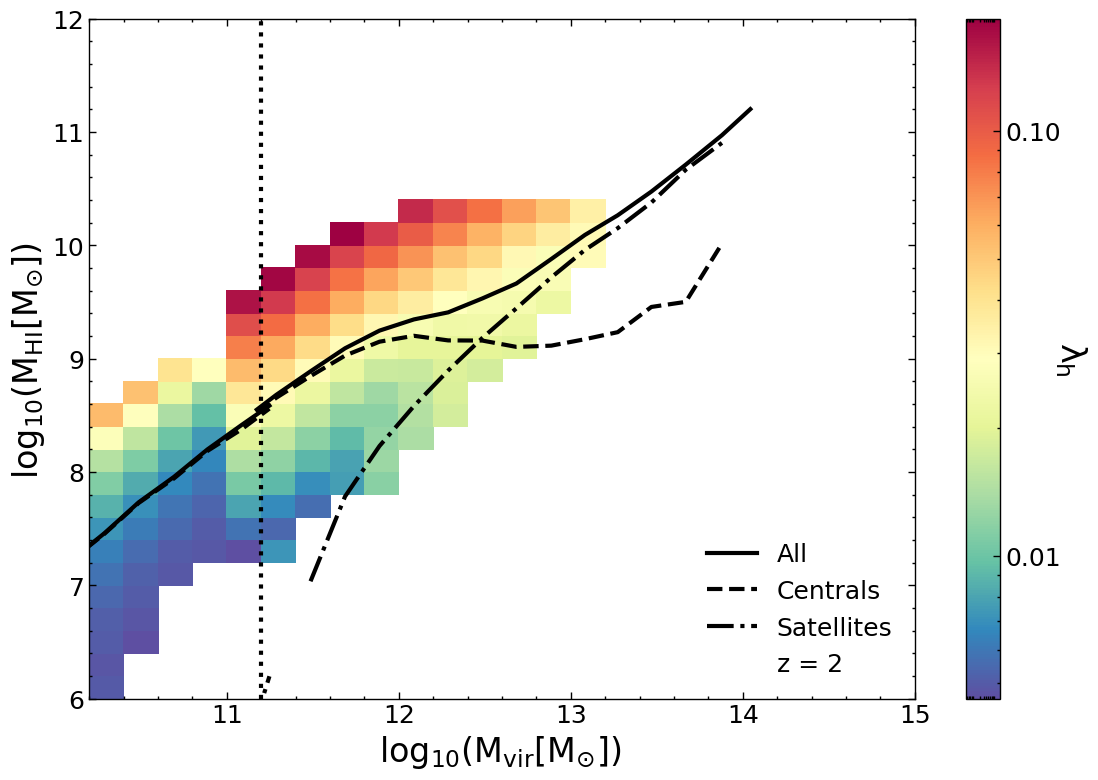} }\\[-3.1ex]
\subfloat{
    \includegraphics[width=0.5\textwidth]{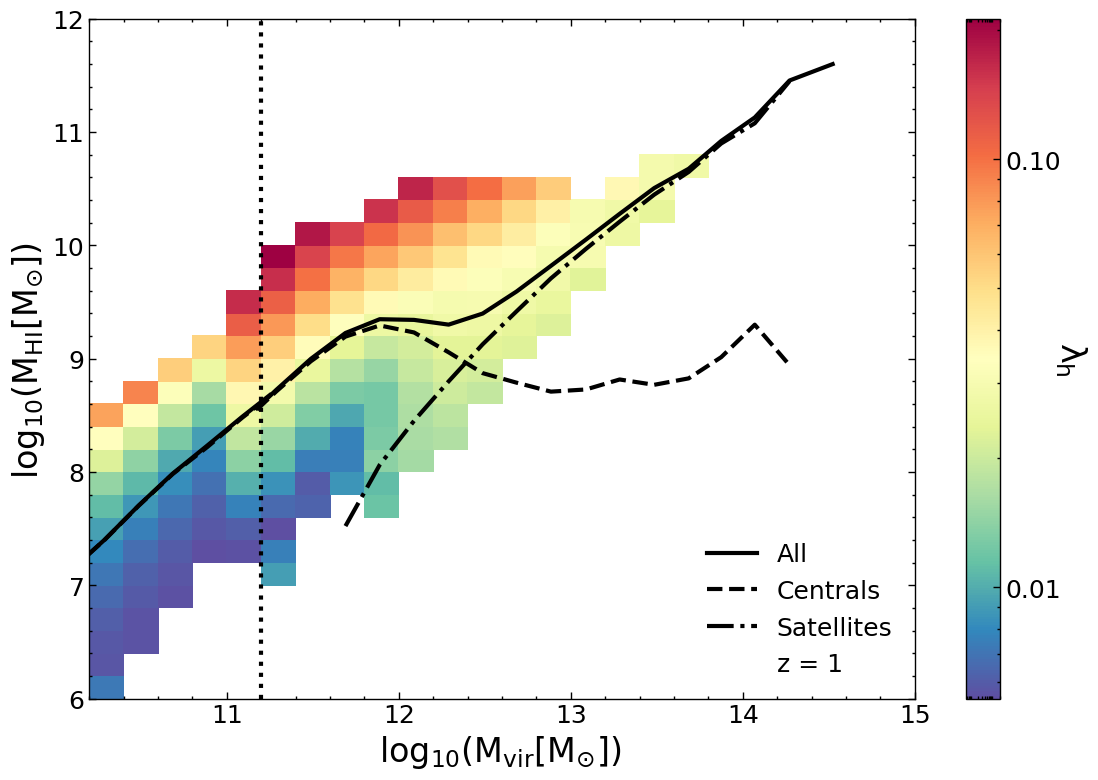} }\\[-3.1ex] 
\subfloat{
    \includegraphics[width=0.5\textwidth]{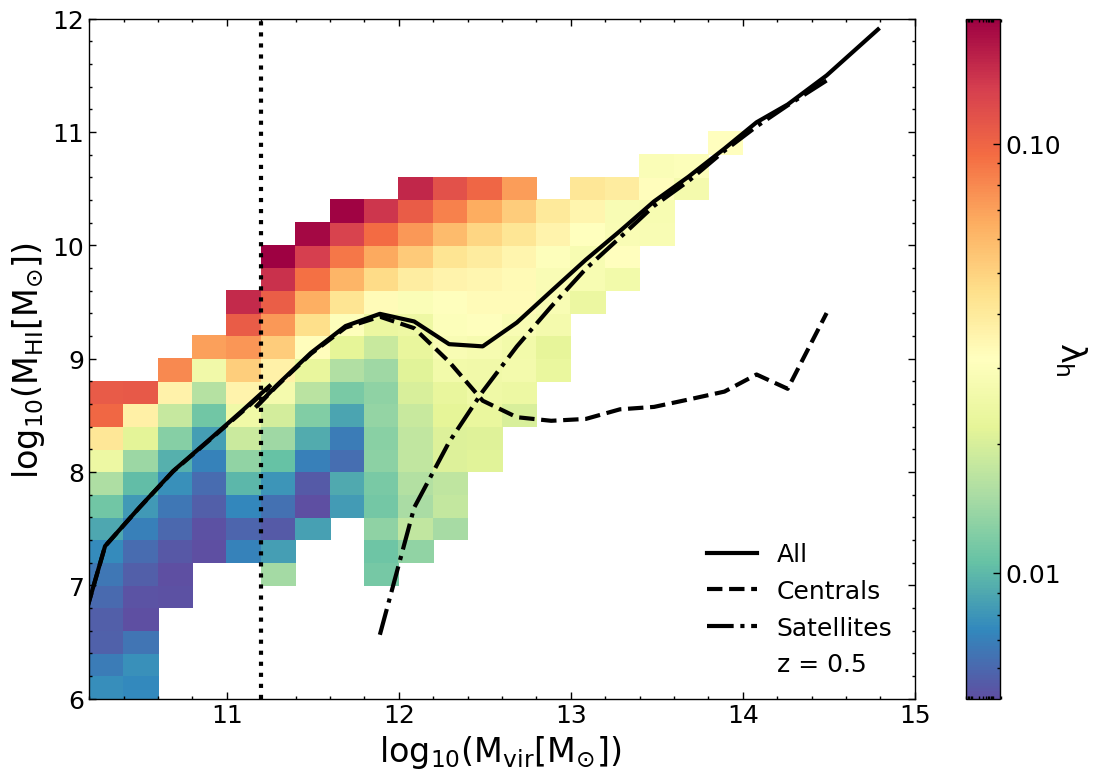}} 
\caption{As in Figure~\ref{fig:age_50-appendix} but here bins are coloured by the median halo's spin parameter, as labelled in the colour bar. There is a strong correlation between the \hi\ mass and the spin parameter at fixed halo mass for haloes with $M_{\rm vir} < 10^{12}\,\rm M_{\odot}$ at $z=0.5$, with the $M_{\rm vir}$ threshold being \solarValue{12.5}\ and $\sim$ \solarValue{13} for $z=1$ and $2$, respectively. Haloes with higher spin parameters are \hi-richer than their counterparts.}
\label{fig:spin parameter-appendix}
\end{figure}

\begin{figure}
\centering
\subfloat{
    \includegraphics[width=0.5\textwidth]{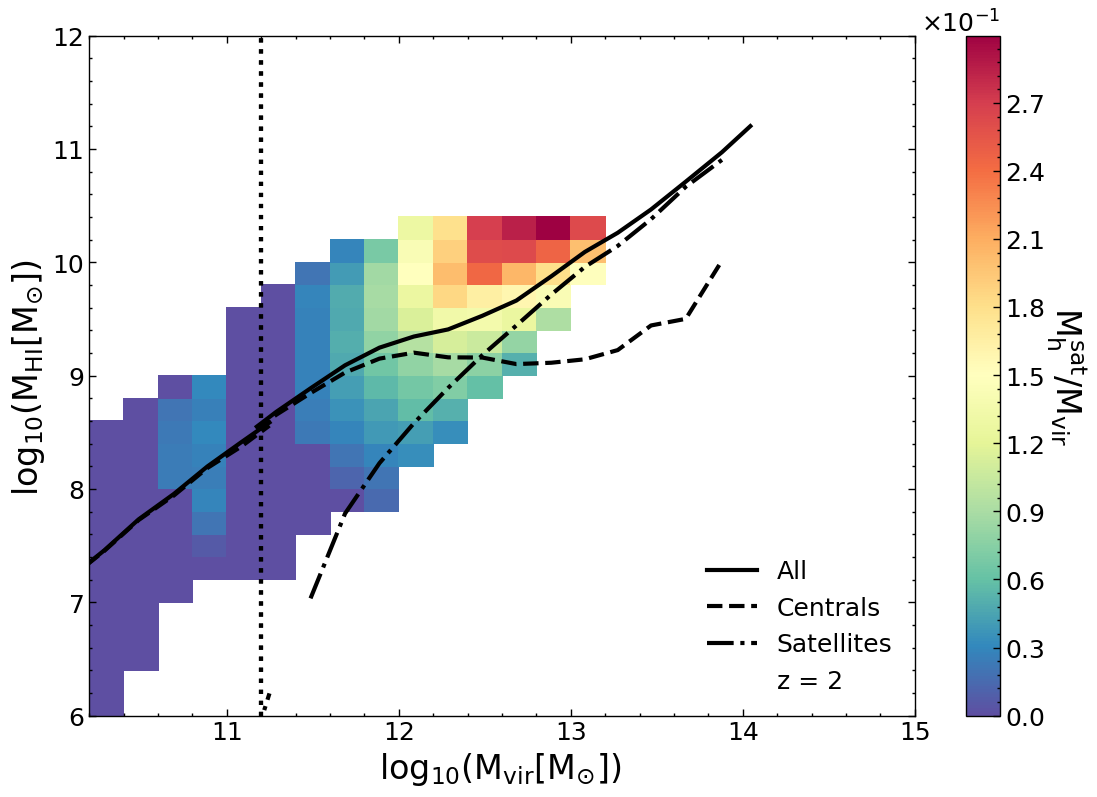} }\\[-3.1ex]
\subfloat{
    \includegraphics[width=0.5\textwidth]{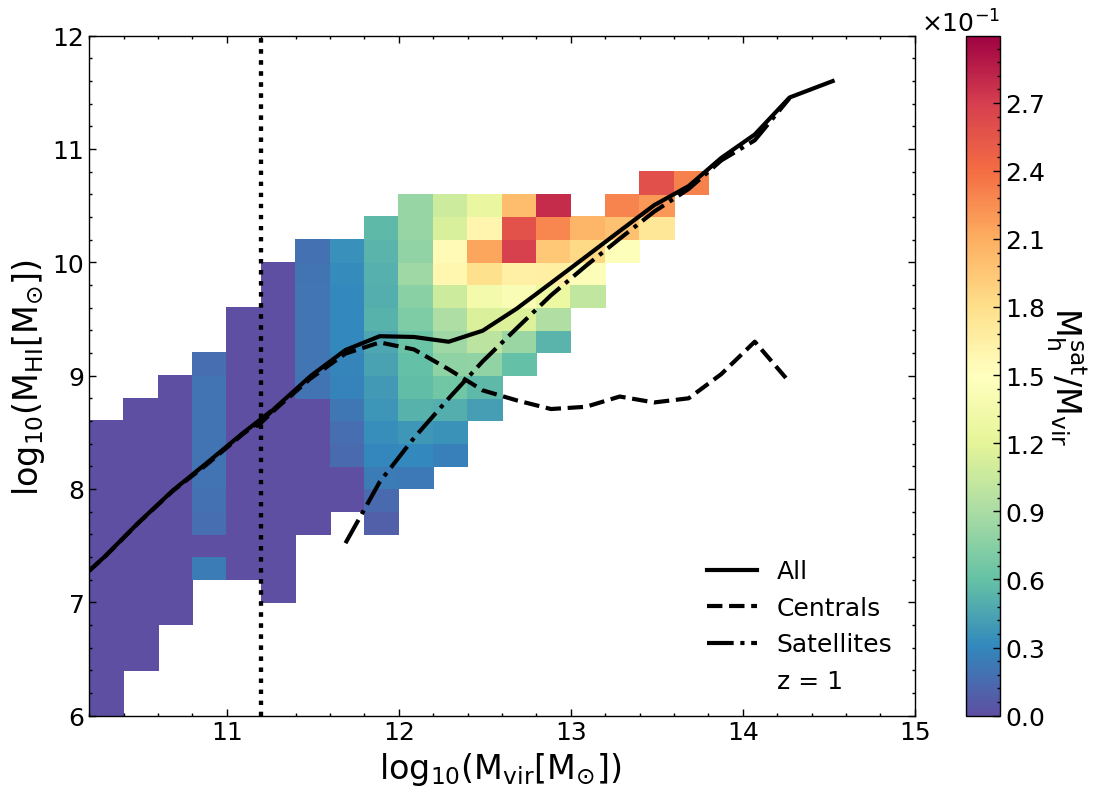} }\\[-3.1ex] 
\subfloat{
    \includegraphics[width=0.5\textwidth]{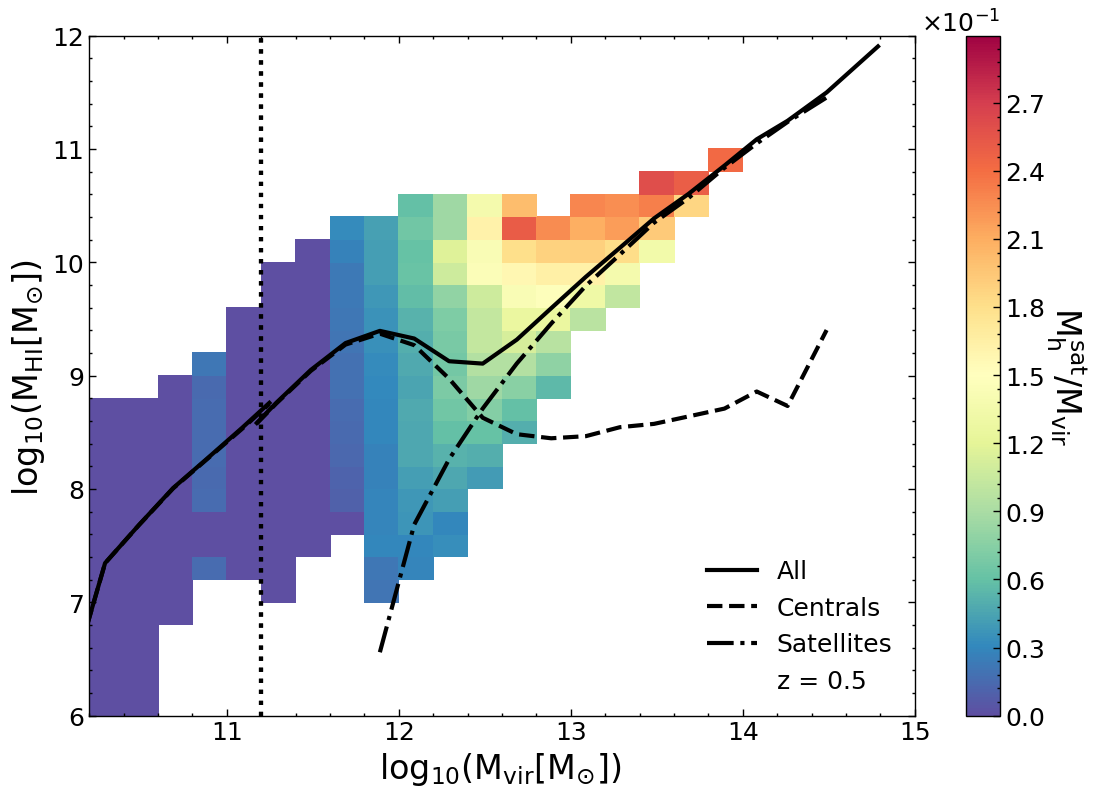}} 
\caption{Similar to the earlier plots (see Figure~\ref{fig:age_50-appendix} and \ref{fig:spin parameter-appendix}), here the contribution of \hi\ contained in satellites to the total \hi\ in the halo containing them. As we reach to higher virial masses, we can see that satellites contain most of the \hi\ in the haloes, irrespective to which redshift it is being observed at.}
\label{fig:fraction_satellite-appendix}
\end{figure}

As stated in Section \ref{subsec:redshift_dependence}, as we move to higher redshifts we find the \textit{Transition Region} getting noticeably smaller in dynamic range, with the scatter around the relation decreasing as well. We have showed earlier in Section \ref{sec:Model-development}, that the residual fits for the HIHM relation are redshift dependent, though the halo properties comprising the residual fits remain the same through out the redshift range in consideration.

Figure \ref{fig:age_50-appendix} shows the \subsuperscript{M}{HI}{}--\subsuperscript{M}{vir}{} relation at $z=0.5,\ 1$ and $2$, colouring each bin with the median formation age. 
As had been seen for the $z=0$ case, \agefifty\ does not show a very strong trend at low mass region, though a slight trend is noticeable in transition and higher mass regions, with  (relatively) younger haloes having higher \hi\ than their older counterparts, throughout the redshift range in consideration. 

As we move towards halo spin parameter in Figure \ref{fig:spin parameter-appendix}, we find that the spin parameter is strongly correlated with the scatter of low-mass region in the HIHM relation. One interesting aspect of the correlation seen is that as we move to higher redshifts, we find the spin parameter correlation extending to higher halo masses than seen in the lower redshift range. As opposed to halo spin parameter showing strong correlation with haloes of masses \subsuperscript{M}{vir}{} $<$ \solarValue{12} at $z=0$, we find the correlation goes as far as halo mass range of \subsuperscript{M}{vir}{} $<$ \solarValue{13} at $z=2$. This is in agreement to our assessment that as we move to higher redshifts, the \textit{Transition Region} gets smaller and move towards higher halo masses. 

Similar to Figures \ref{fig:age_50-appendix} and \ref{fig:spin parameter-appendix}, when we look at the evolution of the \fracmvir trend with redshift in Figure \ref{fig:fraction_satellite-appendix}, we find it more or less similar to what was seen at $z=0$: the higher the value of \fracmvir\ the  higher the \hi\ mass in the halo. This is due to the fact that, as we move to higher halo masses, the number of satellites in those haloes increases, and thus does the total \hi\ contribution of the satellites. 

The evolution of the scatter around the HIHM relation, especially for the \textit{transition region}, through the redshift points to the fact that the flaring of scatter in the transition region at $z=0$ can be related to the AGN feedback efficiency adopted by the model. As we go higher in redshift, AGN feedback becomes less important leading to a decrease in the scatter around the transition region. This effect is also evident in the noticeable bump that is prominent in the $z=0$ and $0.5$, is smoothed out by the time we reach $z=2$.  

Therefore, from Figures \ref{fig:age_50-appendix}, \ref{fig:spin parameter-appendix} and \ref{fig:fraction_satellite-appendix}, it is clear that the trends of $z=0$ persist towards at higher redshifts, which means that we can use the same secondary parameters to fit the scatter around the HIHM relation at different redshifts. 


\newpage
\section{Parameter Fits}
\label{appendix:Spline-Fits}

In Section \ref{subsect:middlemass}, we pointed out that the dependence of the median relation parameters of the quintic polynomial fit for the transition region is hard to parametrise as a function of redshift, and thus we tabulate the coefficients in Table \ref{tab:transition-parameters}. The equation for estimating the \hi\ in the transition region is as follows:  
\begin{equation}
    f_{\rm M_{\rm HI}}(M_{\rm vir},z) = 9 + \sum_{i=0}^{n}\ a_{i}(z) \left(\log_{10}(M_{\rm vir}) - 11.8\right)^{i}.
    \label{eq:transition_fit_parameters}
\end{equation}
where $n=5$, with \subsuperscript{a}{1}{TR}$=0$.

\begin{figure}
\centering
\subfloat{
    \includegraphics[width=0.5\textwidth]{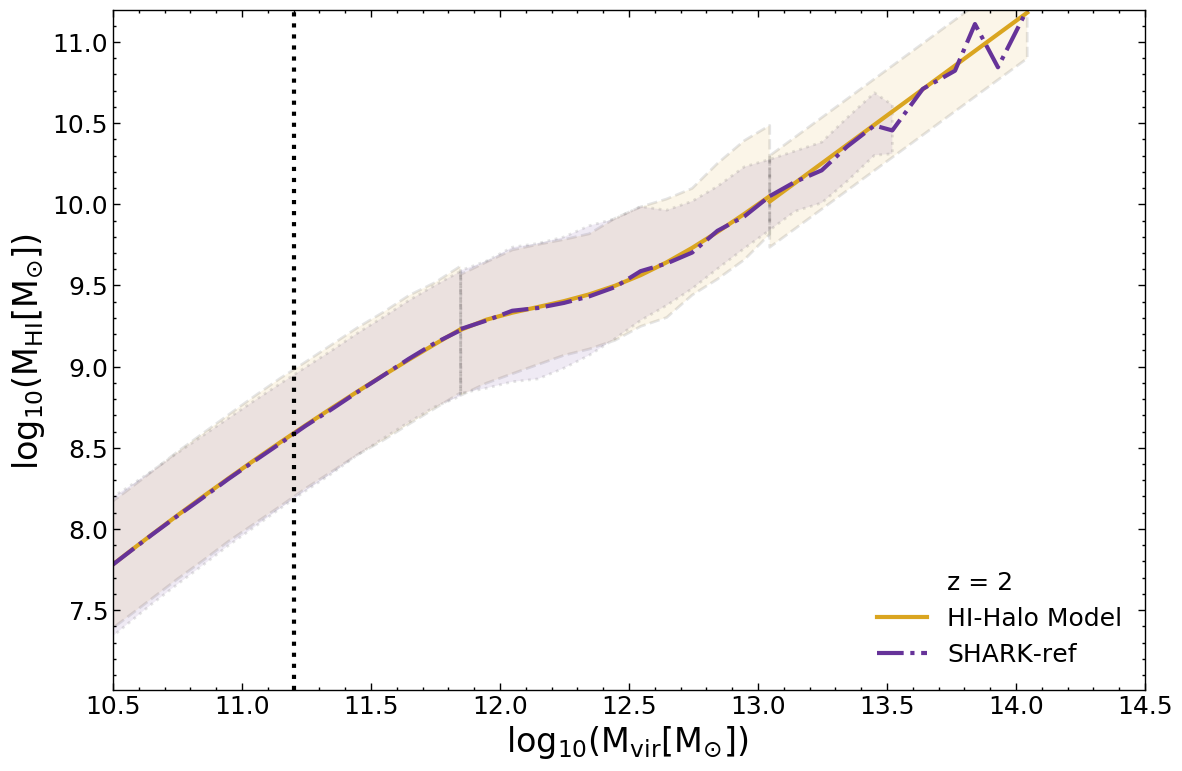} }\\[-3.1ex]
\subfloat{
    \includegraphics[width=0.5\textwidth]{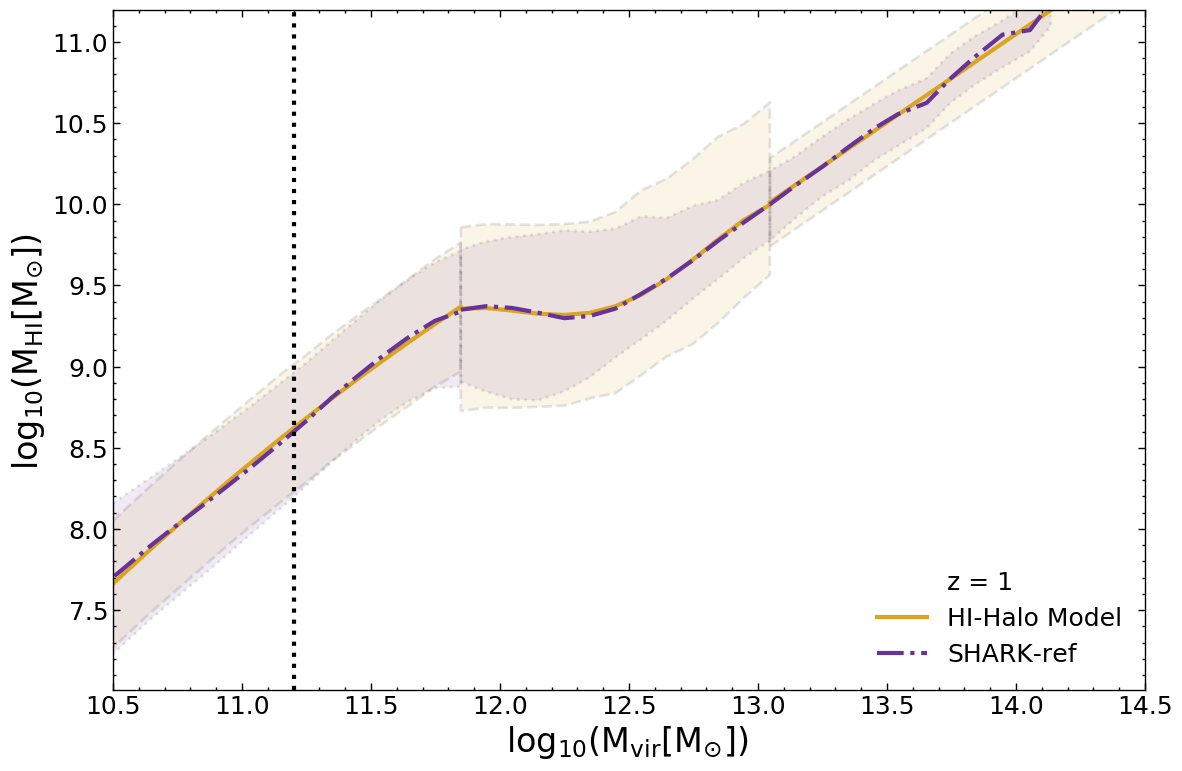} }\\[-3.1ex] 
\subfloat{
    \includegraphics[width=0.5\textwidth]{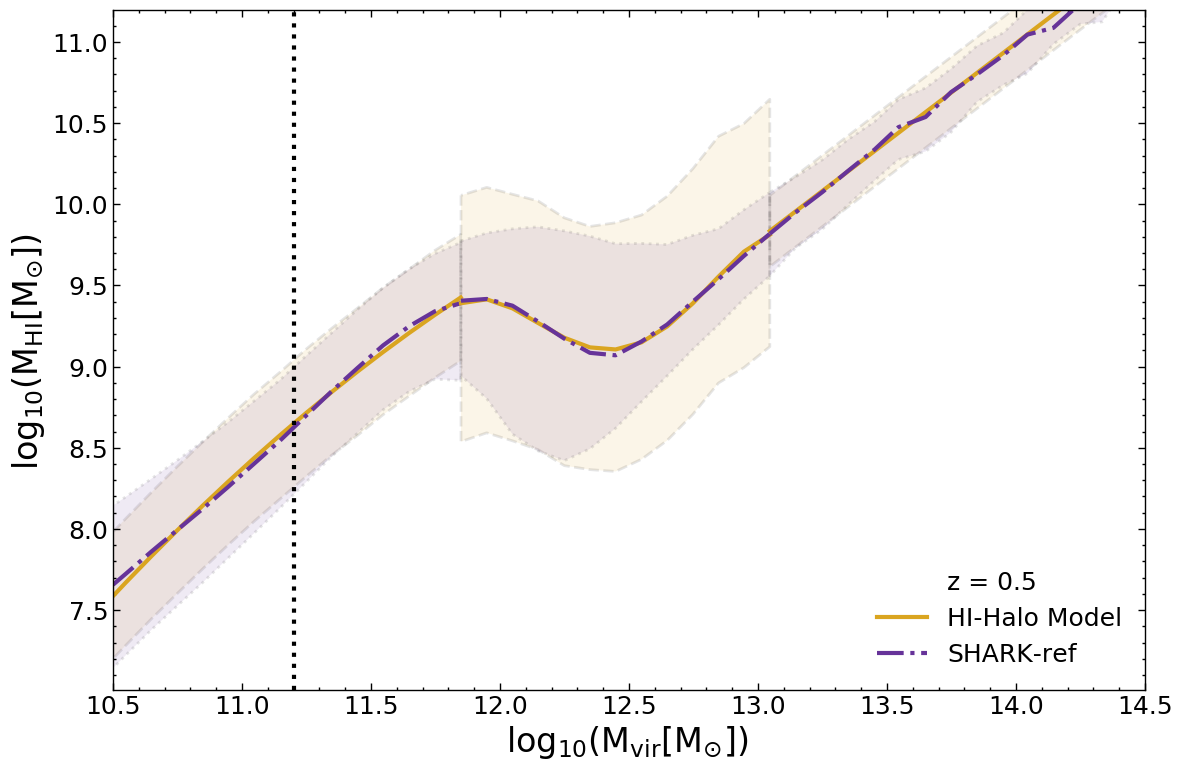}} 
\caption{Overall \hi\ content of haloes as a function of halo mass for \shark-ref (dot-dashed line), and predicted by our numerical model (solid line) at $z=0.5,\ 1$ and $2$, as labelled. The shaded regions represent the \subsuperscript{16}{}{th}--\subsuperscript{84}{}{th} percentile ranges of the distributions. A decrease in the scatter around the transition region is seen as we move towards higher redshifts.}
\label{fig:model-HI-appendix}
\end{figure}

Figure~\ref{fig:model-HI-appendix} compares the true \hi\ content of \shark-ref haloes at $z=2$, $z=1$ and $z=0.5$ with the outcome of applying our numerical HIHM scaling relation to the same underlying halo population (see Equations~\ref{eq:2-part-equation} to \ref{scatterhighmas}).
Figure~\ref{fig:model-HI-appendix} showcases that as we move towards higher redshift the scatter around the \hi\ relation decreases considerably for the transition region, and the shape also evolves into a monotonically increasing relation by the time we reach $z=2$. 

We find that the vertical scatter around the HIHM relation obtained from our numerical model decreases in a similar manner, and can be described by the following functions of redshift, with parameters that depend on the mass region,
\begin{equation}
    \sigma_{\rm low} = 0.189 - 0.017 z,
    \label{eq:low-appendix}
\end{equation}
\begin{equation}
    \sigma_{\rm TR} = 0.138 + 0.771e^{-z},
    \label{eq:TR-appendix}
\end{equation}
\begin{equation}
    \sigma_{\rm high} = 0.185 + 0.142e^{-z},
    \label{eq:high-appendix}
\end{equation}
with $z$ being the redshift. Here, ``low'', ''TR'' and ``high'' refer to the low-mass, transition and high-mass regions, respectively. This also shows that our numerical model becomes more reliable in the transition region as the redshift increases.


\begin{table}

\centering
\caption{Parameters for the quintic polynomial fit for the transition region.}
\begin{tabular}{|l| @{\extracolsep{\fill}} l|l|l|l|l|}
\hline

z & $a^{\rm TR}_{0}$ & $a^{\rm TR}_{2}$ & $a^{\rm TR}_{3}$ & $a^{\rm TR}_{4}$ & $a^{\rm TR}_{5}$\\

\hline
2         & 0.247   & 2.033    & -4.390  & 4.303   & -1.409   \\
1.96      & 0.249   & 1.810    & -3.651  & 3.517   & -1.140   \\
1.91      & 0.260   & 1.319    & -2.065  & 1.758   & -0.496   \\
1.86      & 0.261   & 1.448    & -2.718  & 2.604   & -0.826   \\
1.77      & 0.269   & 1.393    & -2.848  & 3.008   & -1.054   \\
1.73      & 0.275   & 1.180    & -2.365  & 2.641   & -0.962   \\
1.68      & 0.286   & 0.768    & -1.120  & 1.303   & -0.476   \\
1.64      & 0.297   & 0.454    & -0.309  & 0.493   & -0.183   \\
1.6       & 0.304   & 0.235    & 0.170   & 0.168   & -0.117   \\
1.56      & 0.311   & 0.072    & 0.448   & 0.029   & -0.103   \\
1.51      & 0.316   & -0.237   & 1.390   & -0.908  & 0.198    \\
1.43      & 0.328   & -0.376   & 1.411   & -0.582  & -0.017   \\
1.4       & 0.329   & -0.228   & 0.683   & 0.394   & -0.415   \\
1.36      & 0.331   & -0.155   & 0.265   & 0.930   & -0.615   \\
1.32      & 0.335   & -0.023   & -0.363  & 1.675   & -0.876   \\
1.28      & 0.338   & -0.112   & -0.127  & 1.436   & -0.791   \\
1.21      & 0.347   & -0.635   & 1.419   & -0.158  & -0.237   \\
1.17      & 0.354   & -0.763   & 1.753   & -0.536  & -0.080   \\
1.14      & 0.357   & -0.796   & 1.725   & -0.426  & -0.138   \\
1.1       & 0.369   & -1.179   & 2.698   & -1.265  & 0.101    \\
1.07      & 0.370   & -1.169   & 2.527   & -0.984  & -0.020   \\
1         & 0.376   & -1.128   & 1.968   & -0.080  & -0.413   \\
0.97      & 0.380   & -1.332   & 2.515   & -0.598  & -0.246   \\
0.94      & 0.386   & -1.410   & 2.483   & -0.415  & -0.335   \\
0.91      & 0.389   & -1.606   & 2.973   & -0.868  & -0.189   \\
0.88      & 0.388   & -1.479   & 2.227   & 0.162   & -0.603   \\
0.85      & 0.393   & -1.666   & 2.628   & -0.170  & -0.500   \\
0.82      & 0.398   & -1.852   & 2.961   & -0.399  & -0.441   \\
0.79      & 0.403   & -1.862   & 2.682   & 0.075   & -0.643   \\
0.76      & 0.408   & -2.031   & 2.939   & -0.051  & -0.625   \\
0.73      & 0.411   & -2.108   & 2.989   & -0.005  & -0.661   \\
0.71      & 0.412   & -2.037   & 2.434   & 0.782   & -0.976   \\
0.68      & 0.417   & -2.144   & 2.568   & 0.711   & -0.954   \\
0.65      & 0.424   & -2.457   & 3.239   & 0.202   & -0.823   \\
0.62      & 0.428   & -2.428   & 2.926   & 0.611   & -0.970   \\
0.6       & 0.434   & -2.317   & 2.216   & 1.536   & -1.317   \\
0.57      & 0.438   & -2.307   & 1.885   & 2.024   & -1.508   \\
0.55      & 0.442   & -2.393   & 1.881   & 2.166   & -1.580   \\
0.52      & 0.442   & -2.144   & 0.902   & 3.222   & -1.932   \\
0.5       & 0.442   & -1.976   & 0.071   & 4.227   & -2.292   \\
0.47      & 0.448   & -2.020   & -0.083  & 4.526   & -2.416   \\
0.45      & 0.451   & -2.040   & -0.175  & 4.677   & -2.475   \\
0.43      & 0.455   & -2.059   & -0.333  & 4.918   & -2.565   \\
0.4       & 0.460   & -2.223   & 0.039   & 4.573   & -2.448   \\
0.38      & 0.466   & -2.177   & -0.460  & 5.301   & -2.742   \\
0.36      & 0.466   & -2.233   & -0.294  & 5.070   & -2.640   \\
0.34      & 0.473   & -2.352   & -0.178  & 5.045   & -2.642   \\
0.32      & 0.476   & -2.309   & -0.602  & 5.616   & -2.855   \\
0.3       & 0.476   & -2.214   & -1.053  & 6.138   & -3.039   \\
0.27      & 0.479   & -2.395   & -0.664  & 5.812   & -2.942   \\
0.25      & 0.481   & -2.364   & -0.927  & 6.117   & -3.039   \\
0.23      & 0.480   & -2.152   & -1.763  & 7.053   & -3.368   \\
0.21      & 0.486   & -2.135   & -1.968  & 7.281   & -3.435   \\
0.19      & 0.489   & -2.004   & -2.542  & 7.935   & -3.665   \\
0.18      & 0.490   & -1.951   & -2.868  & 8.322   & -3.798   \\
0.16      & 0.492   & -2.209   & -2.055  & 7.386   & -3.441   \\
0.14      & 0.493   & -1.922   & -3.068  & 8.454   & -3.801   \\
0.12      & 0.495   & -1.711   & -4.016  & 9.611   & -4.235   \\
0.1       & 0.501   & -1.919   & -3.551  & 9.211   & -4.111   \\
0.08      & 0.506   & -2.079   & -3.089  & 8.683   & -3.909   \\
0.07      & 0.508   & -1.973   & -3.669  & 9.425   & -4.192   \\
0.05      & 0.514   & -2.043   & -3.732  & 9.636   & -4.293   \\
0.03      & 0.517   & -2.372   & -2.641  & 8.373   & -3.818   \\
0.02      & 0.518   & -2.231   & -3.149  & 8.923   & -4.013   \\
0         & 0.524   & -2.529   & -2.355  & 8.173   & -3.774   \\ 
\hline
\label{tab:transition-parameters}
\end{tabular}
\end{table}

\bsp    
\label{lastpage}
\end{document}